\documentclass[letterpaper,11pt]{article}

\usepackage[linktocpage=true]{hyperref}
\usepackage{amsmath}
 \usepackage{cite}

\usepackage[normalem]{ulem}
\usepackage{xspace}
\usepackage{slashed}
\usepackage[section]{placeins}
\usepackage{xcolor}
\usepackage{footmisc}
\usepackage{subcaption}

\usepackage{graphicx,color}  %
\usepackage{bm}  %
\usepackage{amssymb, graphics, amsmath}
\usepackage{multirow}

\newcommand{\be}{\begin{equation}}
\newcommand{\ee}{\end{equation}}
\newcommand{\bea}{\begin{eqnarray}}
\newcommand{\eea}{\end{eqnarray}}
\newcommand{\ben}{\begin{eqnarray*}}
\newcommand{\een}{\end{eqnarray*}}

\newcommand{\bma}{\begin{pmatrix}}
\newcommand{\ema}{\end{pmatrix}}

\renewcommand{\Im}[1]{{{\rm Im\,} #1}} 

\def\lixo#1{}

\def\slashchar#1{\setbox0=\hbox{$#1$}           
  \dimen0=\wd0                                    
  \setbox1=\hbox{/} \dimen1=\wd1                  
  \ifdim\dimen0>\dimen1                           
    \rlap{\hbox to \dimen0{\hfil/\hfil}}            
    #1                                             
  \else                                          
    \rlap{\hbox to \dimen1{\hfil$#1$\hfil}}        
    /                                           
 \fi}                                           %

\newcommand{\sq}{^{2}}

\newcommand{\nn}{\nonumber}

\newcommand{\tev}{~\text{TeV}}
\newcommand{\gev}{~\text{GeV}}
\newcommand{\ev}{~\text{eV}}
\newcommand{\mev}{~\text{MeV}}

\newcommand{\fbi}{~\text{fb}^{-1}}

\newcommand{\onbb}{0\nu\beta\beta}
\newcommand{\mvmin}{m_{\nu\text{min}}}
\newcommand{\mnmax}{m_{N\text{max}}}

\begin{document}

\begin{titlepage}

\begin{flushright}
ACFI-T22-08
\end{flushright}

\vspace{2.0cm}

\begin{center}

{\LARGE  \bf 
Light Sterile Neutrinos, Left-Right Symmetry, and $0\nu\beta\beta$ Decay
}

\vspace{2cm}

{ \bf Jordy de Vries$^{a,b}$, Gang Li$^{c,d}$, Michael J. Ramsey-Musolf$^{e,d,f}$, Juan Carlos Vasquez$^{g,d}$} 
\vspace{0.5cm}

\vspace{0.25cm}
{\large 
$^a$ 
{\it Institute for Theoretical Physics Amsterdam and Delta Institute for Theoretical Physics, University of Amsterdam, Science Park 904, 1098 XH Amsterdam, The Netherlands.}}

{\large 
$^b$ 
{\it Nikhef, Theory Group, Science Park 105, 1098 XG, Amsterdam, The Netherlands.}}

{\large 
$^c$ 
{\it School of Physics and Astronomy, Sun Yat-sen University, Zhuhai 519082, P.R. China.}}

{\large 
$^d$ 
{\it 
Amherst Center for Fundamental Interactions, Department of Physics, University of Massachusetts, Amherst, MA 01003, USA.}
}

{\large 
$^e$ 
{\it 
Tsung-Dao Lee Institute and School of Physics and Astronomy,
Shanghai Jiao Tong University, 800 Dongchuan Road, Shanghai, 200240 China.}
}

{\large 
$^f$ 
{\it 
Kellogg Radiation Laboratory, California Institute of Technology, Pasadena, CA 91125 USA.}
}

{\large 
$^g$ 
{\it 
Department of Physics and Astronomy, Science Center,   Amherst College, Amherst, MA 01002, USA.}
}

\end{center}

\vspace{0.2cm}

\begin{abstract}
\vspace{0.1cm}
We investigate neutrinoless double beta ($0\nu\beta\beta$) decay rates in minimal left-right symmetric models in presence of relatively light right-handed neutrinos. 
By use of an effective field theory approach, we systematically include all contributions in the model as well as the dependence of the decay amplitude on the masses of right-handed neutrinos.  In type-I and type-II seesaw scenarios, we analyze the impact of right-handed neutrinos heavier than about 10~MeV,
showing that this effect can lead to a detection of $0\nu\beta\beta$ decay in the next-generation experiments even for the normal hierarchy and a relatively large right-handed scale set by the mass of hypothetical right-handed gauge bosons. Finally, we comment on a possible connection between light right-handed neutrinos and the strong CP problem. 

\end{abstract}

\vfill
\end{titlepage}

\tableofcontents

\bigskip

\section{Introduction}

One outstanding puzzle in particle physics is how neutrino masses are generated. Neutrino masses can be explained by adding right-handed neutrino fields that do not transform under any of the Standard Model (SM) gauge symmetries. Such neutrinos, often called sterile neutrinos, can couple to left-handed neutrinos and the Higgs field through a Yukawa interaction. After electroweak  symmetry breaking a Dirac neutrino mass is generated in the same way as for  other SM fermions. However, nothing forbids a Majorana mass term for the sterile neutrinos.
The presence of both Dirac and Majorana mass terms leads to neutrino mass eigenstates that are Majorana states and the associated violation of lepton number ($L$).

The Majorana mass
of the sterile neutrino, dubbed $M_N$, is essentially a free parameter and there are no clear guidelines to what values $M_N$, if present at all, takes in nature. In the canonical type-I seesaw mechanism~\cite{Minkowski:1977sc,Gell-Mann:1979vob,Yanagida:1979as,Glashow:1979nm,Mohapatra:1979ia,Schechter:1980gr}, the active neutrino masses are given by $\sim y_D^2 v^2/M_N$, where $y_D$ is the Yukawa coupling and $v\simeq 246$ GeV, the Higgs vacuum expectation value. For Yukawa couplings $y_D \sim \mathcal O(1)$ this predicts extremely heavy right-handed neutrinos that barely interact with SM fields. Such neutrinos are therefore called sterile neutrinos. However, Yukawa couplings in the SM can be much smaller (see e.g. the mass ratio $m_e/m_t \sim 3 \cdot 10^{-6}$) leading to relatively light sterile neutrinos. This leads to interesting phenomenological scenarios where light sterile neutrinos can be probed in a large range of experiments \cite{Drewes:2013gca,Dasgupta:2021ies,Abdullahi:2022jlv}. Furthermore, light sterile neutrinos could account for dark matter and/or baryogenesis~\cite{Boyarsky:2018tvu,Asaka:2005pn}.

Sterile neutrinos might only appear to be sterile at relatively low energies. 
In left-right symmetric models~\cite{Pati:1974yy,Mohapatra:1974gc,Senjanovic:1975rk,Mohapatra:1979ia,Mohapatra:1980yp}, right-handed neutrinos are charged under $SU(2)_R$ gauge symmetry and interact with right-handed gauge bosons $W_R$. Considering present limits, from high-energy collider experiments~\cite{ATLAS:2018dcj,ATLAS:2019isd,CMS:2018agk} to low-energy precision measurements~\cite{Li:2020wxi}, the mass of $W_R$ lies well above the electroweak scale ($m_{W_R} > 5$ TeV) 
such that right-handed interactions are feeble at low energies and the label `sterile' is still appropriate. 
However, in left-right symmetric models the masses of sterile neutrinos are not necessarily of the same order as the right-handed gauge bosons. They could be much lower, similar to how most SM fermions are much lighter than electroweak gauge bosons. 
The presence of light sterile neutrinos with right-handed interactions can strongly affect neutrinoless double beta  ($0\nu\beta\beta$)  decay rates \cite{Bamert:1994qh,Blennow:2010th,Mitra:2011qr,deGouvea:2011zz,Barry:2011wb,Li:2011ss,Ghosh:2013nya,Girardi:2013zra,Barea:2015zfa,Bolton:2019pcu,Jha:2021oxl,Asaka:2020wfo,Dekens:2020ttz}.

The $\onbb$ decay rate is enhanced by long-distance pion-exchange contributions at hadronic scales if there exists a non-zero mixing between the left-handed gauge boson $W_L$ and right-handed gauges boson $W_R$ \cite{Prezeau:2003xn}. Because of this enhancement, there are good prospects \cite{Li:2020flq} for a positive signal in next-generation $\onbb$ experiments if the lightest sterile neutrino mass lies below the electroweak scale and the right-handed scale is not too far above existing limits. This holds even in the context of a  future precise determination of the sum of neutrino masses from cosmological measurements.
The analysis of Ref.~\cite{Li:2020flq} only considered the region of sterile neutrino masses above the hadronic scale $(\sim 1\gev)$.
In this work we extend our previous studies~\cite{Li:2020flq,Cirigliano:2018yza} to investigate the observability of $0\nu\beta\beta$ in more detail.

We calculate  the $0\nu\beta\beta$ decay rate as a function of sterile neutrino masses ranging from the pion mass to TeV scales) in minimal left-right symmetric models (mLRSM). We apply  an EFT framework that systematically describes the $0\nu\beta\beta$ decay rate in different regimes of sterile neutrino masses \cite{Prezeau:2003xn,Cirigliano:2017djv,Cirigliano:2018yza,Dekens:2020ttz}. By using chiral EFT for the low-energy description we correctly incorporate the symmetries of QCD in the evaluation of the neutrino-mass dependence of hadronic and nuclear matrix elements, going beyond earlier work. The EFT framework automatically includes all $0\nu\beta\beta$ contributions once the Wilson coefficients are obtained from a matching calculation and the neutrino masses are specified.
We divide the phenomenology into two scenarios, namely type-I and type-II seesaw dominance scenarios, and
discuss the phenomenology after considering all possible constraints.

Interestingly, it has been argued that light sterile neutrinos in left-right symmetric models
may provide a solution to the strong CP problem
~\cite{Kuchimanchi:2014ota,Senjanovic:2020int}. Let us consider the mLRSM with a generalized parity ($\mathcal{P}$) as the left-right symmetry~\cite{Pati:1974yy,Mohapatra:1974gc,Senjanovic:1975rk,Senjanovic:1978ev,Mohapatra:1979ia,Mohapatra:1980yp}. 
Recall that the $\bar{\theta}$ parameter of the Standard Model is written as $\bar{\theta}=\theta+\text{arg det}(M_uM_d)$, where $\theta$ is the parameter governing the strength of the $G{\tilde G}$ interaction in the QCD Lagrangian and $M_u,M_d$ are the up and down quark mass matrices.
In the $\mathcal{P}$-symmetric mLRSM  it is usually assumed that
parity is broken spontaneously at some energy above the electroweak scale and the bare theta term is negligible i.e. $\theta\ll \text{arg det}(M_uM_d)$.
This makes $\bar{\theta}$ calculable  after the  spontaneous breaking of parity, $\bar \theta_{\text{tree}} =m_t/(2m_b)  \tan (2 \beta) \sin \alpha$ at tree level~\cite{Maiezza:2010ic,Maiezza:2014ala},  where $\alpha$ is a spontaneous CP phase and $\tan \beta$ a ratio of vacuum expectation values (vevs) appearing in the model. 
The recent global analysis performed in Ref.~\cite{Ramsey-Musolf:2020ndm} requires  $\bar \theta_{\text{tree}} < 10^{-9}$, which means the strong CP problem in the SM is essentially transferred to the necessity of a small spontaneous CP  phase $\alpha \ll 1$. However, this condition is not sufficient. 
In Refs.~\cite{Kuchimanchi:2014ota,Senjanovic:2020int} it was pointed out that loop corrections in the scalar potential  involving charged lepton and sterile neutrinos inside the loop na\"ively give an additional contribution to $\bar{\theta}$. This contribution scales as
\begin{equation}
\label{eq:mnu_theta}
\bar \theta_{\text{loop}} \propto \sqrt{\frac{M_\nu}{m_W}}\times
\sqrt{\frac{M_N}{m_W }}\times \left(\frac{M_N}{m_{W_R}}\right)^2\;,
\end{equation}
where $M_{\nu}$ denotes the active neutrino mass, $M_N$ the sterile neutrino mass, and $m_{W}$ is the SM $W$ boson mass. 
Three mass ratios determine the small scale of $\bar\theta_{\text{loop}}$: (1) $(M_\nu/m_W)^{1/2} \lesssim 1\times 10^{-6}$ for $M_\nu\lesssim 0.1$ eV; (2) $(M_N/m_W)^{1/2}$; and (3) $(M_N/m_{W_R})^2$. This implies that for a fixed value of $m_{W_R}$, the sterile masses $M_N$ cannot be too large. For example, setting $m_{W_R} \simeq 5 $ TeV would require $M_N \lesssim 100$ GeV. If the above argument is correct it would provide a tantalizing connection between electric dipole moment (EDM) experiments that constrain $\bar \theta$ and searches for light sterile neutrinos. In this work, we therefore scrutinize the link in Eq.~\eqref{eq:mnu_theta} and find, unfortunately, no confirmation that a large correction to $\bar \theta$ is actually induced, at least not for the specific choices for the  right-handed leptonic mixing matrix studied in this work. Thus, the validity of the proposal of Refs.~\cite{Kuchimanchi:2014ota,Senjanovic:2020int} 
and the corresponding motivation for light sterile neutrinos remains an interesting open question.

This paper is organized as follows. In Sec.~\ref{sec:mLRSM} we discuss the minimal left-right symmetry model under consideration and discuss the contribution to the theta term with light sterile neutrinos. In Sec.~\ref{sec:EFT} we describe our calculation of $0\nu\beta\beta$ decay rates using an EFT approach. In Sec.~\ref{sec:pheno}, we discuss the phenomenology of $\onbb$ decay in the type-I and type-II seesaw dominance of the mLRSM focusing on the scenarios with light sterile neutrinos, and the connection with the strong CP problem. We conclude in Sec.~\ref{sec:conclusion}.

\section{Minimal left-right symmetric model}
\label{sec:mLRSM}
\subsection{Lagrangian in the weak and mass eigenstates}
We begin by reviewing the 
mLRSM~\cite{Pati:1974yy,Mohapatra:1974gc,Senjanovic:1975rk,Senjanovic:1978ev,Mohapatra:1979ia,Mohapatra:1980yp}. The model is based on the extended  gauge group 
\begin{align*}
G_{LR}\in SU(2)_L\times SU(2)_R\times U(1)_{B-L}\,
\end{align*}
in addition to the $SU(3)_c$ of the strong interaction. The fermion sector consists of left- and right-handed doublets
\begin{align}
L_{L,R}=
\begin{pmatrix}
\nu\\
e
\end{pmatrix}_{L,R}\;,\quad
Q_{L,R}=
\begin{pmatrix}
u\\
d
\end{pmatrix}_{L,R}\;,
\end{align}
that appear for each of the three generations of fermions. Right-handed (RH) neutrinos appear in a doublet and are thus charged under $SU(2)_R$ and couple to right-handed gauge bosons. The scalar sector of the model is extended with respect to the SM. In the minimal model it contains two scalar triplets
\begin{align}
\Delta_{L,R}=
\begin{pmatrix}
\delta_{L,R}^{+}/\sqrt{2} & \delta_{L,R}^{++}\\
\delta_{L,R}^{0}& -\delta_{L,R}^{+}/\sqrt{2}
\end{pmatrix}
\end{align}
that belong to the representations $\Delta_{L}\in (3,1,2)$ and $\Delta_R\in (1,3,2)$ under $G_{LR}$, and a single bi-doublet
\begin{align}
\Phi = \begin{pmatrix}
\phi_1^0 & \phi_2^+\\
\phi_1^- & \phi_2^0 
\end{pmatrix}
\end{align}
that transforms under both $SU(2)_L$ and $SU(2)_R$ as $\Phi \in (2,2^*,0)$. 

The $G_{LR}$ group is broken by the vevs of the scalar fields,
\begin{align}
\langle \Phi \rangle =
\begin{pmatrix}
\kappa/\sqrt{2} & 0\\
0 & \kappa^\prime e^{i\alpha}/\sqrt{2}
\end{pmatrix}\;,\quad 
\langle \Delta_R\rangle = 
\begin{pmatrix}
0 & 0 \\
v_R/\sqrt{2} & 0
\end{pmatrix}\;,\quad
\langle \Delta_L\rangle = 
\begin{pmatrix}
0 & 0 \\
v_L e^{i\theta_L}/\sqrt{2} & 0
\end{pmatrix}\;,
\end{align}
where all parameters are real \cite{Deshpande:1990ip}. In a first step, the spontaneous breaking of 
$G_{LR} \rightarrow SU(2)_L \times U(1)_Y$ is induced by the vev of the right-handed scalar triplet, $v_R$, that is taken to lie well above the electroweak scale $v = 246\gev$ in order to explain the absence of right-handed charged currents in various experiments. The $SU(2)_L \times U(1)_Y \rightarrow U(1)_{\rm QED}$ is spontaneously broken by the vevs of the bidoublet where $\sqrt{\kappa^2 + \kappa^{\prime\,2}}=v$. The vev $v_L$ generates a contribution to the neutrino masses and is discussed below. The phases $\alpha$ and $\theta_L$ induce CP violation.

The spontaneous breaking of the $SU(2)_{L,R}$ gauge symmetries leads to massive electroweak gauge bosons. For our purposes, the charged gauge bosons are the most important. We define mass eigenstates $W_{1,2}^{\pm}$ through
\begin{align}
\begin{pmatrix}
W_L^{\pm} \\
W_R^{\pm}
\end{pmatrix}
=\begin{pmatrix}
\cos\zeta & -\sin\zeta e^{\mp i\alpha}\\
\sin\zeta e^{\pm i\alpha} & \cos\zeta
\end{pmatrix}
\begin{pmatrix}
W_1^{\pm}\\
W_2^{\pm}
\end{pmatrix}\,,
\end{align}
in terms of the mixing parameter  $\tan\zeta = {\kappa \kappa^\prime }/{v_R^2}$.
The charged gauge boson masses
\begin{align}
m_{W_1}\simeq m_{W_L} = \dfrac{g_Lv}{2}\equiv m_W\,,\quad m_{W_2}\simeq m_{W_R} = \dfrac{g_R v_R}{\sqrt{2}} \,,
\end{align}
where $g_{L,R}$ are the gauge couplings of $SU(2)_{L,R}$. 
If the left-right symmetry is imposed, $g_L=g_R$ equal to the SM $SU(2)_L$ gauge coupling $g$, such that $m_{W}/m_{W_R} \simeq v/(\sqrt{2} v_R) \ll 1$. For convenience, we introduce 
\begin{align}
\lambda = \dfrac{m_{W_1}^2}{m_{W_2}^2}\;,\quad
\tan\beta =\dfrac{\kappa^\prime}{\kappa}\equiv \xi\;,
\end{align}
such that the mixing parameter can be expressed as
\begin{align}
\label{eq:LRmixing}
\tan\zeta = \lambda \sin(2\beta)\;.
\end{align}

The charges of the fermions and scalar allow for several Yukawa interactions
\begin{align}
\label{eq:Lag-Yukawa}
\mathcal{L}_Y= & -\bar{Q}_L (\Gamma \Phi + \tilde{\Gamma} \tilde{\Phi}) Q_R -\bar{L}_L (\Gamma_l \Phi + \tilde{\Gamma}_l \tilde{\Phi}) L_R  -(\bar L_L^c i\tau_2 \Delta_L Y_L L_L + \bar L_R^c i\tau_2 \Delta_R Y_R L_R ) + \text{h.c.} 
\end{align}
where $\tilde{\Phi} = \tau_2 \Phi^* \tau_2$ in terms of the Pauli matrix $\tau_2$, $\Psi^c_{L,R} = P_{R,L}\Psi^c$ with $\Psi^c = C \bar \Psi^T$ in terms of the charge conjugation matrix $C=-C^{-1}=-C^\dagger = - C^T$ and the left- and right-handed projectors $P_{R,L} = (1\pm \gamma_5)/2$.

After the electroweak symmetry breaking (EWSB), the leptonic Yukawa interactions give rise to neutrino masses through the seesaw mechanism~\cite{Minkowski:1977sc,Gell-Mann:1979vob,Yanagida:1979as,Glashow:1979nm,Mohapatra:1979ia}
\begin{align}
\mathcal{L}_{\nu} &= -\dfrac{1}{2} (\bar\nu_L^c, \bar\nu_R)  
\begin{pmatrix}
M_L^\dagger & M_D^* \\
M_D^\dagger & M_R^\dagger
\end{pmatrix}
\begin{pmatrix}
\nu_L\\
\nu_R^c
\end{pmatrix}
+\text{h.c.}\nonumber\\
&= -\dfrac{1}{2} (\bar{\nu}_L, \bar{\nu}_R^c) 
\begin{pmatrix}
M_L & M_D \\
M_D^T & M_R
\end{pmatrix}
\begin{pmatrix}
\nu_L^c\\
\nu_R
\end{pmatrix}
+\text{h.c.}
\end{align}
in terms of the symmetric $6\times 6$ matrix~\cite{Schechter:1981cv,Xing:2011zza},
\begin{align}
M_n \equiv 
\begin{pmatrix}
M_L & M_D \\
M_D^T & M_R
\end{pmatrix}\,,
\end{align} 
where the Dirac and Majorana masses are given by
\begin{align}
M_D & = (\kappa \Gamma_l + \kappa^\prime \tilde{\Gamma}_l e^{-i\alpha})/\sqrt{2}\;,\\
M_L &= \sqrt{2} Y_L^\dagger v_L e^{-i\theta_L}\;,\\
M_R & =\sqrt{2} Y_R v_R\;.
\end{align}

There are two possible discrete symmetries that can be imposed between the left and right sectors~\cite{Maiezza:2010ic}. 
 When a generalized parity ($\mathcal{P}$) is the left-right symmetry, the fields transform as
\begin{align}
L_L \leftrightarrow L_R\;,\quad  \Phi\leftrightarrow \Phi^\dagger\;,\quad \Delta_L \leftrightarrow \Delta_R\;.
\end{align}
When a generalized charge conjugation ($\mathcal{C}$) is the left-right symmetry, the fields transform as
\begin{align}
L_L \leftrightarrow L_R^c\;,\quad  \Phi\leftrightarrow \Phi^T\;,\quad \Delta_L \leftrightarrow \Delta_R^*\;.
\end{align}
They lead to
\begin{align}
\mathcal{P}:\quad &  \Gamma_l = \Gamma_l^\dagger\;,\quad \tilde{\Gamma}_l = \tilde{\Gamma}_l^\dagger\;,\quad Y_L=Y_R \;,\nn \\
\mathcal{C}:\quad & \Gamma_l = \Gamma_l^T\;,\quad \tilde{\Gamma}_l = \tilde{\Gamma}_l^T\;,\quad Y_L=Y_R^\dagger \;,
\end{align}
from which we obtain that
\begin{align}
\label{eq:majorana_pc}
\mathcal{P}:\quad & M_L = v_L/v_R M_R^{\dagger} e^{-i \theta_L}  \;, \nn \\
\mathcal{C}:\quad & M_L = v_L/v_R M_R e^{-i \theta_L} \;.
\end{align}
The Dirac mass matrix $ M_D = M_D^T$  in case of $\mathcal{C}$, while $ M_D = M_D^\dagger$ in case of $\mathcal{P}$ if $\kappa^\prime/\kappa\sin\alpha=0$.

For our purposes it will prove useful to work in the mass basis instead of the flavor basis as the  $0\nu\beta\beta$ decay amplitudes are non-trivial functions of the neutrino masses. We diagonalize the neutrino mass matrix through a $6\times6$ unitary matrix $U$,
\begin{equation}\label{mdiagU}
m_\nu = \mathrm{diag}(m_1,\,m_2\dots ,\,m_6) = U^T M_n^\dagger U\,,
\end{equation}
and define mass eigenstate $\nu = (\nu_1, \cdots, \nu_6)^T \equiv N_m + N_m^c$ through
\begin{align}
N_m = 
\begin{pmatrix}
\nu_L^\prime \\
\nu_R^{\prime c}
\end{pmatrix}
= U^\dagger 
\begin{pmatrix}
\nu_L \\
\nu_R^{c}
\end{pmatrix}\,.
\end{align}
The neutrinos $\nu_{1,2,3}$ are active neutrinos, while $\nu_{4,5,6}$ are sterile neutrinos.
We write the mixing matrix as
\begin{align}\label{Udef}
U = 
\begin{pmatrix}
U_{\text{PMNS}} & S \\
T & U_R
\end{pmatrix}
\end{align}
with $U_{\mathrm{PMNS}}$ being the usual $3\times 3$ neutrino-mixing matrix, i.e. Pontecorvo-Maki-Nakagawa-Sakata (PMNS) matrix, and $S$, $T$, and $U_R$ being additional $3\times3$ matrices. Here, we have assumed a basis where the charged lepton mass matrix is diagonal following Ref.~\cite{Barry:2013xxa}.
In comparison with the notations in Refs.~\cite{Senjanovic:2011zz,Nemevsek:2012iq,Senjanovic:2016vxw}, $U_{\text{PMNS}}=V_L$ and $U_R^* = V_R$.  

The left- and right-handed fields can now be written as
\begin{align}
\label{eq:mass_state}
\nu_L &= (P U)P_L \nu = U_{\text{PMNS}} \nu_L^\prime + S \nu_R^{\prime c} \,,&  \nu_L^c &= (P U^*)P_R \nu = U_{\text{PMNS}}^* \nu_L^{\prime c} + S^* \nu_R^{\prime}\,,\nn\\
\nu_R &= (P_s U^*)P_R \nu = T^* \nu_L^{\prime c} + U_R^* \nu_R^{\prime}\,,&   \nu_R^c &= (P_s U)P_L \nu = T \nu_L^\prime + U_R \nu_R^{\prime c} \,.
\end{align}
where $P_{L,R}$ are the usual projectors in spinor space, while $P$ and $P_s$ are $3\times 6$ projectors in flavor space 
 \begin{equation}
\label{eq:projector}
P = \begin{pmatrix}\mathcal I_{3\times 3} & 0_{3 \times 3}  \end{pmatrix}\,,\qquad
P_s = \begin{pmatrix} 0_{3\times 3} & \mathcal I_{3 \times 3}  \end{pmatrix}\, .
 \end{equation}
The matrices $S$ and $T$ can be expressed as~\cite{Schechter:1981cv,Korner:1992zk,Grimus:2000vj,Hettmansperger:2011bt} 
\begin{align}
\label{eq:st-matrix}
S = R U_R\;,\qquad T = -R^\dagger U_{\text{PMNS}}\;,
\end{align} 
where the matrix $R = M_D M_R^{-1}$ to leading order gives rise to the active-sterile mixing. Note that in Eq.~\eqref{Udef} we have neglected the non-unitary effects in the active neutrino mixing, which are in the form of $(1-\eta) U_{\rm PMNS}$ with $\eta\equiv R R^\dagger /2$. The experimental constraints on $\eta_{\alpha\beta}$ with $\alpha,\beta=e,\mu,\tau$ can be found in Ref.~\cite{Fernandez-Martinez:2016lgt}, which is smaller than $10^{-5}-10^{-3}$ depending on $\alpha$ and $\beta$. 

Following Ref.~\cite{Mitra:2011qr}, the full $6\times 6$ matrix can be decomposed as
\begin{align}
U = \mathcal{U}_1 \mathcal{U}_2 
\end{align}
with 
\begin{align}
\mathcal{U}_1 = \begin{pmatrix}
1 & R \\
-R^\dagger & 1
\end{pmatrix}\;,\quad 
\mathcal{U}_2 = \begin{pmatrix}
U_{\text{PMNS}} & 0\\
0 & U_R
\end{pmatrix}
\end{align}
Then
\begin{align}
\mathcal{U}_1^\dagger M_n \mathcal{U}_1^* = \begin{pmatrix}
M_\nu & 0\\
0 & M_N
\end{pmatrix} \,,
\end{align}
with 
\begin{align}
\label{eq:seesaw}
M_\nu &= M_L - M_D M_R^{-1} M_D^T\;,\nn\\
M_N &= M_R\;.
\end{align}
The block-diagonal matrix is diagonalized through
\begin{align}
\label{eq:diag2}
\mathcal{U}_2^\dagger \begin{pmatrix}
M_\nu & 0\\
0 & M_N
\end{pmatrix} \mathcal{U}_2^* = 
\begin{pmatrix}
\widehat{M}_\nu & 0\\
0 & \widehat{M}_N
\end{pmatrix}
\end{align}
where $\widehat{M}_\nu\equiv \mathrm{diag}(m_1,m_2,m_3)$ and $\widehat{M}_N \equiv \mathrm{diag}(m_4,m_5,m_6)$.
\subsection{Theta term and light sterile neutrinos}\label{theta}

In the mLRSM under the assumption of $\mathcal{P}$,
it has been argued that neutrinos can be connected to the strong CP problem~\cite{Senjanovic:2020int}\footnote{If $\mathcal{C}$ is taken as the left-right symmetry, the $\bar{\theta}$ parameter is arbitrary and no strict connection can be obtained, see discussions in Ref.~\cite{Senjanovic:2020int}.}. We briefly discuss this proposed connection here to set the stage for later analysis. We write $\bar{\theta}=\theta+\text{arg det}(M_uM_d)$, where $\theta$ is the coefficient  of  $G{\tilde G}$ term in the QCD Lagrangian, $G$ and $\tilde G$ denote the gluon field strength and its dual, respectively, and $M_u$, $M_d$ correspond to the up and down quark mass matrices. 

Assuming $\theta\ll \text{arg det}(M_uM_d)$, 
the tree-level contribution to the theta term from spontaneous symmetry breaking is given by~\cite{Maiezza:2014ala}, 
\begin{align}
\bar{\theta}_{\text{tree}} = \frac{m_t}{2m_b}\tan (2\beta)\sin\alpha \;,
\end{align}
where $m_b$ and $m_t$ are bottom and top quark masses, respectively. 

Ref.~\cite{Kuchimanchi:2014ota} identified additional loop contributions  to  $\bar{\theta}$ involving leptons. Ref.~\cite{Senjanovic:2020int} argued that these loop corrections to the $\bar{\theta}$ parameter imply an upper limit on the sterile neutrino masses, once the right-handed gauge boson $W_R$ and active neutrino masses are fixed.   This is the line of reasoning we will follow in this work.  The expression for $\bar{\theta}$ term up to one-loop level is given by  $\bar{\theta} = \bar\theta_{\text{tree}} + \bar\theta_{\text{loop}}$, where~\cite{Kuchimanchi:2014ota,Senjanovic:2020int} 
\begin{align}
\label{thetaQCDLR}
\bar\theta_{\text{loop}} 
& =  \frac{1}{4 \pi^{2}} \frac{m_{t}}{m_{b}} \frac{1}{v_R^2(\kappa^2 - \kappa'^2) }\Im{\operatorname{Tr}\left(M_{R}^{\dagger} M_{R}\left[M_D, M_{l} \right]\right)} \ln \frac{\sqrt{2} M_{P l}}{v_{R}}
\end{align}
Here, $M_{Pl}$ is the Plank mass and $M_l \equiv (\kappa \tilde\Gamma_l + \kappa^\prime \tilde{\Gamma}_l e^{i\alpha})/\sqrt{2}$  is the mass matrix for the charged leptons.

In the mLRSM,  the above expression is a function of the Majorana and Dirac neutrino mass matrices. As we shall see in Sec.~\ref{CPconnection}, in the $\mathcal{P}$-symmetric mLRSM the Dirac mass matrix $M_D$ might be written in terms of the active and sterile neutrino masses $\widehat M_{\nu}$ and $\widehat M_{N}$, $M_D = U_{\text{PMNS}}\sqrt{- \widehat{M}_N \widehat{M}_\nu} U_{\text{PMNS}}^\dagger$ for the type-I seesaw dominance and $U_R = U_{\text{PMNS}}^*$.  
Thus, barring accidental cancellations between the terms $\bar\theta_{\text{tree}}$ and $\bar\theta_{\text{loop}}$ or the vanishing of $\bar\theta_{\text{loop}}$, the EDM experiments would set {\it upper} bounds on the spontaneous phase $
\alpha \ll 1$ as well as the sterile neutrino masses.
We discuss this in more detail in Sect.~\ref{CPconnection}.

\section{Neutrinoless double beta decay in effective field theory}
\label{sec:EFT}
We now turn to the calculation of $0\nu\beta\beta$ decay rates for a wide range of neutrino masses.  The calculation involves many  energy scales, ranging from the MeV $Q$-value of nuclear isotopes all the way to scale of lepton number violation (LNV) that can lie well above the electroweak scale. In the mLRSM in particular, we have the right-handed scale $v_R$ and the associated masses of the right-handed gauge bosons and scalars. In addition, we have the masses of sterile neutrinos $m_i$, for $i=4,5,6$ that could be of the order of the right-handed scale as well, but can also be significantly lower.

The separation of scales suggests the use of effective field theory (EFT). Refs.~\cite{Prezeau:2003xn,Cirigliano:2017djv,Cirigliano:2018yza,Dekens:2020ttz} developed an end-to-end EFT framework that connects high-scale sources of LNV to $0\nu\beta\beta$ decay rates. We focus on the case where $m_i \ll v_R$. The remaining beyond-the-SM (BSM) degrees of freedom  are integrated out at the right-handed scale and matched to the neutrino-extended Standard Model EFT ($\nu$SMEFT)~\cite{Lehman:2014jma,Liao:2016qyd}. This EFT includes all effective operators that are invariant under the SM gauge group and contain SM fields as well as singlet $\nu_R$ fields. The higher-dimensional operators are then evolved to the electroweak scale where heavy SM fields are integrated out. At even lower energies, around $\Lambda_\chi \sim \rm GeV$, the resulting EFT operators are matched to hadronic lepton-number-violating operators by the use of chiral EFT, the low-energy EFT of QCD.

\subsection{EFT formulae}\label{EFTform}

We consider the case $m_i \ll v_R$ corresponding to relatively light sterile neutrinos. The mLRSM is matched to the $\nu$SMEFT Lagrangian at the right-handed scale $v_R$. For masses $m_i \sim v \gg \Lambda_\chi$, the sterile neutrinos are integrated out together with heavy SM fields at their explicit mass scale. Similarly, for $v > m_i > \Lambda_\chi$ sterile neutrinos can be integrated out. However, in the case where $m_i \leq \Lambda_\chi$, sterile neutrinos must be kept as explicit degrees of freedom at very low energies. This requires an extension of chiral EFT to include explicit sterile neutrinos that was developed in Ref.~\cite{Dekens:2020ttz}. We apply the developed EFT framework to efficiently describe the $0\nu\beta\beta$ decay phenomenology of the mLRSM for light sterile neutrino masses. We briefly describe the EFT formulae here and refer to Refs.~\cite{Prezeau:2003xn,Cirigliano:2017djv,Cirigliano:2018yza,Dekens:2020ttz} for further details. 

Direct and indirect searches for right-handed gauge bosons and scalars put stringent limits on their masses that must lie well above the TeV scale \cite{ATLAS:2018dcj,ATLAS:2019isd,CMS:2018agk,Beall:1981ze,Maiezza:2010ic,Bertolini:2014sua,Dekens:2021bro}. We integrate out the $W_R$ boson at its threshold and match to effective $\nu$SMEFT operators. They are explicitly invariant under the SM gauge group but not under $G_{LR}$. Important operators are
\begin{equation}\label{nuSMEFT}
\mathcal L_{6,\nu\mathrm{SMEFT}}= \bar e_R \gamma^\mu  C_R^{(6)} \nu_R\,\bar u_R \gamma_\mu d_R + \bar e_R \gamma^\mu  C_L^{(6)} \nu_R\, \left[i (D_\mu \varphi)^\dagger \tilde \varphi\right] + \mathrm{h.c.}\;,
\end{equation}
where $C_{R,L}^{(6)}$ are $3\times3$ matrices in lepton-flavor space scaling as two inverse powers of the heavy scale, and   we focused on couplings to first-generation quarks. 
The matching relations are given by
\begin{equation}
\label{eq:matching_dim6}
C_{R}^{(6)}(m_{W_R}) = -\frac{1}{v_R^2}V^R_{ud}\,,\qquad C_{L}^{(6)}(m_{W_R}) = 2 \frac{\xi e^{-i \alpha}}{1+\xi^2}\frac{C_R^{(6)}}{V^R_{ud}}\,.
\end{equation}
In addition, integrating out scalar fields leads to $\Delta L=2$ dimension-nine operators that do not contain neutrinos. These are discussed at the end of this subsection. In Eq.~\eqref{eq:matching_dim6}, $V_{ud}^R$ denotes the $(1,1)$ element of right-handed quark mixing matrix. In the $\mathcal{P}$-symmetric mLRSM, the generalized parity $\mathcal{P}$ is taken as the left-right symmetry, $V^R$ is equal to the left-handed quark mixing matrix, i.e. the Cabibbo-Kobayashi-Maskawa (CKM) matrix, in the limit of $\xi \sin\alpha \to 0$~\cite{Senjanovic:2014pva,Senjanovic:2015yea}. This is a good approximation as will be discussed in Sec.~\ref{sec:pheno}. In the $\mathcal{C}$-symmetric mLRSM, the generalized charge conjugation $\mathcal{C}$ is taken as the left-right symmetry, and $V^R$ is equal to the complex conjugate of the CKM matrix up to additional overall phases~\cite{Maiezza:2010ic}.

The subsequent discussion depends on the masses of the sterile neutrinos. We first consider the case $m_{i} < m_W$.
 The operators in Eq.~\eqref{nuSMEFT} have vanishing QCD anomalous dimensions and thus do not evolve between $m_{W_R}$ and $m_W$ up to small electroweak corrections.
At the electroweak scale, we integrate out the $W$ boson and match to an $SU(3)_c \times U(1)_{\rm QED}$-invariant effective Lagrangian. We find the effective operators 
\begin{equation}\label{Lweak}
\mathcal L_{6,\nu\mathrm{LEFT}} = \frac{2 G_F}{\sqrt{2}} \left\{ \bar u_L \gamma_\mu d_L \left[ \bar e_L \gamma^\mu C^{(6)}_{{\rm VLL}} \nu  +  \bar e_R \gamma^\mu C^{(6)}_{{\rm VLR}} \nu\right]+ \bar u_R \gamma_\mu d_R\,  \bar e_R \gamma^\mu C^{(6)}_{{\rm VRR}} \nu  \right\}\,,
\end{equation}
where we work in the neutrino mass basis and the relations of $\nu$ to the fields in the flavor basis are given by Eq.~\eqref{eq:mass_state}, and $G_F=(\sqrt{2} v^2)^{-1}$ denotes the Fermi constant. The matching coefficients are
 \begin{eqnarray}
\label{eq:dim6WCs}
C^{(6)}_{{\rm VLL}}(m_W)  &=& -2 V_{ud}\,PU\,,\nn\\
C^{(6)}_{{\rm VLR}}(m_W)  &=& V_{ud}\,\left(v^2  C_{L}^{(6)}(m_{W_R})\right)\,P_s U^*\,,\nn\\
C^{(6)}_{{\rm VRR}}(m_W)  &=& \left(v^2  C_{R}^{(6)}(m_{W_R})\right)\,P_s U^*\,.
 \end{eqnarray}
The above expressions accommodate all neutrino-exchange contributions to the $\onbb$ decay in the mLRSM.  Although the interactions described by $\mathcal L_{6,\nu\mathrm{LEFT}}$ conserves total lepton number, they contribute to the $\onbb$ decay process, in which lepton number is violated by the Majorana masses $m_i$ entering propagators of exchanged, virtual neutrinos. The operators in Eq.~\eqref{Lweak} have vanishing QCD anomalous dimensions and  do not evolve between $m_{W}$ and $m_i$  for $m_i$ larger than $\Lambda_\chi$.

 Mass eigenstates with $m_i < \Lambda_\chi$ are kept as explicit degrees of freedom.   The charged currents in Eq.~\eqref{Lweak} are matched to a chiral EFT Lagrangian at scales around $m_0\simeq 2$ GeV. This matching is discussed in detail in Ref.~\cite{Dekens:2020ttz} and the associated $0\nu\beta\beta$ decay contributions are discussed in the next subsection. Mass eigenstates with $m_i > \Lambda_\chi$ are integrated out at their respective thresholds. This procedure induces a set of $\Delta L=2$ six-fermion operators 
\begin{align}\label{dim9}
\mathcal{L}_{9,\text{LEFT}} &=\dfrac{1}{v^5} \bigg[ C_{1R}^{(9)} O_{1}+ C_{1R}^{(9)\prime} O_{1}^{\prime} + C_{4R}^{(9)} O_{4} + C_{5R}^{(9)} O_{5} \bigg]\bar{e}_R C \bar{e}_R^T\nn\\
&+\dfrac{1}{v^5}\bigg[ C_{1L}^{(9)} O_{1} + C_{1L}^{(9)\prime} O_{1}^{\prime}\bigg]\,\bar{e}_L C \bar{e}_L^T\,,
\end{align}
where we followed the operator definitions of Ref.~\cite{Cirigliano:2018yza} 
\begin{align}
O_{1} &=  \bar{q}_L^\alpha  \gamma_\mu \tau^+ q_L^\alpha \ \bar{q}_L^\beta  \gamma^\mu \tau^+ q_L^\beta\;,\nn\\
O_{1}^{\prime} &=  \bar{q}_R^\alpha  \gamma_\mu \tau^+ q_R^\alpha \ \bar{q}_R^\beta  \gamma^\mu \tau^+ q_R^\beta\;,\nn\\
O_{4} &=  \bar{q}_L^\alpha  \gamma_\mu \tau^+ q_L^\alpha \ \bar{q}_R^\beta  \gamma^\mu \tau^+ q_R^\beta\;,\nn\\
O_{5} &=  \bar{q}_L^\alpha  \gamma_\mu \tau^+ q_L^\beta \ \bar{q}_R^\beta  \gamma^\mu \tau^+ q_R^\alpha\;.
\end{align}
Here $\alpha$, $\beta$ are color indices and $\tau^+$ is an isospin ladder operator converting a down quark to an up quark.
The matching coefficients are given by 
\begin{align}\label{dim9match}
C_{1R}^{(9)}(m_i) &= -\dfrac{v}{2} \left(C_{{\rm VLR}}^{(6)}\right)_{ei}^2 \dfrac{1}{m_i}\;,\nn\\
C_{1R}^{(9)\prime}(m_i) &= -\dfrac{v}{2} \left(C_{{\rm VRR}}^{(6)}\right)_{ei}^2 \dfrac{1}{m_i}\;,\nn\\
C_{4R}^{(9)}(m_i) &= -v \left(C_{{\rm VLR}}^{(6)}\right)_{ei}\left(C_{{\rm VRR}}^{(6)}\right)_{ei} \dfrac{1}{m_i}\;,\nn\\
C_{5R}^{(9)}(m_i) &= 0\;,\nn\\
C_{1L}^{(9)}(m_i) &= -\dfrac{v}{2} \left(C_{{\rm VLL}}^{(6)}\right)_{ei}^2 \dfrac{1}{m_i}\;,\nn\\
C_{1L}^{(9)\prime}(m_i) &= 0\;.
\end{align}
We evolve the operators from the scale where they are induced, $m_i$, to the low-energy scale $m_0=2\gev$ where we match to chiral EFT by using one-loop QCD anomalous dimensions \cite{Cirigliano:2018yza} 
\begin{align}\label{runningO1}
C_{1R}^{(9)}(m_0)&= C_{1R}^{(9)}(m_i) \eta^{2/\beta_0}\;,\nn\\
C_{1R,1L}^{(9)\prime}(m_0)&= C_{1R,1L}^{(9)\prime}(m_i) \eta^{2/\beta_0}\;,
\end{align}
and 
\begin{align}\label{running}
\begin{pmatrix}
C_{4R}^{(9)}(m_0) \\
C_{5R}^{(9)}(m_0)
\end{pmatrix}
= 
\begin{pmatrix}
\eta^{1/\beta_0} & 0\\
\dfrac{1}{3}\left(\eta^{-8/\beta_0}-\eta^{1/\beta_0}\right) & \eta^{-8/\beta_0}
\end{pmatrix}
\begin{pmatrix}
C_{4R}^{(9)}(m_i) \\
C_{5R}^{(9)}(m_i)
\end{pmatrix}\,,
\end{align}
where $\eta \equiv \alpha_s(m_{i})/\alpha_s(m_0)$, and $\beta_0=(11N_c-2n_f)/3$ with $N_c=3$. $n_f$ denotes the number of active quark flavors which depends on $m_i$. Additional four-quark two-lepton operators appear with additional derivatives that lead to suppressed contributions but can depend on different combinations of Wilson coefficients such as $ \left(C_{{\rm VLL}}^{(6)}\right)_{ei}\left(C_{{\rm VLR}}^{(6)}\right)_{ei} $.

We briefly discuss what changes in the above description for $m_i > m_W$. In principle, we should integrate out the sterile neutrinos before integrating out the SM electroweak gauge bosons. Some of the resulting SMEFT operators have non-vanishing QCD anomalous dimensions and thus evolve between $m_i$ and $m_W$. In practice, the evolution above the electroweak scale is only a minor correction and most of the evolution occurs at relatively low energy scales where the QCD coupling is larger. We therefore neglect these complications and apply Eqs.~\eqref{dim9}-\eqref{running} for $m_{W_R} > m_i>m_W$ as well.

The dimension-nine Lagrangian in Eq.~\eqref{dim9} also receives contributions from the exchange of doubly charged scalar $\delta_{R}^{--}$, which can be easily incorporated by considering additional contributions to the Wilson coefficients~\cite{Cirigliano:2018yza} that get added to the above expressions 
\begin{align}\label{scalar}
\delta C_{1R}^{(9)\prime}(m_{\delta_R}) &= - \dfrac{v^5}{v_R^4} \left(V^R_{ud}\right)^2 \dfrac{ m_i}{m_{\delta_R}^2}(P_s U^*)_{ei}^2\;,\nn\\
\delta C_{1R}^{(9)}(m_W) &= 4 \left(\dfrac{V_{ud}}{V_{ud}^R}\right)^2 \left( \dfrac{\xi e^{-i\alpha}}{1+\xi^2} \right)^2 C_{1R}^{(9)\prime}(m_{\delta_R}) \;,\nn\\
\delta C_{4R}^{(9)}(m_W) &= 4\left(\dfrac{V_{ud}}{V_{ud}^R}\right) \left( \dfrac{\xi e^{-i\alpha}}{1+\xi^2} \right) C_{1R}^{(9)\prime}(m_{\delta_R})\;.
\end{align}
The evolution of the above Wilson coefficients to $m_0$ can be read from Eqs.~\eqref{runningO1} and \eqref{running} after substituting $m_i$  by $m_{\delta_R}$ for $\delta C_{1R}^{(9)\prime}$ or by $m_W$ for the remaining coefficients.

\subsection{Decay rates}
\label{sec:NLDBD}

After obtaining the appropriate matching conditions and evolution factors to low scales, the $0\nu\beta\beta$ decay rates of different isotopes can be directly read from expressions derived in Ref.~\cite{Dekens:2020ttz}. The total amplitude is written as 
\bea\label{amplitude}
\mathcal A = \frac{g_A^2 G_F^2 m_e}{\pi R_A} &\times& \Big[ \mathcal A_L\,\bar u(k_1)  P_R u^c(k_2)  + \mathcal A_R\,\bar u(k_1)  P_L u^c(k_2) + \mathcal A_M\,\bar u(k_1)  \gamma^0\gamma^5 u^c(k_2) \nn\\
&&+ \mathcal A_{E}\,\bar u(k_1) \gamma^0 u^c(k_2) \dfrac{E_1 -E_2}{m_e} + \mathcal A_{m_e}\,\bar u(k_1)  u^c(k_2)  \Big]\,,
\eea
where $u(k_i)$ denotes spinors of the outgoing electrons, and
\begin{align}
\label{eq:amplitude_DBD}
\mathcal A_{L,R} &= \sum_{i=1}^{n_L} \mathcal A_{L,R,M} (m_i) + \sum_{i=n_L+1}^{6} \mathcal A_{L,R}^{(9)} (m_i)\;,\\
\mathcal A_{M,E,m_e} &= \sum_{i=1}^{n_L} \mathcal A_{M,E,m_e} (m_i)+ \sum_{i=n_L+1}^{6} \mathcal A_{M,E,m_e}^{(10)} (m_i)\;.
\end{align}
Here, $n_L$ denotes the number, ranging between 3 and 6, of neutrino masses below $\Lambda_\chi$. Heavier neutrinos lead to contact interactions which are described by four-quark two-electron dimension-nine operators for $\mathcal A_{L,R}$ while an additional derivative is required for $\mathcal A_{M,E,m_e}$ leading to effective dimension-ten operators, see the discussion below Eq.~\eqref{running} and appendix D of Ref.~\cite{Dekens:2020ttz}. As such, the contributions from $\mathcal A_{M,E,m_e}^{(10)} (m_i)$ scale as $1/m_i^2$ for $m_i > \Lambda_\chi$ and play a marginal  role.

The inverse half-life of the $0\nu\beta\beta$ decay then becomes
\bea\label{eq:T1/2}
\left(T^{0\nu}_{1/2}\right)^{-1} &=& g_A^4 \left[ G_{01} \, \left( |\mathcal A_{L}|\sq + |\mathcal A_{R}|\sq \right)
- 2 (G_{01} - G_{04}) \textrm{Re} \mathcal A_{L}^* \mathcal A_{R} \right]\,\nn\\
&+& G_{09} |\mathcal A_M|^2 + G_{06}{\rm Re}\Big[ (\mathcal{A}_L- \mathcal{A}_R)\mathcal{A}_M^* \Big]\nn\\
&-& 2G_{03} {\rm Re}\Big[ (\mathcal{A}_L+ \mathcal{A}_R)\mathcal{A}_{E}^* + 2\mathcal{A}_{m_e} \mathcal{A}_E^* \Big] + 4 G_{02} |\mathcal{A}_E|^2 \nn\\
&+& 2G_{04} \Big[ |\mathcal{A}_{m_e}|^2 + {\rm Re}\big[(\mathcal{A}_L+ \mathcal{A}_R)\mathcal{A}_{m_e}^*\big]  \Big]
\;,
\eea
where $G_{01}$, $G_{02}$ $G_{03}$, $G_{04}$, $G_{06}$ and $G_{09}$ are electron phase space factors. We have tabulated them for various isotopes in Table~\ref{Tab:phasespace}.

\begin{table}
\center
\begin{tabular}{|c|cccc|}
\hline
\hline
\cite{Horoi:2017gmj}	    & $^{76}$Ge & $^{82}$Se & $^{130}$Te & $^{136}$Xe \\ 

\hline
$G_{01}$    & 0.22 & 1. & 1.4 & 1.5 \\
$G_{02}$    & 0.35 & 3.2 & 3.2 & 3.2 \\
$G_{03}$    & 0.12 & 0.65 & 0.85 & 0.86 \\
$G_{04}$    & 0.19 & 0.86 & 1.1 & 1.2 \\
$G_{06}$    & 0.33 & 1.1 & 1.7 & 1.8 \\
$G_{09}$    & 0.48 & 2. & 2.8 & 2.8 \\\hline
\hline
$Q/{\rm MeV} $ \cite{Stoica:2013lka} & 2.04& 3.0&2.5 & 2.5 \\
\hline\hline
\end{tabular}
\caption{Phase space factors in units of $10^{-14}$ yr$^{-1}$ obtained in Ref.~\cite{Horoi:2017gmj}. The last row shows the $Q$ value of various isotopes.}
\label{Tab:phasespace}
\end{table}

\subsubsection{Contributions from light neutrinos}

The subamplitudes in Eq.~\eqref{eq:T1/2} were derived in Ref.~\cite{Dekens:2020ttz} and given by
\begin{eqnarray}\label{Anu}
\mathcal A_L(m_i) &=&  -\frac{m_i}{4 m_e} \left[\mathcal M_V(m_i)+\mathcal M_A(m_i)\right] \left( C^{(6)}_{\rm VLL} \right)^2_{ei}+\mathcal A_L^{(\nu)}(m_i)\,,\nn\\
\mathcal A_R(m_i) &=& -\frac{m_i}{4 m_e} \bigg\{ \mathcal M_V(m_i) \left( C^{(6)}_{\rm VRR} + C^{(6)}_{\rm VLR} \right)^2_{ei} 
					+  \mathcal M_A(m_i) \left( C^{(6)}_{\rm VRR} - C^{(6)}_{\rm VLR} \right)^2_{ei}\bigg\}\nn\\
			      && + 	\mathcal A_R^{(\nu)}(m_i)	\,,\nn\\
\mathcal{A}_M(m_i) &=& -\dfrac{m_N}{2m_e}\mathcal{M}_{VA} (m_i) \left( C^{(6)}_{\rm VLL}\right)_{ei} \left(C^{(6)}_{\rm VLR}\right)_{ei} + \mathcal{A}_{M}^{(\nu)}(m_i)\;,\nn\\
\mathcal{A}_{E}(m_i)&=& -\dfrac{1}{2}\mathcal{M}_{E,L} (m_i) \left( C^{(6)}_{\rm VLL}\right)_{ei} \left(C^{(6)}_{\rm VLR}\right)_{ei} + \mathcal{A}_{E, L}^{(\nu)}(m_i)\;,\nn\\
&& -\dfrac{1}{2} \mathcal{M}_{E,R} (m_i) \left( C^{(6)}_{\rm VLL}\right)_{ei} \left(C^{(6)}_{\rm VRR}\right)_{ei} + \mathcal{A}_{E,R}^{(\nu)}(m_i)\;,\nn\\
\mathcal{A}_{m_e}(m_i)&=& -\dfrac{1}{2}\mathcal{M}_{m_e,L} (m_i) \left( C^{(6)}_{\rm VLL}\right)_{ei} \left(C^{(6)}_{\rm VLR}\right)_{ei} + \mathcal{A}_{m_e, L}^{(\nu)}(m_i)\;,\nn\\
&& -\dfrac{1}{2} \mathcal{M}_{m_e,R} (m_i) \left( C^{(6)}_{\rm VLL}\right)_{ei} \left(C^{(6)}_{\rm VRR}\right)_{ei} + \mathcal{A}_{m_e,R}^{(\nu)}(m_i)\;.
\eea
Terms with superscript $(\nu)$ denote contributions from hard neutrinos with virtual momenta $k\geq \Lambda_
\chi$ that are integrated out in the low-energy chiral EFT description \cite{Cirigliano:2018hja}. 
The remaining terms arise from the exchange of so-called potential neutrinos that appear as explicit degrees of freedom at low energies.  $\mathcal M_{V,A}(m_i)$  are combinations of nuclear matrix elements (NMEs), which are expressed\footnote{We have neglected a tiny contribution from tensor magnetic NMEs.}  as 
\begin{eqnarray}\label{eq:NME_mi}
\mathcal M_{V}(m_i) &=&  -\frac{g_V^2}{g_A^2} M_F(m_i) + M^{MM}_{GT}(m_i)\,, \nn \\ 
\mathcal M_{A}(m_i) &=&  M^{AA}_{GT}(m_i) + M^{AP}_{GT}(m_i)  +M^{PP}_{GT}(m_i)  +  M^{AP}_{T}(m_i) + M^{PP}_{T}(m_i)\,,
\end{eqnarray}
and similarly
\bea\label{eq:NME_lameta}
\mathcal{M}_{VA}(m_i) &=&2\dfrac{g_A}{g_M} \Big[M_{GT}^{MM}(m_i)\Big]\;,\nn \\
 \mathcal M_{m_e,L}(m_i) &=&  \frac{1}{6}\bigg\{ \frac{g^2_V}{g_A^2} M_F(m_i)  - \frac{1}{3} \Big[ M_{GT}^{AA}(m_i) - 4 M^{AA}_{T}(m_i)\bigg]\nn\\
 && - 3 \Big[ M^{AP}_{GT}(m_i) + M^{PP}_{GT}(m_i) + M^{AP}_{T}(m_i) + M^{PP}_{T}(m_i) \Big] \bigg\} \, , \nn \\
 \mathcal M_{m_e,R} (m_i) &=&  \frac{1}{6}\bigg\{   \frac{g^2_V}{g_A^2} M_F(m_i)  + \frac{1}{3} \Big[ M_{GT}^{AA}(m_i) - 4 M^{AA}_{T}(m_i)\Big] \nn\\
 &&+  3 \Big[ M^{AP}_{GT}(m_i) + M^{PP}_{GT}(m_i) + M^{AP}_{T}(m_i) + M^{PP}_{T}(m_i)\Big] \bigg\} \,,\nn\\
\mathcal{M}_{E,L}(m_i) &=& -\dfrac{1}{3} \bigg\{ \dfrac{g_V^2}{g_A^2} M_F(m_i) + \dfrac{1}{3}\Big[ 2 M_{GT}^{AA}(m_i) + M_T^{AA}(m_i) \Big] \bigg\}\;,\nn\\
\mathcal{M}_{E,R}(m_i) &=& -\dfrac{1}{3} \bigg\{ \dfrac{g_V^2}{g_A^2} M_F(m_i) - \dfrac{1}{3}\Big[ 2 M_{GT}^{AA}(m_i) + M_T^{AA}(m_i) \Big] \bigg\}\;. 
\eea

The NMEs depend on  the mass of the neutrino that is being exchanged.  This dependence can be neglected for active neutrinos as their masses lie well below nuclear scales. The mass dependence is much more important for sterile neutrinos and should in principle be calculated explicitly. For most NMEs this has not been done. Instead we will use interpolation formulae that have been checked against data where available, to interpolate between the small and large neutrino mass limits \cite{Dekens:2020ttz}. For example, 
\be\label{intsimple}
M_{F}(m_i) = M_{F,\, sd}\frac{m_\pi^2}{m_i^2 + m_\pi^2 \frac{M_{F,\, sd}}{M_F}}\,.
\ee
and equivalent expressions hold for $M^{AA}_{GT}(m_i)$, $M^{AP}_{GT}(m_i)$, $M^{PP}_{GT}(m_i)$, $M^{AP}_{T}(m_i)$, and $M^{PP}_{T}(m_i)$, by replacing $M_F$ and $M_{F\,sd}$ by the corresponding NMEs. For $m_i \ll m_\pi$ with $m_\pi$ being the pion mass, Eq.~\eqref{intsimple} simply becomes $M_F$ and for $m_i \gg m_\pi$ this drops as $(m_\pi^2/m_i^2)M_{F,\,sd}$.  The magnetic NME is slightly more complicated and given by
\begin{align}
M_{GT}^{MM}(m_i) &=\dfrac{g_M^2}{ 6g_A^2} \Big[ \dfrac{m_\pi^2}{m_N^2} M_{GT,sd}^{AA} -\dfrac{m_i^2}{m_N^2} M_{GT}^{AA}(m_i) \Big]\;,
\end{align}
which ensures that for for $m_i \gg m_\pi$, we obtain the scaling $M_{GT}^{MM}(m_i) \sim m_i^{-2}$ as anticipated below Eq.~\eqref{eq:amplitude_DBD3}. 
The neutrino-mass independent NMEs $M_F$, $M_{F,sd}$, $M_{GT,sd}^{AA}$, etc., are tabulated in Table~\ref{tab:NME} for various isotopes and nuclear methods. 

\begin{table}
\center
$\renewcommand{\arraystretch}{1.5}
\begin{array}{l||rrr|rr|rr |rr}
 \text{NMEs} & \multicolumn{3}{c|}{\text{}^{76} \text{Ge}} & \multicolumn{2}{c|}{\text{}^{82} \text{Se}} & \multicolumn{2}{c|}{ \text{}^{130} \text{Te}} & \multicolumn{2}{c}{  \text{}^{136} \text{Xe}}  \\
& \text{\cite{Hyvarinen:2015bda}} &   \text{\cite{Menendez:2017fdf}}  
& \text{\cite{Barea:2015kwa}} & \text{\cite{Hyvarinen:2015bda}} &    \text{\cite{Menendez:2017fdf}} & \text{\cite{Hyvarinen:2015bda}} &   \text{\cite{Menendez:2017fdf}} & \text{\cite{Hyvarinen:2015bda}} &   \text{\cite{Menendez:2017fdf}} \\
 \hline
 M_F 			   & $-$1.74    &  $-$0.59 	& $-$0.68 	& $-$1.29 	&  $-$0.55	& $-$1.52    	&   $-$0.67	& $-$0.89  	&   $-$0.54 \\
 M_{GT}^{AA} 		   & 5.48       &  3.15	    	& 5.06 		& 3.87 		&  2.97		& 4.28    	&   2.97	& 3.16   	&   2.45 \\
 M_{GT}^{AP} 		   & $-$2.02    & $-$0.94	& $-$0.92 	& $-$1.46    	& $-$0.89   	& $-$1.74 	&  $-$0.97	& $-$1.19   	&  $-$0.79  \\
 M_{GT}^{PP} 		   & 0.66  	&  0.30		& 0.24 		& 0.48          &  0.28 	& 0.59   	&   0.31   	& 0.39   	&   0.25\\
 M_{GT}^{MM} 		   & 0.51 	& 0.22		& 0.17 		& 0.37       	& 0.20 		& 0.45 		&  0.23 	& 0.31 		&  0.19 \\
 M_T^{AP} 	   	   & $-$0.35 	& $-$0.01	& $-$0.31 	& $-$0.27    	& $-$0.01 	& $-$0.50	&     0.01	& $-$0.28 	&     0.01	\\
 M_T^{PP} 	 	   & 0.10     	&    0.00	& 0.09  	& 0.08       	& 0.00 		& 0.16 		&  $-$0.01 	& 0.09 		&  $-$0.01  \\\hline
 M_{F,\, sd}  	   	   & $-$3.46 	& $-$1.46	& $-$1.1  	& $-$2.53   	& $-$1.37 	& $-$2.97 	&  $-$1.61 	& $-$1.53   	&  $-$1.28	\\
 M^{AA}_{GT,\, sd}  	   &    11.1 	& 4.87		& 3.62		& 7.98    	& 4.54 		& 	10.1 	&  5.31 	&    5.71    	&  4.25  \\
M^{AP}_{GT,\, sd}	   & $-$5.35 	& $-$2.26 	& $-$1.37 	& $-$3.82    	& $-$2.09 	& $-$4.94 	&  $-$2.51 	& $-$2.80  	&  $-$1.99  \\
M^{PP}_{GT,\, sd}	   & 1.99 	& 0.82		& 0.42		& 1.42      	& 0.77 		& 1.86 		&  0.92 	& 1.06  	&   0.74\\
M^{AP}_{T,\, sd}	   & $-$0.85 	&  $-$0.05	&  $-$0.97	& $-$0.65    	&  $-$0.05	& $-$1.50 	&   0.07	& $-$0.92  	&   0.05		\\  
M^{PP}_{T,\, sd} 	   & 0.32 	&  0.02		&  0.38		& 0.24       	&  0.02		& 0.58 		&   $-$0.02	& 0.36  	&   $-$0.02 \\
\end{array}$
\caption{
Nuclear matrix elements as obtained in the quasi-particle random phase approximation (QPRA)  \cite{Hyvarinen:2015bda}, shell model  \cite{Menendez:2017fdf}, and interacting boson model (IBM)  \cite{Barea:2015kwa}
for different isotopes.}
\label{tab:NME}
\end{table}

What remains is the description of $\mathcal A_{L,R,M,E,m_e}^{(\nu)}(m_i)$ which are more subtle. $\mathcal A_{L,R,M}^{(\nu)}(m_i)$ describes the contribution from hard neutrinos. These are light neutrinos that carry large virtual momentum and thus cannot, despite being light, be resolved in the low-energy chiral EFT. Their importance was emphasized for the standard mechanism (the exchange of three light Majorana neutrinos) in Ref.~\cite{Cirigliano:2018hja}. The low-energy contributions from hard neutrinos come in the form of $\Delta L=2$ neutrinoless operators, for instance in the form of $nn\rightarrow pp + ee$ or $\pi\pi\rightarrow ee$ contact interactions, with coupling constants that depend on $m_i$. While considerable progress has been made, not all \lq\lq low-energy constants\rq\rq\ (LECs) are known at present. 

We follow Ref.~\cite{Dekens:2020ttz} and write 
\begin{equation}
\mathcal A_{L}^{(\nu)}(m_i) = -\frac{m_i}{2m_e}\frac{m_\pi^2 g_\nu^{NN}(m_i)}{g_A^2}\left(C_{{\rm VLL}}^{(6)}\right)^2_{ei}\,,
\end{equation}
where $g_\nu^{NN}(m_i)$ is an a priori unknown LEC that describes a neutrino-mass dependent local $nn\rightarrow pp + ee$ transition. Ideally, $g_\nu^{NN}(m_i)$ is calculated on the lattice but this is not possible yet. Fortunately,  using chiral and isospin symmetry,  $g_\nu^{NN}(m_i=0)$ can be related to charge-independence-breaking nucleon-nucleon interactions for which data does exist \cite{Cirigliano:2019vdj ,Richardson:2021xiu}. Model calculations of the LEC $g_\nu^{NN}(0)$ were found to be in good agreement with these extractions \cite{Cirigliano:2020dmx,Cirigliano:2021qko}, and implemented in nuclear structure calculations in Refs.~\cite{Wirth:2021pij,Jokiniemi:2021qqv} confirming their importance. Based on these findings we extract 
\begin{equation}
g_\nu^{NN}(m_i=0) = -(3.6\pm1.8)\,\rm{fm}^2\,,
\end{equation}
which is not far from the initial expectation $|g_\nu^{NN}(m_i=0)| \sim F_\pi^{-2} \sim 4.5$ fm$^2$ \cite{Cirigliano:2018hja}. To get conservative limits we set $g_\nu^{NN}(m_i=0) = -1.8$ fm$^2$, the smallest value within the range.

 Little is known about how this coupling varies as a function of $m_i$. To get a smooth description of $0\nu\beta\beta$ decay amplitudes as a function of $m_i$ the sum of the long-distance and hard-neutrino contributions must converge to the $\mathcal A_{L,R,M}^{(9)}(m_i)$, which  describes the contributions from sterile neutrinos integrated out at the quark level, for sufficiently large $m_i$. From this demand, interpolation formulae were derived \cite{Dekens:2020ttz}
\begin{eqnarray}\label{naiveinterNN}
g^{\rm NN}_{\nu} (m_i) &=& \frac{g^{\rm NN}_{\nu}(0) }{1  -  m_i^2 \frac{g^{\rm NN}_{\nu}(0)}{2}\left[\theta(m_0-m_i)\hat g_{1}^{\rm NN}(m_0) +\theta(m_i-m_0)\hat g_{1}^{\rm NN}(m_i)\right]^{-1}   }\,,
\end{eqnarray}
where
\begin{equation}
 \hat{g}^{\rm NN}_1(m_i) = \left( g^{\rm NN}_{1}(m_i) - \frac{1}{4} (1 + 3 g_A^2) \right)\,,
\end{equation}
and
\begin{eqnarray}\label{gNNRGE}
g_{1}^{\rm NN}(m_i) &=& g_{1}^{\rm NN}(m_0) - \frac{3}{4} \frac{\alpha_s}{\pi} \left( 1 - \frac{1}{N_c}\right)g_{1}^{\rm  NN}(m_0) \log \frac{m_i^2}{m_0^2}\,.
\end{eqnarray}
which takes into account the QCD evolution of the dimension-nine operators. The only unknown is then the value of $g_{1}^{\rm  NN}(m_0)$ which is the 
matrix element of a local four-quark-two-electron operator that has not been calculated with nonperturbative techniques. We follow Ref.~\cite{Dekens:2020ttz} and set $g_{1}^{\rm  NN}(m_0) = \frac{1}{4} (1 + 3 g_A^2)+1$. This ensures that the interpolation formula is smooth and is consistent with $g^{\rm NN}_{\nu} (m_i=0)$ as well as naive-dimensional-analysis estimates in the large $m_i$ limit. 

The next subamplitude we need is $ \mathcal A_R(m_i)$ which depends on both $C^{(6)}_{{\rm VLR}}$, and $C^{(6)}_{{\rm VRR}}$. 
From Eq.~\eqref{Anu},  $\mathcal A_R(m_i)$ involves the left-right mixing, which 
in the large $m_i$ limit generates the dimension-nine operators that induce a leading-order $\pi\pi ee$ operator. The same is true for the light $m_i$ case but then from hard-neutrino exchange. The hard-neutrino exchange part becomes
 \begin{equation}\label{hardR}
 \mathcal A_R^{(\nu)}(m_i) = \frac{m_\pi^2}{m_e v}\left[ \frac{1}{m_\pi^2}C^{\pi\pi}_{iR} M_{PS,\,sd} - \frac{2}{g_A^2} C^{NN}_{iR} M_{F,\,sd}\right]\,,
\end{equation}
where the coefficients are given by\footnote{We neglected a short-distance $nn\rightarrow pp+ee$ contribution proportional to $\left( C_{{\rm VRR}}^{(6)} \right)_{ei}\left( C_{{\rm VLR}}^{(6)} \right)_{ei}$ in the second equation. This contribution is poorly known and since it enters at the same level as the well understood $C_{iR}^{\pi\pi}$ contribution, we assume the latter to saturate the hard-neutrino contribution proportional to $\left( C_{{\rm VRR}}^{(6)} \right)_{ei}\left( C_{{\rm VLR}}^{(6)} \right)_{ei}$.} 
\begin{align}\label{CLR}
C_{iR}^{\pi\pi}& = 2\dfrac{m_i v}{F_\pi^2} g_{LR}^{\pi\pi}(m_i) \left( C_{{\rm VRR}}^{(6)} \right)_{ei}\left( C_{{\rm VLR}}^{(6)} \right)_{ei}\;,\nn\\
C^{NN}_{iR} &=  \frac{m_i v}{4} g_\nu^{NN}(m_i)  \bigg[\left(C_{{\rm VRR}}^{(6)}\right)^2_{ei}+\left(C_{{\rm VLR}}^{(6)}\right)^2_{ei}\bigg]\;. 
\end{align}
The pionic term is well known as it has been calculated on the lattice for large $m_i$ \cite{Nicholson:2018mwc} and is connected to the electromagnetic contribution to the pion mass splitting for $m_i=0$. We write
\begin{eqnarray}\label{naiveinter0}
g^{\pi \pi}_{\rm LR} (m_i)&=& \frac{g_{\rm LR}^{\rm \pi \pi}(0) }{1  -  m_i^2 \frac{4 g_{\rm LR}^{\rm \pi \pi}(0)}{F_\pi^2}\left[\theta(m_0-m_i)g^{\pi\pi}_{4}(m_0) +\theta(m_i-m_0)g^{\pi\pi}_{4}(m_i)\right]^{-1}   }\,,
\end{eqnarray}
 where $g_{\rm LR}^{\rm \pi \pi}(0) \simeq 0.8 F_\pi^2$ and $g^{\pi\pi}_{4}(m_0)= -(1.9\pm 0.2)$ GeV$^2$, and 
 \begin{equation}\label{gpipi7}
g_{4}^{\pi\pi}(m_i) = g_4^{\pi\pi}(m_0) + \frac{3}{4} \frac{\alpha_s}{\pi} \left( g_5^{\pi\pi}(m_0) - \frac{1}{N_c} g_{4}^{\pi\pi}(m_0)\right) \log \frac{m_i^2}{m_0^2}\,.
\end{equation} 
The renormalization group evolution (RGE) mixes $O_{4R}^{(9)}$ and  $O_{5R}^{(9)}$~\cite{Cirigliano:2018yza} and that is why the RGE of $g^{\pi\pi}_{4}$ involves the LEC $g^{\pi\pi}_{5}(m_0)=-(8.0\pm0.6)$ GeV$^2$ \cite{Nicholson:2018mwc}. The form of Eq.~\eqref{naiveinter0} agrees very well with a recent explicit lattice-QCD computation of the $m_i$ dependence of the pionic LEC \cite{Tuo:2022hft}. 

Finally, we have $\mathcal{A}_{M,E,m_e}^{(\nu)}$ which contributes through $\Delta L=2$ pion-nucleon-lepton and nucleon-nucleon-lepton vertices. Unfortunately, very little is known about the corresponding LECs and we therefore set $\mathcal{A}_{M,E,m_e}^{(\nu)}=0$ and take the long-distance pieces in Eq.~\eqref{Anu} as an approximation for the magnetic piece. Because $\mathcal{A}_{M,E,m_e}$ correspond to corrections that are already suppressed in the chiral perturbation theory power counting, see Ref.~ \cite{Cirigliano:2017djv}, which can be seen by the $m_\pi^2/m_N^2$ scaling of $M_{GT}^{MM}(m_i)$ and the missing $1/m_e$ in $\mathcal A_{E,m_e}$ compared to the other subamplitudes, this approximation does not lead to large uncertainties. 

The largest uncertainties of our expressions arise from the uncertainties in the NMEs and the short-distance corrections proportional to $g_\nu^{NN}(m_i)$. Combined they lead to an uncertainty in the $0\nu\beta\beta$ decay rate of roughly an order of magnitude \cite{Dekens:2020ttz,Graf:2022lhj}. We stress that any computation of $0\nu\beta\beta$ rates in the literature suffers from a similar uncertainty, and this uncertainty is implied in all results presented below. 

\subsubsection{Contributions from heavy neutrinos}
Neutrinos with masses $m_i > \Lambda_\chi$  are integrated out at the quark level as described in Eqs.~\eqref{dim9}-\eqref{dim9match}. The induced dimension-nine operators generate $\Delta L=2$ $\onbb$ decay operators in the chiral EFT Lagrangian. The associated contributions to $0\nu\beta\beta$ decay rates were calculated in Refs.~\cite{Prezeau:2003xn,Cirigliano:2018yza}. We do not give explicit expressions here as they are automatically obtained through the interpolation formulae in the previous subsection. That is, for $m_i \geq \Lambda_\chi$ the interpolation formulae ensure that the expressions for the long-distance and hard-neutrino amplitudes in $\mathcal{A}_{L,R}^{(\nu)}(m_i)$ automatically match onto the corresponding  $\mathcal{A}_{L,R}^{(9)}(m_i)$. For example, for $m_i > m_0 = 2$ GeV, the interpolation formula of the LEC $g_{LR}^{\pi\pi}(m_i)$ evaluates to $-F_\pi^2g_4^{\pi\pi}(m_i)/(4m_i^2)$ which would correspond to lepton-number-violating pionic interactions induced by the dimension-nine operators 
as identified in Ref.~\cite{Prezeau:2003xn}.
In this way, by modifying Eq.~\eqref{eq:amplitude_DBD} to
\begin{align}
\label{eq:amplitude_DBD2}
\mathcal A_{L,R} \equiv \sum_{i=1}^{6} \mathcal A_{L,R} (m_i) \;,
\end{align}
we include contributions from heavy neutrinos through the interpolation formulae. The RGE of the local dimension-nine operators is incorporated through the RGE of the LECS $ g_\nu^{NN}(m_i)$ and $g^{\pi \pi}_{\rm LR} (m_i)$ in Eqs.~\eqref{gNNRGE} and \eqref{gpipi7}, respectively.

The remaining subamplitudes $\mathcal{A}_{M,E,m_e}^{(\nu)}(m_i)$ are suppressed in the large-$m_i$ regime. The construction of the associated dimension-ten operators and the associated chiral Lagrangian has not been performed and we approximate the contributions in these regimes through the interpolation formulae for the NMEs in the long-distance contributions. We write
\begin{align}
\label{eq:amplitude_DBD3}
\mathcal A_{M,E,m_e} \equiv \sum_{i=1}^{6} \mathcal A_{M,E,m_e} (m_i) \;,
\end{align}
while setting $A^{(\nu)}_{M,E,m_e} =0$. This essentially corresponds to a factorization estimate for the heavy-neutrino contributions and is relatively uncertain. Eq.~\eqref{eq:amplitude_DBD3} gives the correct $m_i^{-2}$ behavior and the associated contributions are suppressed for large $m_i$.

\section{Phenomenology}
\label{sec:pheno}
We now investigate the $0\nu\beta\beta$ decay phenomenology in two scenarios of the mLRSM: the type-II and type-I seesaw dominance. In the former, the Dirac mass vanishes, $M_D=0$, and the neutrino masses are obtained by diagonalizing the Majorana mass matrices $M_L$ and $M_R$ (related to each other by Eq.~\eqref{eq:majorana_pc}). In the type-I seesaw dominance, the Majorana mass $M_L=0$ vanishes for $v_L=0$ and the Dirac mass matrix $M_D$ is fixed by the physical neutrino masses and the mixing matrices of the active, $U_{\text{PMNS}}$, and sterile, $U_R$, neutrinos. We discuss this in more detail below.

In both cases, the $\onbb$ decay rates depend on the mass of the right-handed gauge bosons, the scalar fields, and the sterile neutrino masses. 
For generalized parity $\mathcal P$ as the left-right symmetry, it is possible to limit right-handed gauge boson mass $m_{W_R}$
from low-energy measurements 
regardless of the sterile neutrino masses. The exact limits depend on how the strong CP problem is addressed
Without a Peccei-Quinn mechanism the limits strengthen considerably to $m_{W_R} \geq 17$ TeV  
and we require $\alpha \ll 1$ as discussed in Sec.~\ref{theta}. The most conservative constraints require an implementation of a Peccei-Quinn mechanism and lead to $m_{W_R} \geq 5.5$ TeV 
\cite{Dekens:2021bro}. The relatively weaker constraints arise because of cancellations between various contributions to kaonic CP violation and require a non-zero value for the combination $2\xi \sin\alpha/(1-\xi^2)\simeq -0.01$.
For generalized charge conjugation $\mathcal{C}$ as the left-right symmetry, there are additional phases such that the contributions to kaonic CP violation and neutron EDM can be made to vanish by appropriate tuning, and no lower limit on $m_{W_R}$ is obtained~\cite{Bertolini:2019out}

Direct searches for the right-handed gauge boson $W_R$ decaying into sterile neutrino at the LHC~\cite{ATLAS:2018dcj,ATLAS:2019isd,CMS:2018agk} 
are able to exclude the $W_R$ mass up to $5\tev$,
along with which the parameter space of sterile neutrino mass above $100\gev$ is also constrained.
Ref.~\cite{Nemevsek:2018bbt} shows that for the sterile neutrino mass below $45\gev$, $m_{W_R} \lesssim 5\tev$ is excluded by recasting the search for a new charged gauge boson in the final state of an electron and missing energy at the LHC Run-2 with the integrated luminosity of $36\fbi $~\cite{ATLAS:2017jbq}. The CMS Collaboration used
 the full LHC Run-2 data of $138\fbi$ and obtained that the new charged gauge boson lighter than
$5.4\tev$ is excluded~\cite{CMS:2022yjm}~\footnote{
Ref.~\cite{Nemevsek:2018bbt} made a projection for the integrated luminosity of $300\fbi$, from which it is expected that $m_{W_R} \lesssim 5.4\tev$ is excluded at the LHC Run-2.}. Besides, an individual bound on the left-right mixing parameter $\zeta\leq 1.25\times 10^{-3}$ can be obtained by requiring the CKM unitarity~\cite{Seng:2018yzq,Li:2020flq}, which is weaker than the constraints from direct searches for $W_R$.

In Ref.~\cite{Li:2022cuq}, it shows that the ratio of vevs $\xi\equiv \tan\beta$ is constrained by the $\rho$ parameter, $\rho = 1 + [\sin^2(2\beta) +  (1-\tan^2\theta_W)^2/2]\lambda$~\cite{Czakon:1999ga}, where $\theta_W$ is the weak mixing angle. By using the experimental value of $\rho$, we obtain $\xi\leq 0.34$ for $m_{W_R}=6\tev$. A larger $\xi$ is allowed by the $\rho$ parameter for a heavier $W_R$. Note that $\xi<0.8$ is required to ensure that no couplings in the mLRSM scalar sector become too large for perturbation theory \cite{Maiezza:2010ic}. In our analysis, we will choose two benchmark values of $\xi=0,0.3$. 

On the other hand, constraints exist from low-energy processes and astrophysics for sterile neutrino masses at the MeV scale \footnote{There also exist constraints on the mass of the lightest sterile neutrino from Big Bang Nucleosynthesis (BBN)~\cite{Nemevsek:2011aa}. A analysis in the context of type-I seesaw models was performed in Ref.~\cite{Sabti:2020yrt}, from which  we can estimate the sensitivities in the mLRSM with the substitution $\mid\theta_{\alpha}\mid^2\to \lambda^2[1+\sin^2(2\beta)] \mid(U_{\rm PMNS})_{\alpha j}\mid^2 $ for $\alpha=e,\mu,\tau$ and $j=1,2,3$. For $m_{W_R}=7\tev$,  roughly the lightest neutrino mass below $200\mev$ can be excluded depending on $(U_{\rm PMNS})_{\alpha j}$ and $\sin(2\beta)$; while for $m_{W_R}=25\tev$,  the lightest neutrino mass down to $10\mev$ is allowed by BBN when the lifetime of the lightest neutrino is very long (reaching about 40 seconds~\cite{Boyarsky:2020dzc}). We will not consider these constraints that depend on the thermal history of the universe. But the main results of this work do not change if the constraints~\cite{Sabti:2020yrt} are taken into account.
}. From Refs.~\cite{Carpentier:2010ue,Li:2020wxi}, the low-energy precision measurements of meson decay branching ratios and lepton flavor universality constrain $m_{W_R}$ for the sterile neutrino mass below the pion mass. For the sterile neutrino mass $\sim 80\mev$, the constraint is the most stringent, which is $m_{W_R}\gtrsim 3.4\tev$~\cite{Li:2020wxi}.
If the lightest sterile neutrino has a mass below $10$ MeV, severe constraints appear from supernovae cooling \cite{Barbieri:1988av,Nemevsek:2011aa}, which leads to the bound $m_{W_R}\gtrsim 21\tev ~(23\tev)$ for $\xi=0 ~(0.3)$. 

In Sec.~\ref{EFTform}, we saw that $\onbb$ decay also receives contributions from doubly charged scalar $\delta_R^{--}$, which are described by Wilson coefficients in Eq.~\eqref{scalar}. The ratio of these contributions to the contributions from sterile neutrinos is $\sim \max\{m_i^2/m_{\delta_R}^2,|{\bf q}|^2/m_{\delta_R}^2\}$, where $|{\bf q}|\sim m_\pi$ is typical nuclear scale. Direct collider searches for doubly charged scalars  exclude $m_{\delta_R}\leq 870\gev$~\cite{CMS:2017fhs,ATLAS:2017xqs,ATLAS:2021jol}. 
When the left-right symmetry is assumed, the constraints from $\mu\to eee$ and $\mu\to e\gamma$ give $m_i^2/m_{\delta_R}^2\lesssim 0.8\%$ for $m_{W_R}=6\tev$~\cite{Dev:2018sel}. In Ref.~\cite{Maiezza:2016ybz}, it was found that $m_{\delta_R}> 12$~TeV $m_{W_R} = 6$~TeV  from the perspective of perturbativity of the scalar potential. 
Considering these constraints, we therefore neglect the contributions of $\delta_R^{--}$ as they are negligible for the sterile neutrino masses under consideration.

\begin{table}[t]
\caption{The fitted three active-neutrino oscillation parameters taken from Ref.~\cite{Esteban:2020cvm}. The squared mass difference $\Delta m_{ij}^2=m_i^2-m_j^2$, and $\Delta m_{31}^2>(<)~0$ in the normal (inverted) hierarchy.}
\begin{center}
\renewcommand{\arraystretch}{2.5}
\begin{tabular}{c||c|c|c|c|c|c}
\hline 
 & $\sin^2\theta_{12}$ & $\sin^2\theta_{23}$ & $\sin^2\theta_{13}$ &  $\dfrac{\Delta m_{21}^2}{10^{-5}\ev^2}$ &  $\dfrac{|\Delta m_{31}^2|}{10^{-3}\ev^2}$ & $\delta/^\circ$\\ \hline \hline
NH & $0.304^{+0.013}_{-0.012}$ & $0.570^{+0.018}_{-0.024}$ & $0.02221^{+0.00068}_{-0.00062}$ & $7.42^{+0.21}_{-0.20}$ & $2.514^{+0.028}_{-0.027}$ & $195^{+51}_{-25}$ \\ \hline
IH & $0.304^{+0.013}_{-0.012}$ & $0.575^{+0.017}_{-0.021}$ & $0.02240^{+0.00062}_{-0.00062}$ & $7.42^{+0.21}_{-0.20}$ & $2.497^{+0.028}_{-0.028}$ & $286^{+27}_{-32}$
\\ \hline
\end{tabular}
\renewcommand{\arraystretch}{1}
\end{center}
\label{tbl:oscillation_data}
\end{table}

The other constraints are obtained from neutrino oscillation. 
The squared mass difference and mixing angles measured in current neutrino oscillation experiments,  tabulated in Table~\ref{tbl:oscillation_data}, are used as input for the $\onbb$ decay. The active neutrino masses are known once the lightest neutrino mass is chosen and the hierarchy specified. In the normal hierarchy (NH), $m_1<m_2<m_3$, 
\begin{align}
m_2 = \left(m_1^2+\Delta m_{21}^2\right)^{1/2}\;,\quad  m_3 = \left(m_1^2+\Delta m_{31}^2\right)^{1/2}\;,
\end{align}
while in the inverted hierarchy (IH), $m_3<m_1<m_2$,
\begin{align}
m_1 = \left(m_3^2-\Delta m_{31}^2\right)^{1/2}\;,\quad  m_2 = \left(m_3^2+\Delta m_{21}^2-\Delta m_{31}^2\right)^{1/2}\;
\end{align}
The $3\times 3$ PMNS  matrix is defined in the usual way
\begin{equation}
U_{\rm{PMNS}} = \left(\begin{array}{ccc}
1 & 0 & 0 \\
0 & c_{23} & s_{23} \\
0 & -s_{23} & c_{23}
\end{array}\right)\left(\begin{array}{ccc}
c_{13} & 0 & s_{13} e^{-i \delta} \\
0 & 1 & 0 \\
-s_{13} e^{i \delta} & 0 & c_{13}
\end{array}\right)\left(\begin{array}{ccc}
c_{12} & s_{12} & 0 \\
-s_{12} & c_{12} & 0 \\
0 & 0 & 1
\end{array}\right)\left(\begin{array}{ccc}
1 & 0 & 0 \\
0 & e^{i \lambda_1} & 0 \\
0 & 0 & e^{i \lambda_2}
\end{array}\right)\;.
\end{equation}
 The Majorana phases $\lambda_{1,2}$ are unknown and marginalized in calculating the $\onbb$ decay rate.

\subsection{Type-II seesaw dominance} 
In the type-II seesaw dominance scenario, $M_D = 0$, and the neutrino mass matrix becomes
\begin{equation} 
M_n =  \bma M_L &0 \\ 0 &M_R \ema\,\,.
\end{equation}
From Eqs.~\eqref{eq:seesaw} and \eqref{eq:diag2}, we obtain
\begin{align} 
U^\dagger_{\mathrm{PMNS}} M_L U_{\mathrm{PMNS}}^* &= \widehat{M}_\nu \,,&
U^\dagger_{R} M_R U_{R}^* &= \widehat{M}_N\,.
\end{align}
The relations between the Majorana mass terms $M_L$ and $M_R$ are given in Eq.~\eqref{eq:majorana_pc}. One can immediately obtain that the right-handed neutrino mixing matrix $U_R$ is related to the PMNS matrix~\cite{Joshipura:2001ui},
\begin{align}
\mathcal{P}:\quad & U_R = U_{\mathrm{PMNS}}^* e^{-i \theta_L/2} \;, \nn \\
\mathcal{C}:\quad & U_R = U_{\mathrm{PMNS}} e^{i \theta_L/2} \;.
\end{align}
In both cases, the neutrino masses obey the relation
\begin{align}
\widehat{M}_N = \dfrac{v_R}{v_L}\widehat{M}_\nu\;.
\end{align}
More explicitly, for the NH the masses of sterile neutrinos $m_{4,5,6}$ are related to $m_{1,2,3}$ via
\begin{align}\label{massNH}
m_{4,5}=\dfrac{m_{1,2}}{m_3}m_{6}\;.
\end{align} 

The dependence of the sterile neutrino masses $m_{4,5}$ on the lightest neutrino mass in the NH is depicted in Fig.~\ref{fig:mvmin_mvR_NH}. 
 For $m_6 = 10, 100, 1000\gev$ we can see that $m_4\leq 1 \gev$ for the lightest neutrino mass $ m_1 \leq 10^{-2}, 10^{-3}, 10^{-4}$~eV, respectively and sterile neutrinos become explicit degrees of freedom in chiral EFT. 

\begin{figure}[t]
\centering
\includegraphics[width=0.35\columnwidth]{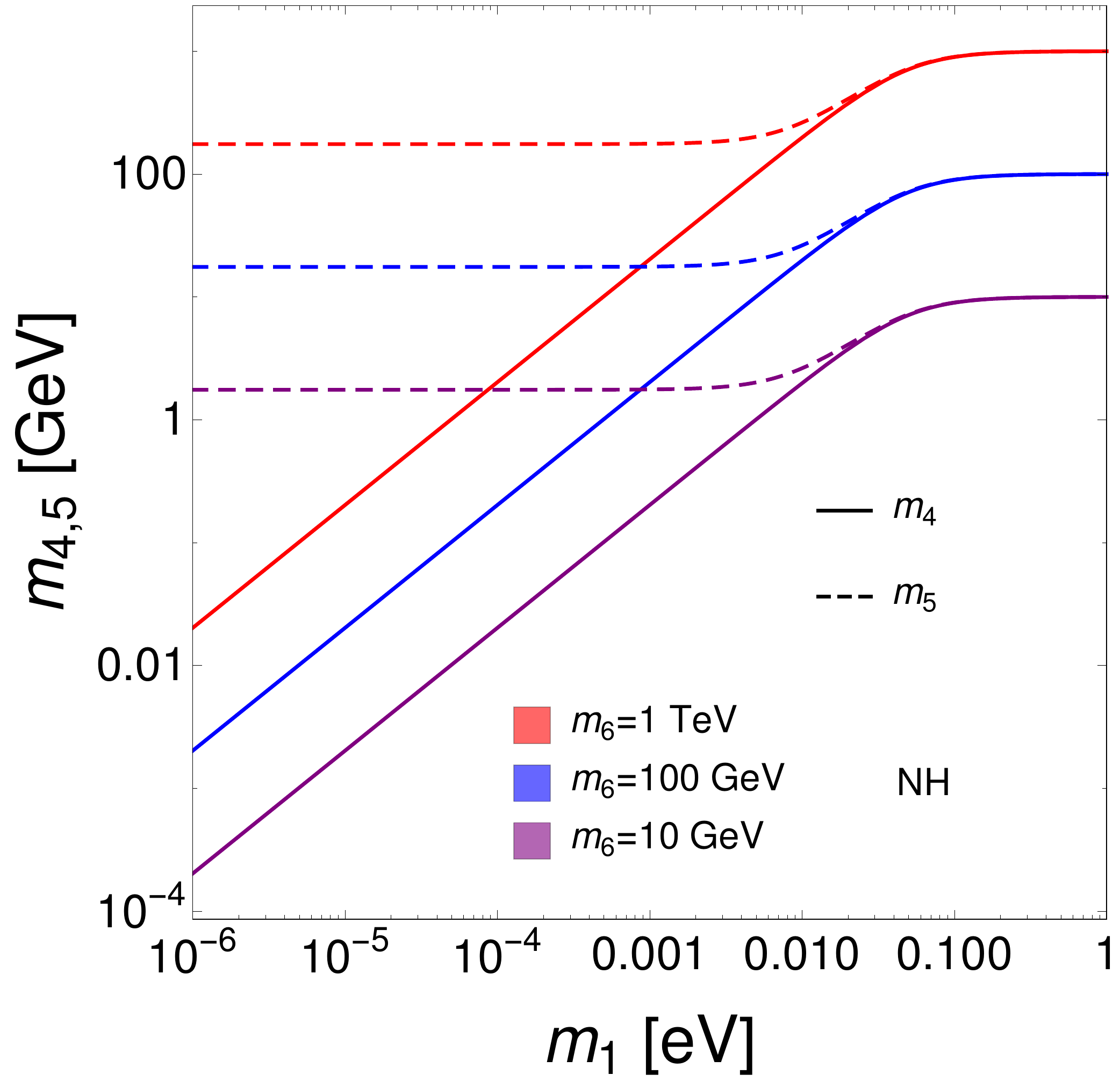}
\caption{Sterile neutrino masses $m_4$ (solid) and $m_5$ (dashed) as a function of the lightest neutrino mass $ m_1$ for the heaviest neutrino mass $m_6=10\gev$ (purple), $100\gev$ (blue) and $1\tev$ (red) in the NH.}
\label{fig:mvmin_mvR_NH}
\end{figure}
\begin{figure}[t]
  \centering
  \subcaptionbox{\label{typeII:a}}{\includegraphics[width=0.3\columnwidth]{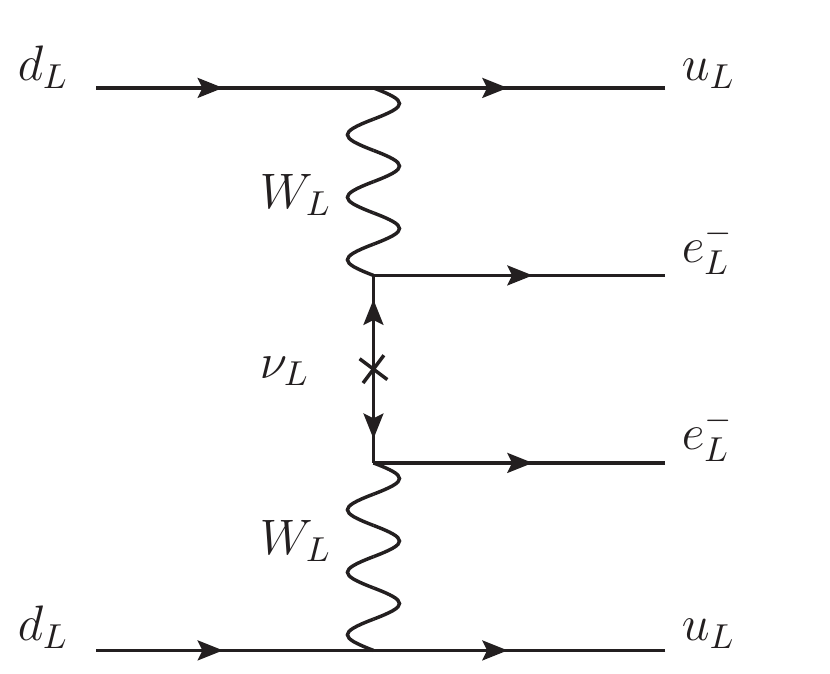}}\hspace{1em}%
  \subcaptionbox{\label{typeII:b}}{\includegraphics[width=0.3\columnwidth]{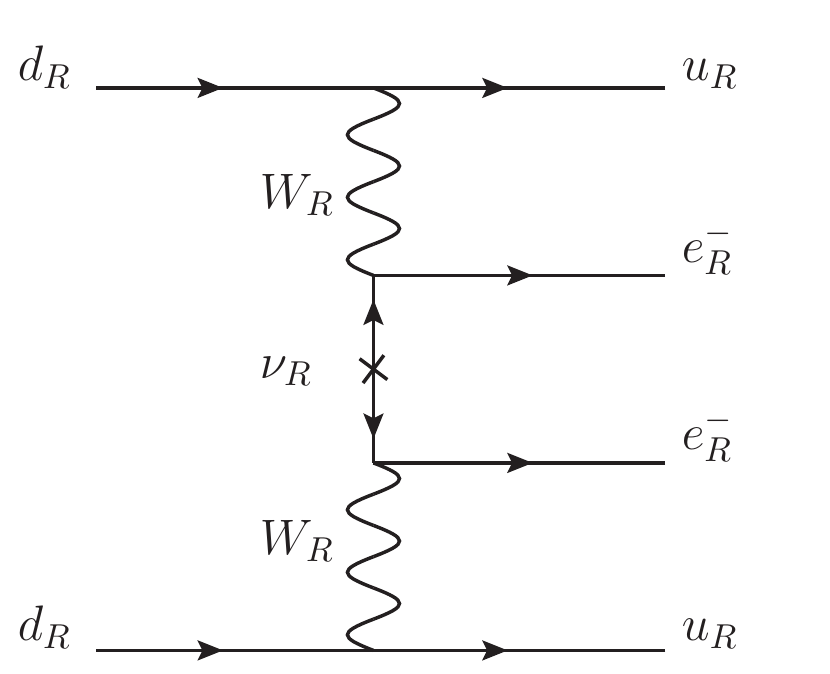}}\hspace{1em}\\
  \subcaptionbox{\label{typeII:c}}{\includegraphics[width=0.3\columnwidth]{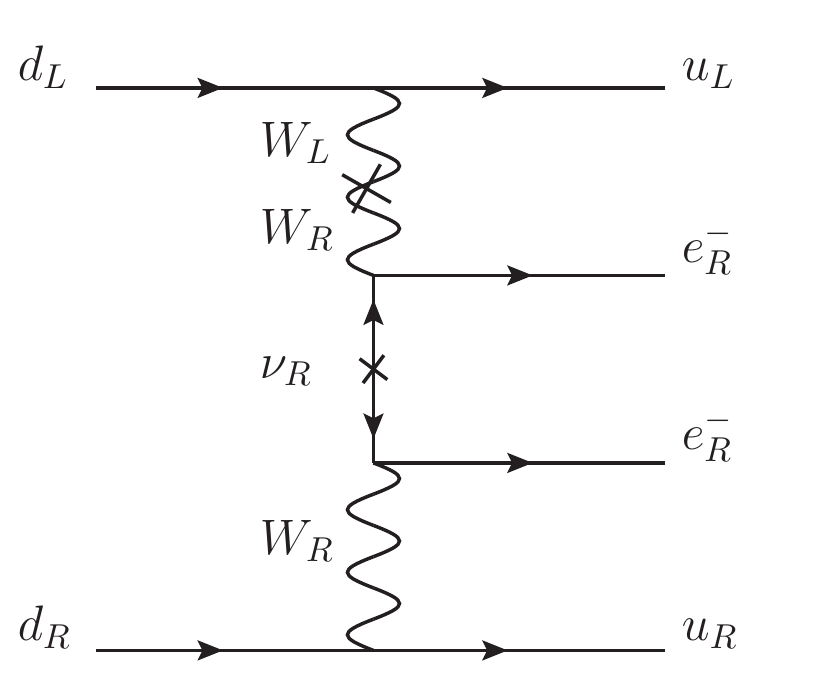}}\hspace{1em}%
  \subcaptionbox{\label{typeII:d}}{\includegraphics[width=0.3\columnwidth]{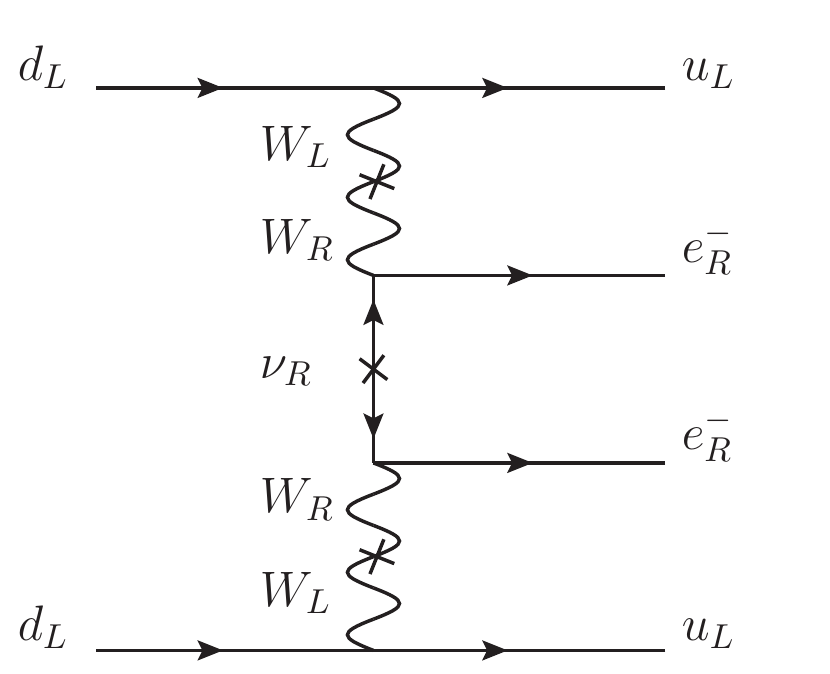}}
\caption{
The diagrams responsible for the $\onbb$ decay independent  of the Dirac mass term. The cross vertices $(\times)$ in the fermion lines of $\nu_L$ and $\nu_R$ denote the insertion of Majorana mass terms $M_L$ and $M_R$, respectively, while that in between $W_L-W_R$ denotes the left-right mixing. In the limit of type-II dominance, there is no mixing between $\nu_L$ and $\nu_R$. 
}
\label{fig:typeII}
\end{figure}

The limit $M_D\rightarrow 0$ simplifies the $0\nu\beta\beta$ decay rate expressions. Several subamplitudes vanish
 \begin{equation}
\mathcal A_M\sim \left(C_{{\rm VLL}}^{(6)}\right)_{ei}\left(C_{{\rm VLR}}^{(6)}\right)_{ei} \sim (P U)_{ei} (P_s U^*)_{ei} \sim \left(\left(U_{\rm{PMNS}} \, 0\right)\left( 0\,U_R^*\right)^T\right)_{ee} =0\,,
\end{equation}
and similar for $\mathcal A_{E,m_e}$. The amplitudes $\mathcal A_{L,R}$ remain: $\mathcal A_{L}$ describes the contributions from the exchange of active neutrinos in the standard mechanism depicted in  Fig.~\ref{fig:typeII}(a), while $\mathcal{A}_R$ comes from sterile neutrino exchange in Fig.~\ref{fig:typeII}(b)(c)(d). 

The inverse half-life now takes the simple form
\bea\label{eq:T1/2TYPE2}
\left(T^{0\nu}_{1/2}\right)^{-1} &=& g_A^4 \left[ G_{01} \, \left( |\mathcal A_{L}|\sq + |\mathcal A_{R}|\sq \right)
- 2 (G_{01} - G_{04}) \textrm{Re} \mathcal A_{L}^* \mathcal A_{R} \right]\,,\eea
and depends on several parameters.
\begin{itemize}
\item The mass of right-handed gauge bosons $m_{W_R}$, the vevs  ratio $\xi$ that tracks the mixing between $W_L$ and $W_R$, and the spontaneous CP phase $\alpha$. 
\item Neutrino masses. In the type-II seesaw scenario, the unknowns are, in addition to the mass ordering, the mass of the lightest neutrino $\mvmin$ and the mass of the heaviest neutrino $m_{N\textrm{max}}$ ($\mvmin=m_1$ and $m_{\textrm{max}}=m_6$ in the NH). Once these are specified, the other masses are given by Eq.~\eqref{massNH}. 
\item Two Majorana phases $\lambda_{1,2}$ in the PMNS matrix and the additional phase $\theta_L$ that appears in the right-handed neutrino mixing matrix.
\end{itemize}
Thus, eight parameters in the type-II seesaw scenario are shown in Tab.~\ref{tab:parameters}. 
We have checked that in this scenario the $\onbb$ decay rates are insensitive to whether $\mathcal{P}$ or $\mathcal{C}$ is chosen, and the value of $\alpha$, for the cases in Tab.~\ref{tab:parameters}, once the other unknown phases $\lambda_{1,2}$ and $\theta_L$ are varied.
Additional parameters appear if we include contributions from the scalar sector. Considering the stringent limits on their masses, we neglect their subleading contributions in what follows. \\
\\

\begin{table}
\centering
\caption{Independent parameters for the cases we consider.}
\begin{tabular}{l|l|l|l}
\hline 
 & seesaw  & $U_R$  & parameters \\
\hline 
\hline 
\multirow{2}{*}{$\mathcal{P}$} & type-II  & $U_R=U_{\text{PMNS}}^* e^{-i\theta_L/2}$  & $\lambda_1,\lambda_2, \theta_L, m_1,m_6,m_{W_R}, \xi,\alpha$ \\
\cline{2-4} 
 & type-I & $U_R=U_{\text{PMNS}}^* $  & $\lambda_1,\lambda_2, \theta_L, m_1, m_4, m_5,m_6,m_{W_R}, \xi, \alpha$  \\
\hline 
\multirow{2}{*}{$\mathcal{C}$} & type-II  & $U_R=U_{\text{PMNS}} e^{i\theta_L/2}$  & $\lambda_1,\lambda_2, \theta_L, m_1,m_6,m_{W_R}, \xi,\alpha$ \\
\cline{2-4} 
 & type-I & $U_R=U_{\text{PMNS}} $  & $\lambda_1,\lambda_2, \theta_L, m_1, m_4, m_5,m_6,m_{W_R}, \xi, \alpha$  \\
\hline 
\end{tabular}
\label{tab:parameters}
\end{table}

\noindent
 \textbf{Analysis I.} We assume that the mass of the heaviest sterile neutrino is relatively large $\mnmax = 1\tev$.
We begin by setting $m_{W_R}=7$ TeV
and pick two values for $\xi$: $\xi=0$ and $\xi=0.3$. 
 The phases $\lambda_{1,2}$ and $\theta_L$ are marginalized to obtain a lower and upper bound for the $0\nu\beta\beta$ decay rates in the mLRSM. We use the shell-model NMEs \cite{Menendez:2017fdf} and a fixed value $g_\nu^{NN}(m_i=0) = -1.8$ fm$^2$. Other choices of LECs or NMEs give qualitatively similar results on the log-log plots but can shift predictions up or down by roughly a factor $3$. 

 The resulting ${}^{136}$Xe lifetime is shown in the left panel of Fig.~\ref{fig:half-life_II_1} as a function of $m_1$ in the NH. In the NH, the lightest sterile neutrino mass $m_4$ is directly correlated with $m_1$, $m_4 = m_1\times (m_{N\textrm{max}}/m_3) \simeq 2\cdot 10^{13}\, m_1$ for $m_1<0.01$ eV. The red (blue) band corresponds to $\xi=0$ ($\xi=0.3$). In contrast, the purple band corresponds to the so-called standard mechanism:  the exchange of three light Majorana neutrinos in the limit of $m_{W_R}\rightarrow \infty$ and no additional sterile neutrinos.

\begin{figure}[t]
\centering
\includegraphics[width=0.32\columnwidth]{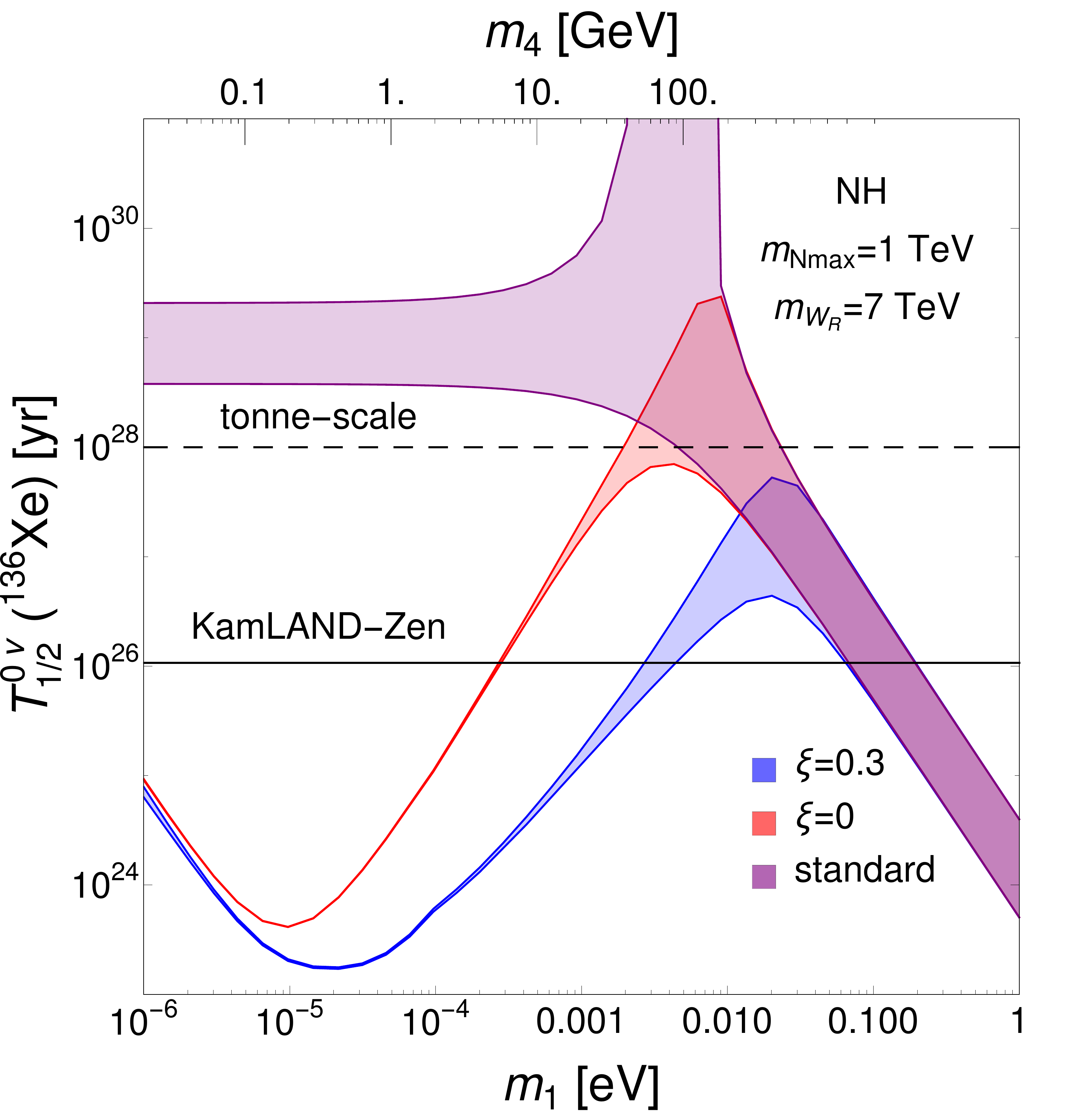}
\includegraphics[width=0.32\columnwidth]{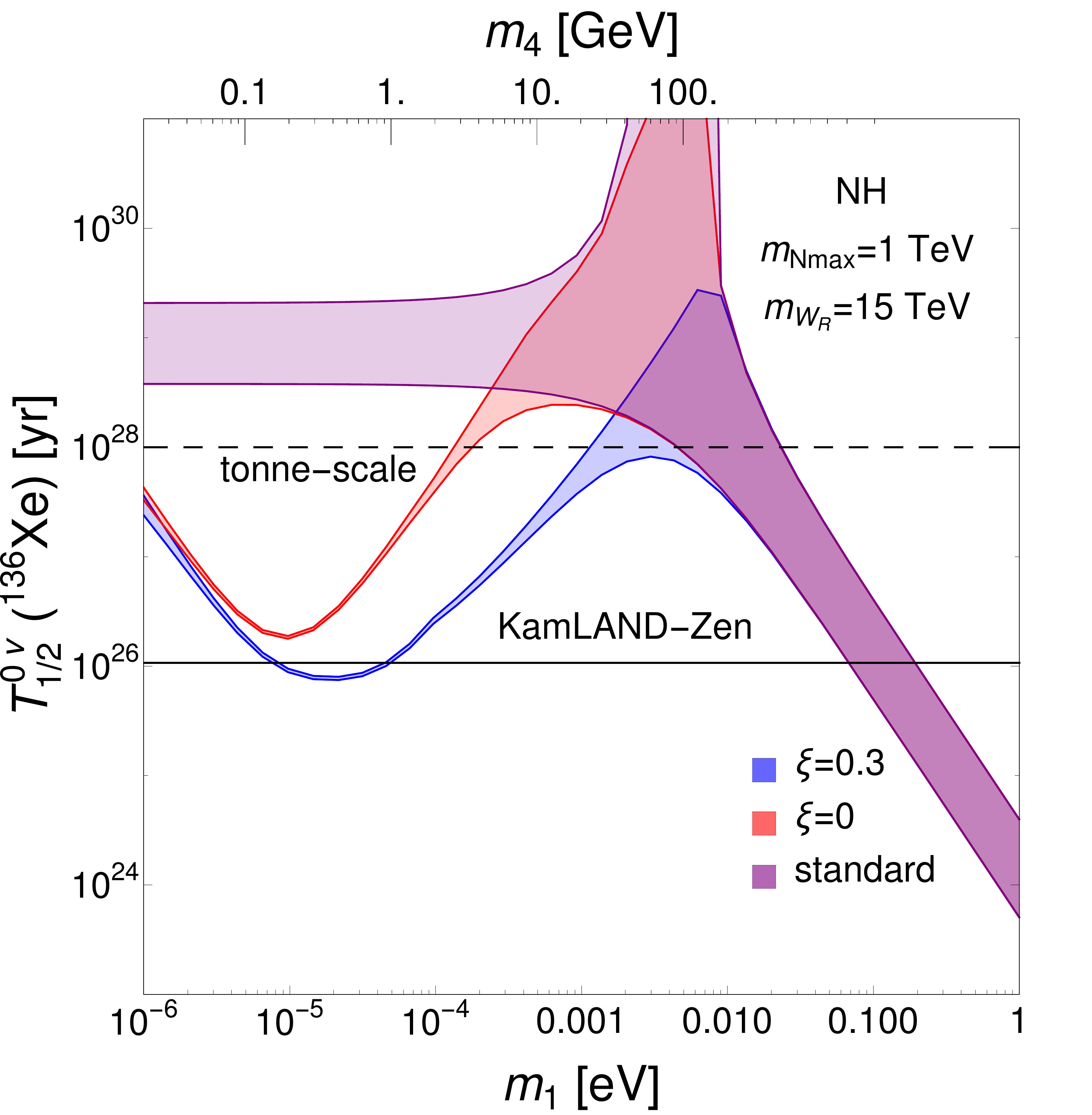}
\includegraphics[width=0.32\columnwidth]{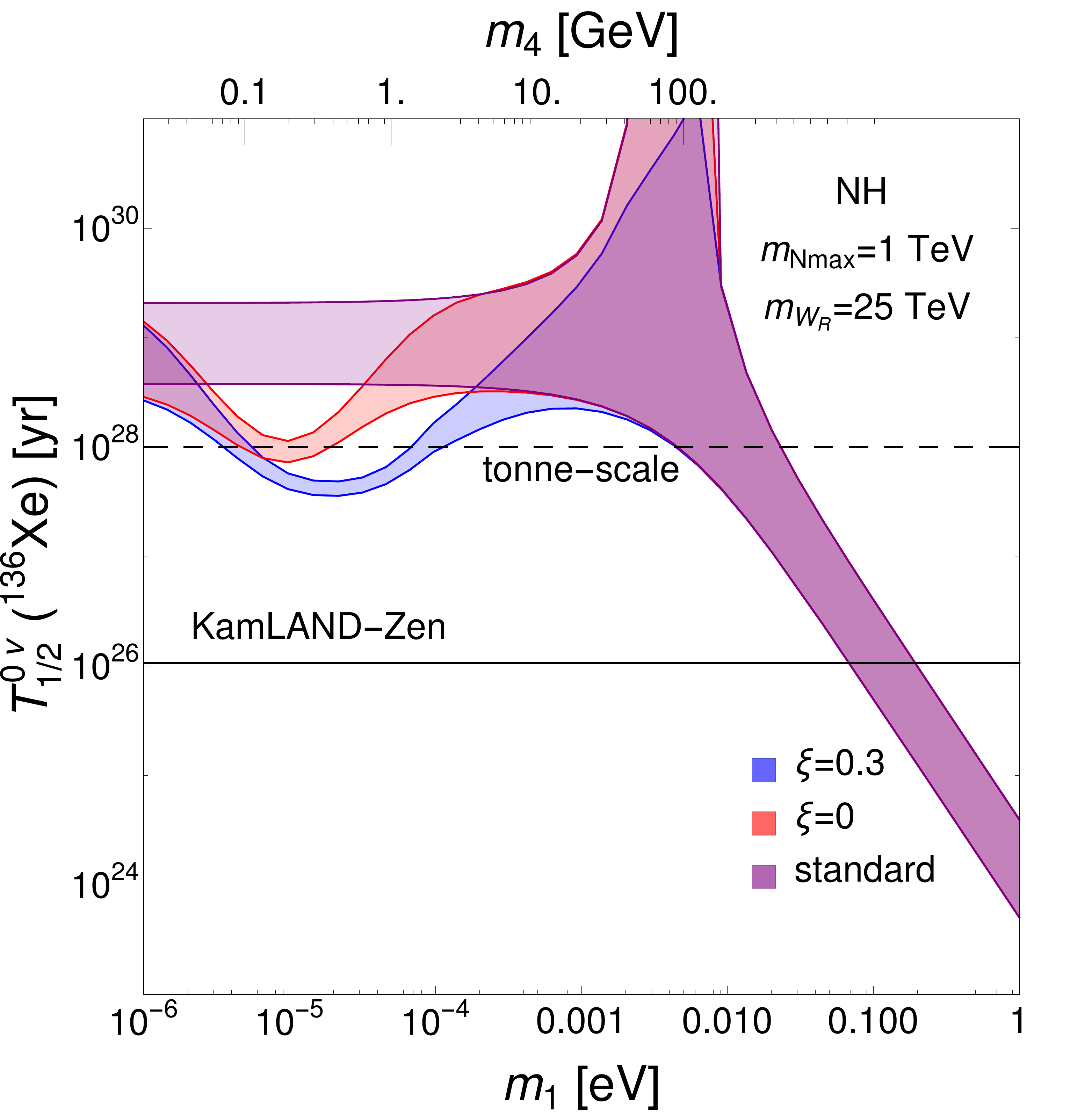}
\caption{
The half-life as a function of the lightest neutrino mass $m_1$ in the type-II dominance scenario for the NH with $m_{N\text{max}}=1\tev$ and $m_{W_R}=7 \tev$ (left), $15\tev$ (middle), $25\tev$ (right). The red and blue regions are obtained with the parameter $\xi=0$ and 0.3, respectively. The purple regions correspond to the standard mechanism contributions. The lightest sterile neutrino mass $m_4$ for given $m_1$ is shown in the upper frames. The solid and dashed black lines denote the current constraint from the KamLAND-Zen experiment and the expected constraint from tonne-scale experiments, respectively. The uncertainty associated to hadronic and nuclear matrix elements is not shown. 
}
\label{fig:half-life_II_1}
\end{figure}

Several things can be learned from the left panel of Fig.~\ref{fig:half-life_II_1}. First of all, for $m_1 > 0.01$ eV, and thus $m_4>200$ GeV, the $0\nu\beta\beta$ decay rate  is dominated by the standard mechanism independent of $\xi$.
For smaller $m_1$, $m_4$, there are larger contributions from non-standard mechanisms~\cite{Li:2020flq}, which, depending on the  sterile neutrinos masses, are captured by the dimension-six or dimension-nine operators in Sec.~\ref{EFTform}. Where the non-standard mechanisms take over depends on the value of $\xi$. For nonzero $\xi$ we have a sizable contribution to $(C^{(6)}_{{\rm VRR}})_{ei} (C^{(6)}_{{\rm VLR}})_{ei}$ (for sterile masses above a few GeV, these contribute through the dimension-nine operator $O_{4} \bar{e}_R C \bar{e}_R^T$ with the matching of coefficients given in Eq.~\eqref{dim9match}). The $0\nu\beta\beta$ decay rate then gets the dominant contribution from $\pi\pi\rightarrow e e$ interactions proportional to $g_{4,5}^{\pi\pi}$ that do not appear for $\xi=0$~\cite{Prezeau:2003xn,Cirigliano:2018yza,Li:2020flq}, see Eqs.~\eqref{hardR}-\eqref{naiveinter0}. Neglecting dimensionless $\mathcal O(1)$ numbers, the ratio of the first and second term in Eq.~\eqref{hardR} is given by
\begin{equation}
\frac{C^{\pi\pi}_{iR}}{ m_\pi^2 C^{NN}_{iR}} \sim \frac{\xi}{m^2_{\pi}f^2_{\pi}} \frac{g^{\pi\pi}_{LR}(m_i)}{g^{NN}_{\nu}(m_i)} 
\xrightarrow[m_4\gg \Lambda_\chi]{} \xi \frac{g_4^{\pi\pi}}{m_\pi^2 g_1^{NN}}\sim \xi\frac{\Lambda_\chi^2}{m_\pi^2}\,,
\end{equation} 
enhanced by two powers in the chiral power counting if $\xi = \mathcal O(1)$. This enhancement largely disappears for $m_4 < \Lambda_\chi$, explaining why the blue and red bands are much closer for small $m_1$.

There is a second turnover point for small enough $m_1$ and $m_4$. The exact point depends on the value of $\xi$ (this is driven by non-trivial mass dependence of the  $g^{\pi\pi}_{LR}(m_i)$, not considered in earlier literature) but occurs around $m_4 \sim m_\pi$. For such small sterile neutrino masses, we can no longer neglect the momentum-dependent term in the neutrino propagator, which scales as $m_{4}/{\bf q}^2$ for $m_{4}\ll |{\bf q}| $, where $|{\bf q}|\sim m_\pi$ is related to the typical nuclear binding momentum. 

For the relatively small value of $m_{W_R}$, $m_{W_R}=7\tev$, there are good prospects for observing a positive signal of $\onbb$ decay for $\mathrm{meV}<m_1 < 0.1$ eV, as pointed out in Ref.~\cite{Li:2020flq}. Next-generation experiments can cover most of the parameter space apart from a remaining sliver around $m_1 =  0.01$ eV for $\xi=0$. The region of $m_1<  \mathrm{meV}$ ($m_4 < 20\gev$) is almost excluded, where the predicted lifetimes are above the existing KamLAND-Zen limits.

In the middle and right panels of Fig.~\ref{fig:half-life_II_1} we increase the mass of the right-handed gauge boson gradually to $m_{W_R} = 15\tev$ and $25\tev$, respectively. In case of $m_{W_R} = 15\tev$, a large portion of the parameter space for $m_1<\mathrm{meV}$ ($m_4<20\gev$) can be probed in next-generation experiments. In case of $m_{W_R} = 25\tev$, the contributions from the exchange of $W_R$ are suppressed, and the $0\nu\beta\beta$ decay rate is dominated by the standard mechanism for $m_1 > 3\,\mathrm{meV}$. Depending on $\xi$, next-generation experiments can make a detection for $100\,\mathrm{MeV}<m_4<2$ GeV. For even higher values of $m_{W_R}$, the $0\nu\beta\beta$ decay rate will be dominated by the standard mechanism regardless of sterile neutrino masses.
\\
\\
 \textbf{Analysis 2.} We now calculate the half-lives of the $\onbb$ decay for much smaller sterile neutrino masses. We set $\mnmax=m_6=10\gev$ and display the results in the NH in Fig.~\ref{fig:half-life_II_2}.  There now exist cancellation between different contributions to $\mathcal{A}_R$ when varying the phases $\lambda_1$ and $\lambda_2$, which lead to dramatically increasing half-lives in specific ranges of $m_1$. These cancellations are similar to the cancellations occurring in the standard mechanism for NH and correspond to a sliver of parameter space\footnote{To demonstrate this, we consider the case of $m_{W_R}=7\tev$ and  $\xi=0.3$ as in the left panel of Fig.~\ref{fig:half-life_II_2}. For $m_1=2\cdot 10^{-4}\ev$, by setting $\alpha=0$ in the $\mathcal{C}$-symmetric mLRSM the amplitudes $\mathcal{A}_R(m_4)=-3.7\cdot 10^{-6}e^{i\theta_L}$, $\mathcal{A}_R(m_5)=-3.2\cdot 10^{-6}e^{i(\theta_L+2\lambda_1)}$, and $\mathcal{A}_R(m_6)=(5.1+4.8i)\cdot 10^{-8}e^{i(\theta_L+2\lambda_2)}$. For $\lambda_1\simeq \pi/2$, there exists  a large cancellation between $\mathcal{A}_R(m_4)$ and $\mathcal{A}_R(m_5)$.\label{footnote:cancellation-II}}.
 \begin{figure}[t]
\centering
\includegraphics[width=0.32\columnwidth]{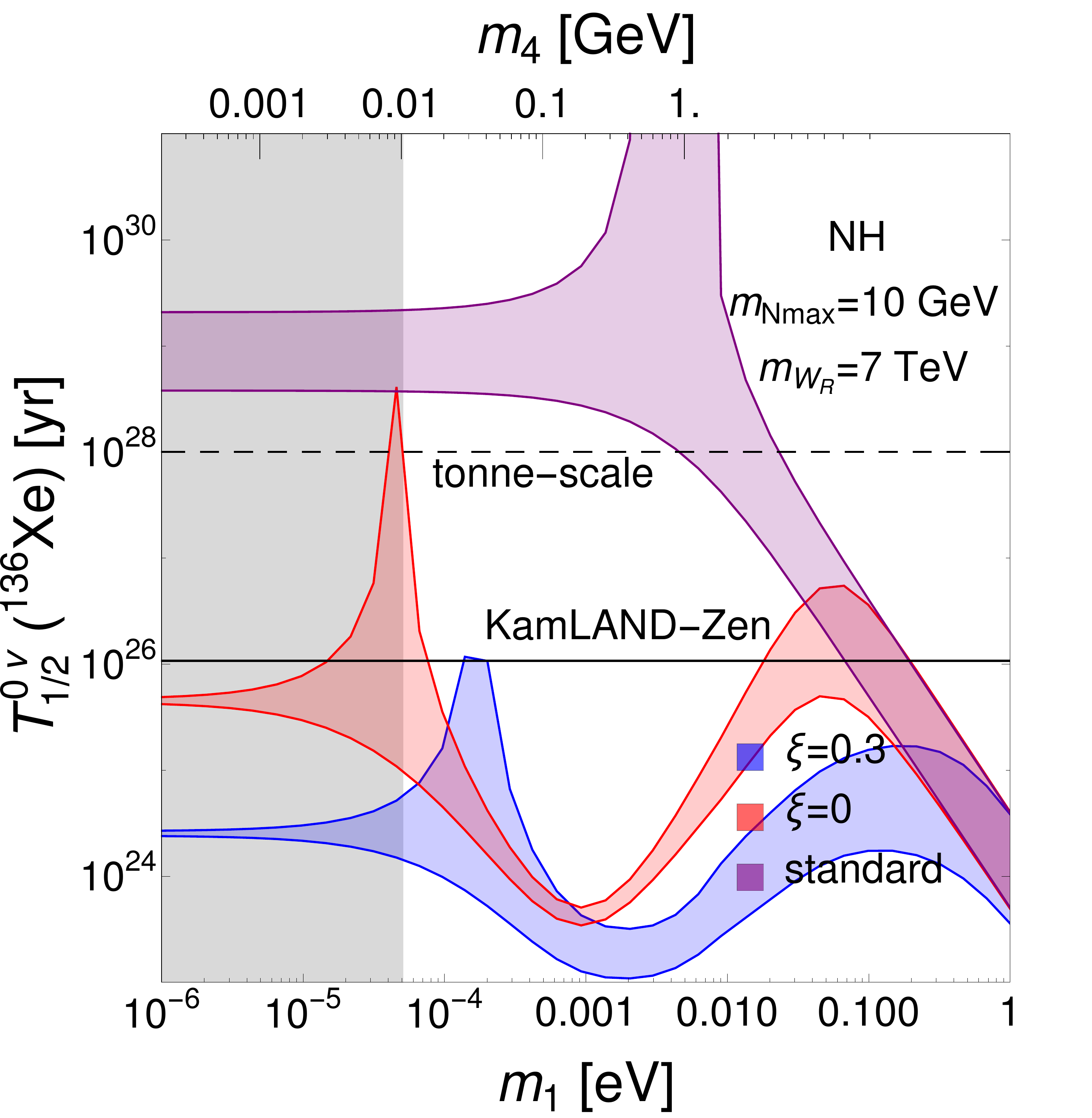}
\includegraphics[width=0.32\columnwidth]{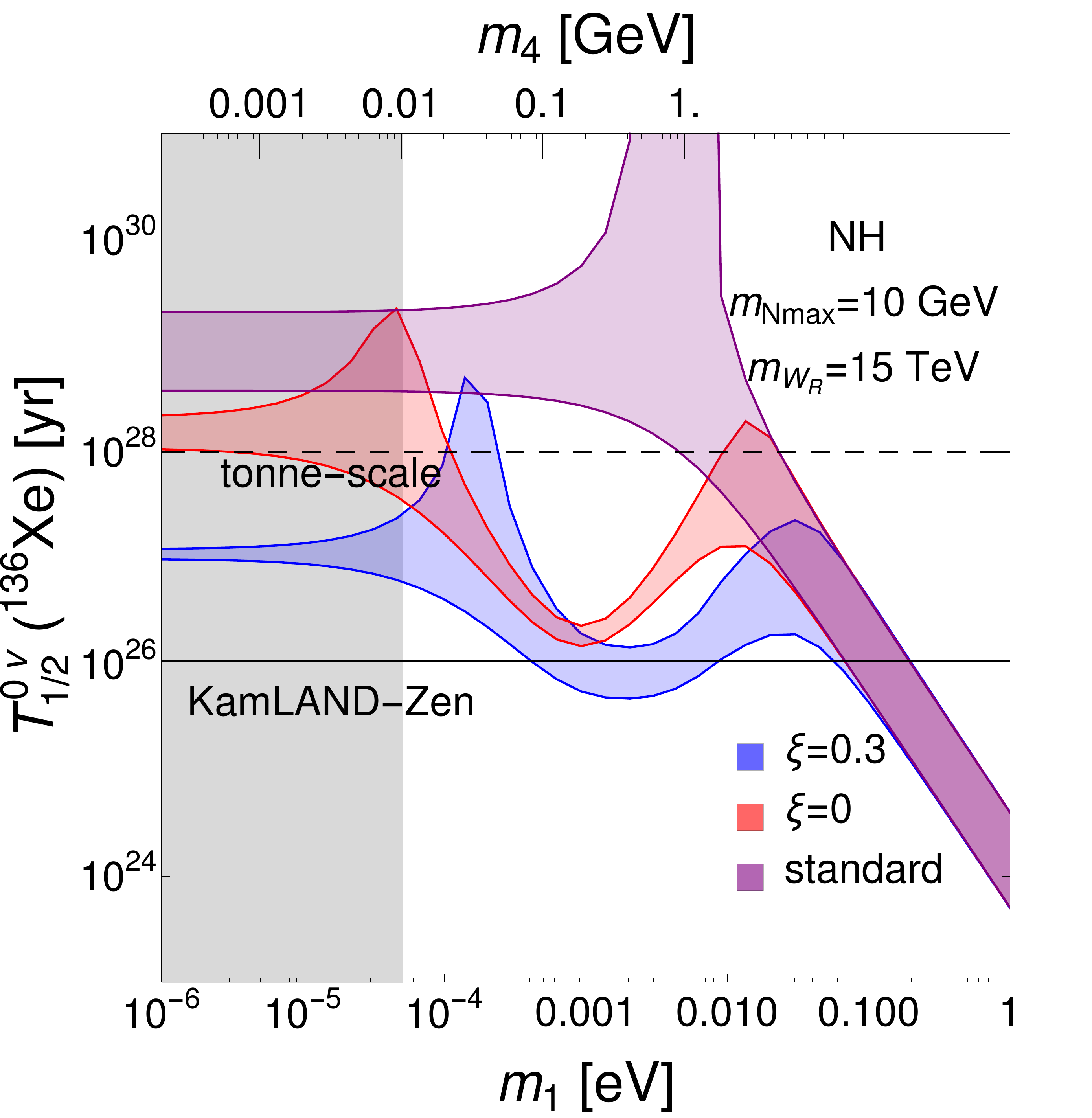}
\includegraphics[width=0.32\columnwidth]{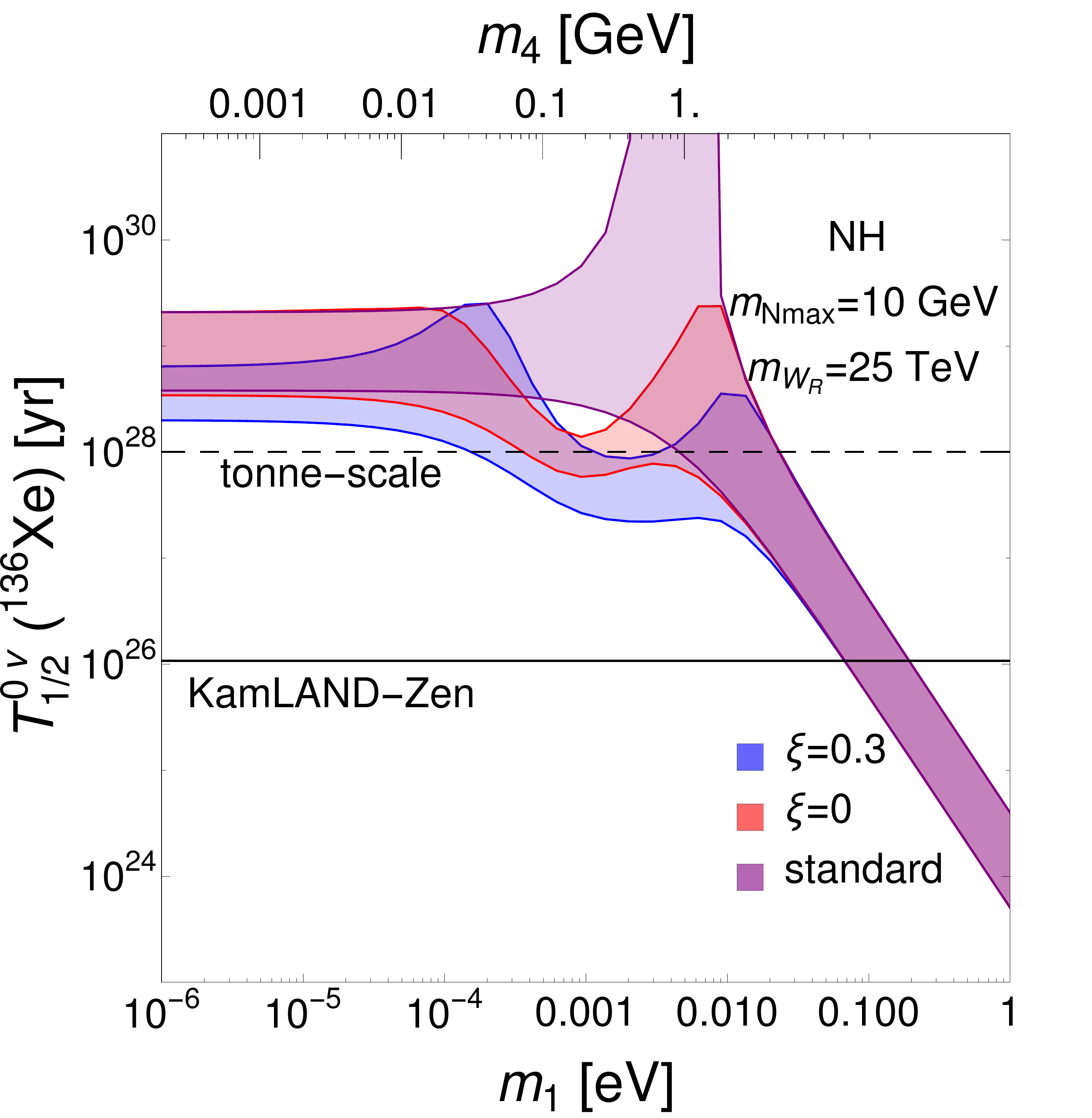}
\caption{
The same as Fig.~\ref{fig:half-life_II_1}, but with $\mnmax=10\gev$. The gray bands for $m_4<10\mev$ are excluded by astrophysics in the left and middle panels. 
}
\label{fig:half-life_II_2}
\end{figure}
 
For $\mnmax=10\gev$, the other two sterile neutrinos are, of course, even lighter.
We require $m_4> 10\mev$ for $m_{W_R}=7,15\tev$ constrained by the supernova cooling~\cite{Barbieri:1988av,Nemevsek:2011aa}.
Consequently, we find that for relatively small $m_{W_R}=7\tev$, the model is ruled out, barring the narrow regions of $m_1$ with cancellation, while for $m_{W_R}=15\tev$, most of the parameter space is in the reach of next-generation experiments. For heavier right-handed gauge bosons, $m_{W_R}=25\tev$, positive signals are still expected in next-generation experiments for $100\mev< m_4 <1\gev$.

The plots in Figs.~\ref{fig:half-life_II_1} and \ref{fig:half-life_II_2} highlight a central point of this work: a $0\nu\beta\beta$ decay detection is possible in a next-generation experiment in a realistic BSM model that avoids direct and indirect searches at colliders and low-energy precision measurements. This conclusion remains even if neutrino masses follow the normal hierarchy\footnote{If active neutrinos are in the inverted hierarchy, a detection of $0\nu\beta\beta$ decay in tonne-scale $\onbb$ decay experiments is always possible barring accidental cancellation.
}.
The contributions of light sterile neutrinos can dominate the standard contributions from active neutrinos, leading to an enhanced $0\nu\beta\beta$ decay rate.

\subsection{Type-I seesaw dominance}
We now explore the phenomenology of $\onbb$ decay in the limit of type-I seesaw dominance: Dirac mass term $M_D \neq 0$ and Majorana mass term $M_L=0$. In this scenario, $\onbb$ decay receives additional contributions compared to the type-II seesaw dominance scenario, which are depicted in Fig.~\ref{fig:typeI}. A few comments are in order. 
Diagram in Fig.~\ref{fig:typeI}(a) gives rise to the standard mechanism contribution for the exchange of active neutrinos,
analogous to the diagram in Fig.~\ref{fig:typeII}(a). The contribution from the exchange of sterile neutrinos was studied in the canonical type-I seesaw models~\cite{Halprin:1983ez,Leung:1984vy,Xing:2009ce,Asaka:2020wfo}.
The diagrams in Fig.~\ref{fig:typeI}(b)(c)
with the exchange of active neutrinos in the mass basis have been discussed in different contexts~\cite{Doi:1982dn,Vergados:1985pq,Pantis:1996py,Suhonen:1998ck,Hirsch:1996qw,Chakrabortty:2012mh,Barry:2013xxa,Dev:2014xea,Stefanik:2015twa,Yang:2021ueh}, which are also called $\lambda$ and $\eta$ diagrams~\cite{Doi:1982dn}  characterized by quark currents with mixed chiralities and purely left-handed chiralities, respectively,
while the diagrams with the exchange of sterile neutrinos (counterparts of the $\lambda$ and $\eta$ diagrams) have drawn less attention.

\begin{figure}[t]
  \centering
  \subcaptionbox{\label{typeI:c}}{\includegraphics[width=0.3\columnwidth]{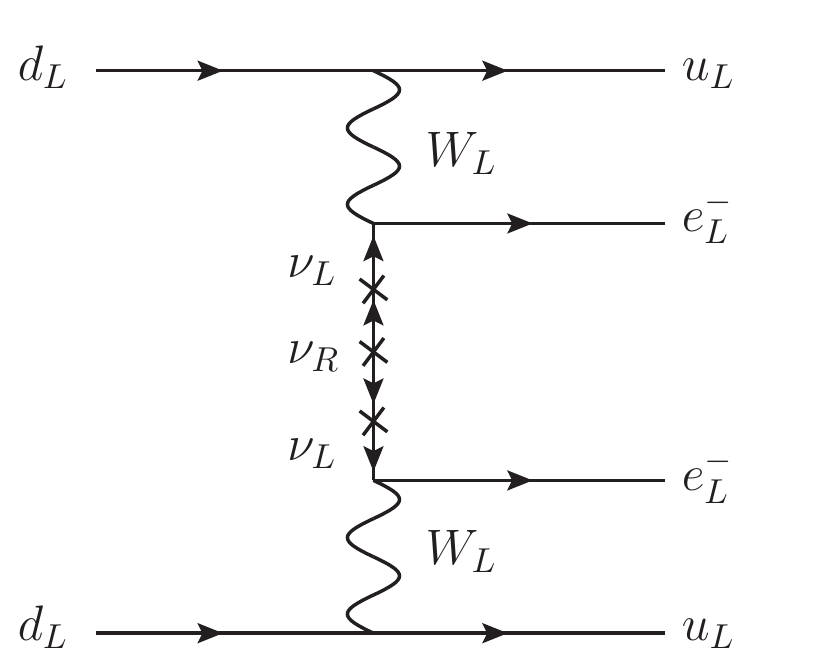}}\hspace{1em}%
  \subcaptionbox{\label{typeI:a}}{\includegraphics[width=0.3\columnwidth]{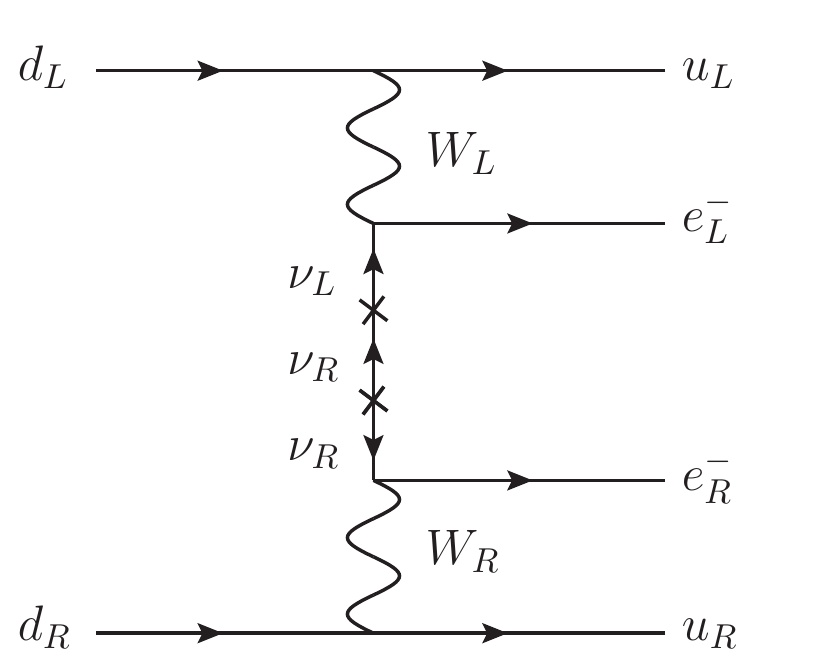}}\hspace{1em}%
  \subcaptionbox{\label{typeI:b}}{\includegraphics[width=0.3\columnwidth]{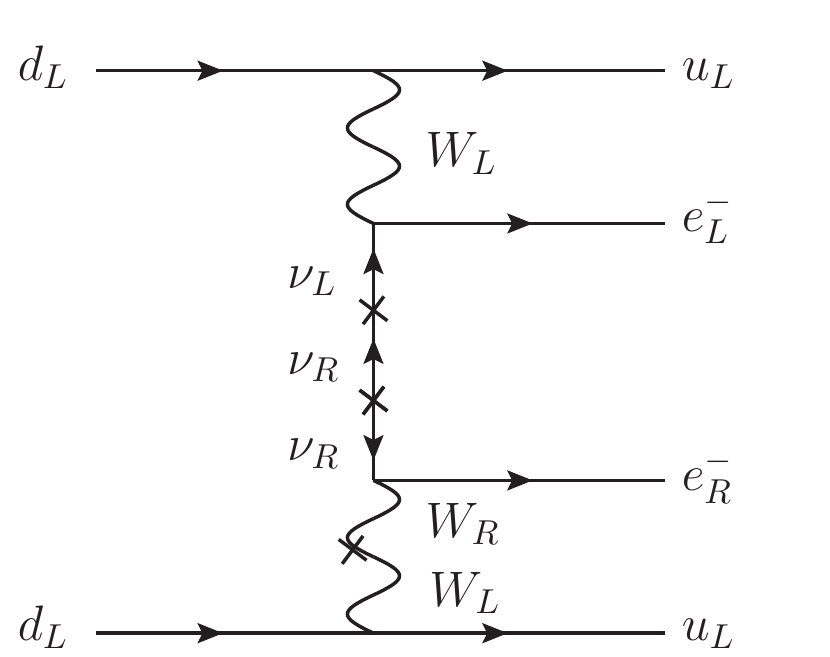}}
\caption{
Contributions to the $\onbb$ decay due to a non-vanishing Dirac mass term $M_D$. 
The cross vertices $(\times)$ in between $\nu_R-\nu_R$, $\nu_L-\nu_R$, and $W_L-W_R$ denote the insertion of Majorana mass term, active-sterile neutrino mixing, and the left-right mixing, respectively. 
}
\label{fig:typeI}
\end{figure}

All of these contributions are easily and automatically evaluated in our EFT framework. 
If the right-handed neutrino $\nu_R$ are heavy with masses $\sim \tev$, they are integrated out above the electroweak scale. The resulting dimension-seven $\Delta L=2$ SMEFT operators~\cite{Lehman:2014jma} are in the forms of 
$O_{Leu\bar{d}\varphi } = \epsilon_{ij} (L_i^T C \gamma_\mu e ) \varphi_j  \bar{d}_R \gamma^\mu u_R$ and $O_{L\varphi De } = \epsilon_{ij} (L_i^T C \gamma_\mu e ) \varphi_j \epsilon_{mn}\varphi_m (D^\mu \varphi)_n  $ corresponding to the diagrams in Fig.~\ref{fig:typeI}(b)(c), respectively, 
as studied in Ref.~\cite{Cirigliano:2018yza}\footnote{The SM Higgs field $\varphi$ in $\epsilon_{ij} (L_i^T C \gamma_\mu e ) \varphi_j$ stems from leptonic Yukawa interactions with the bi-doublet $\Phi$ in Eq.~\eqref{eq:Lag-Yukawa}, while $\epsilon_{mn}\varphi_m (D^\mu \varphi)_n $ gives rises to the SM  $W$ field after the EWSB.}.
If the masses of right-handed neutrinos are below the electroweak scale, they are kept as explicit degrees of freedom, which leads to the dimension-six $\Delta L=0$ $\nu$SMEFT operators in Eq.~\eqref{nuSMEFT}. In both cases, we can use the decay rate expression in Eq.~\eqref{eq:T1/2}
with the contributions from diagrams in Fig.~\ref{fig:typeII}(b)(c)(d) and Fig.~\ref{fig:typeI}(a)(b)(c).

The subamplitudes $\mathcal{A}_L(m_i)$ correspond to the contributions in Fig.~\ref{fig:typeI}(a).
The diagrams in Fig.~\ref{fig:typeI}(b)(c) are described by the subamplitudes  $\mathcal{A}_{M,E,m_e}(m_i)$, where $i=1,2,3$ $(i=4,5,6)$ for active (sterile) neutrinos. 
From Eq.~\eqref{Anu}, $\mathcal{A}_{M}(m_i)$ is proportional to the left-right mixing parameter 
\begin{equation}
\zeta= \frac{m_W^2}{m_{W_R}^2}\frac{ 2\xi}{1+\xi^2}\,,
\end{equation}
and it is enhanced by a factor of $m_N/m_e$ relative to $\mathcal{A}_{E,m_e}(m_i)$. For more details, one can refer to Ref.~\cite{Cirigliano:2017djv}. 
Therefore $\mathcal{A}_{M}(m_i)$ is always much larger than $\mathcal{A}_{E}(m_i)$ and $\mathcal{A}_{m_e}(m_i)$ unless we take the limit $\xi\to 0$.

For concreteness, we assume for the right-handed neutrino mixing matrix $U_R = U_{\text{PMNS}}^*$ and $U_R = U_{\text{PMNS}}$ if the generalized parity $\mathcal{P}$ and generalized charge conjugation $\mathcal{C}$ are taken as the left-right symmetry, respectively. 
For these choices, the Dirac mass matrix $M_D$ in the limit of type-I  dominance, $v_L=0$,  is given by~\cite{Nemevsek:2012iq,Senjanovic:2016vxw}
\begin{align}
\label{MD_P}
\mathcal{P}:\quad M_D &=U_{\text{PMNS}} \widehat{M}_N \sqrt{- \widehat{M}_N^{-1} \widehat{M}_\nu } U_{\text{PMNS}}^\dagger\;, \nn\\
\mathcal{C}:\quad M_D &=U_{\text{PMNS}} \widehat{M}_N \sqrt{ - \widehat{M}_N^{-1} \widehat{M}_\nu } U_{\text{PMNS}}^T\;.
\end{align} 
It is noted that the form of $M_D$ in case of $\mathcal{P}$ is obtained in the limit of $\xi \sin\alpha\to 0$~\cite{Senjanovic:2016vxw}. 
We obtain that the active-sterile neutrino mixing matrix $R=M_D M_R^{-1}$ can be written in the simple forms as follows:
\begin{align}
\mathcal{P}: \quad R &= i U_{\text{PMNS}}R_D
U_{\text{PMNS}}^T\;,\nn\\
\mathcal{C}: \quad R &= i U_{\text{PMNS}} R_D
U_{\text{PMNS}}^\dagger\;,
\end{align}
where 
\begin{align}
\label{eq: RD}
R_D \equiv \begin{pmatrix}
\sqrt{m_1/m_4} & 0 & 0\\
0 & \sqrt{m_2/m_5}  & 0 \\
0 & 0 & \sqrt{m_3/m_6} 
\end{pmatrix}\;
\end{align}
is introduced. It is easy to see that even for  $ \Vert R_D \Vert \sim  \sqrt{0.1{\rm eV}/10{~\rm MeV}}=10^{-4}$, we can safely neglect the non-unitary effects in the active neutrino mixing matrix~\cite{Fernandez-Martinez:2016lgt}. 

Substituting the expression of $R$ into Eq.~\eqref{Udef} and Eq.~\eqref{eq:st-matrix}, we obtain the $6\times 6$ neutrino mixing matrix $U$ in the mLRSM
\begin{align}
\label{eq:Umatrix}
\mathcal{P}: \quad  U &= 
\begin{pmatrix}
U_{\text{PMNS}} & i U_{\text{PMNS}} R_D\\
i U_{\text{PMNS}}^* R_D & U_{\text{PMNS}}^*
\end{pmatrix}\;,\nn \\
\mathcal{C}: \quad  U &= 
\begin{pmatrix}
U_{\text{PMNS}} & i U_{\text{PMNS}} R_D\\
i U_{\text{PMNS}} R_D & U_{\text{PMNS}}
\end{pmatrix}\;.
\end{align}

\begin{figure}[!t]
\centering
\includegraphics[width=0.32\columnwidth]{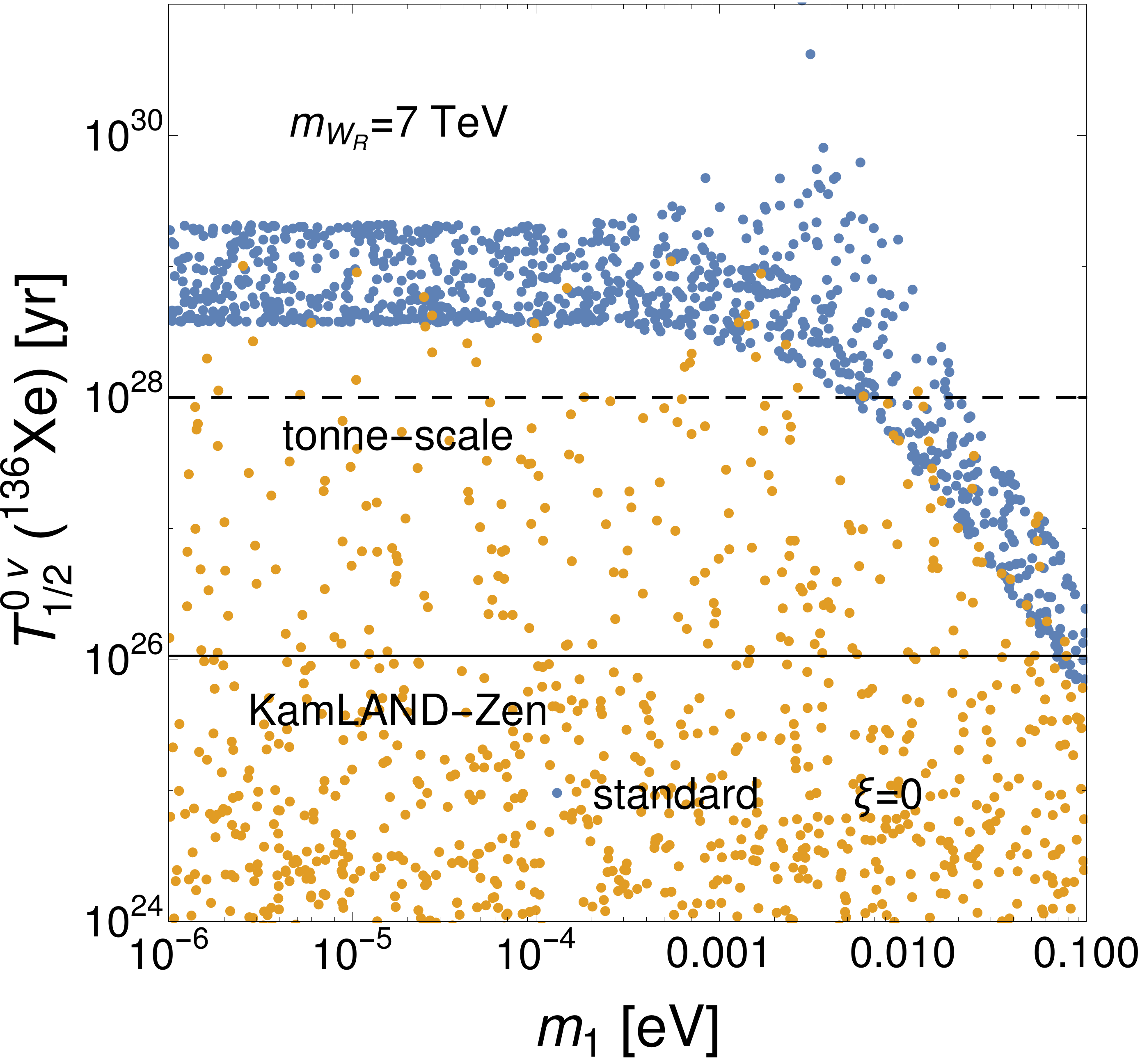}
\includegraphics[width=0.32\columnwidth]{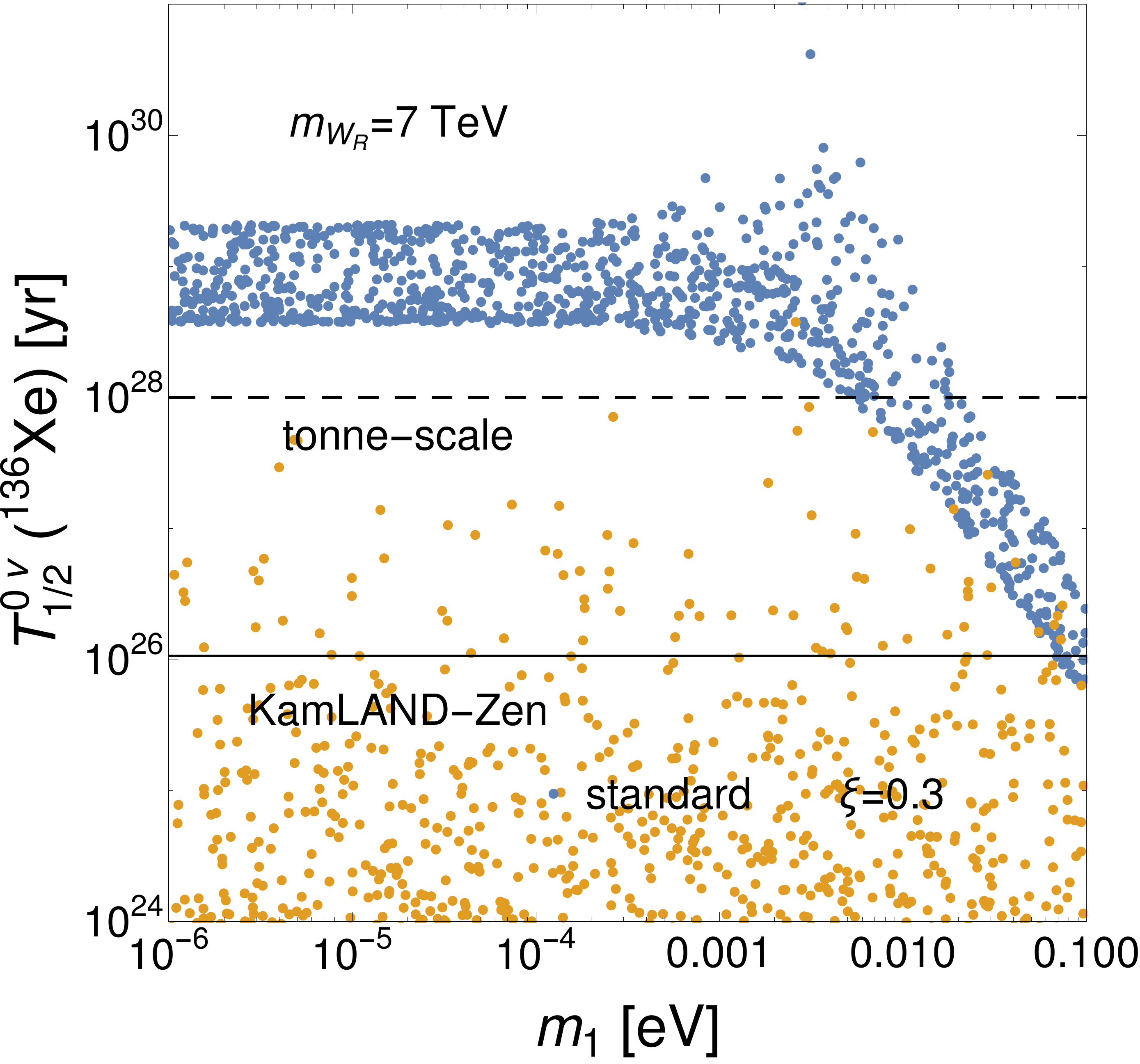}\\
\includegraphics[width=0.32\columnwidth]{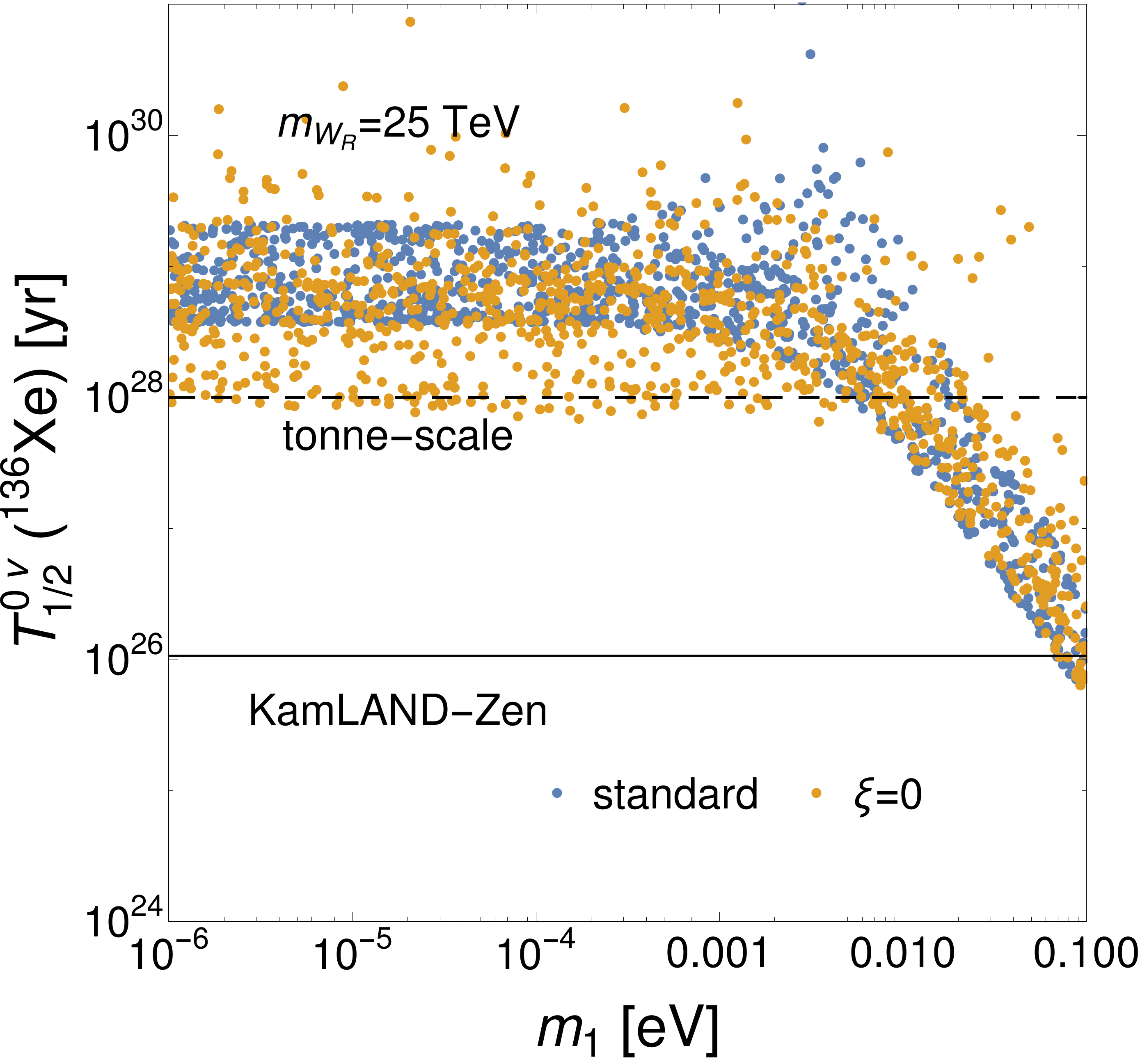}
\includegraphics[width=0.32\columnwidth]{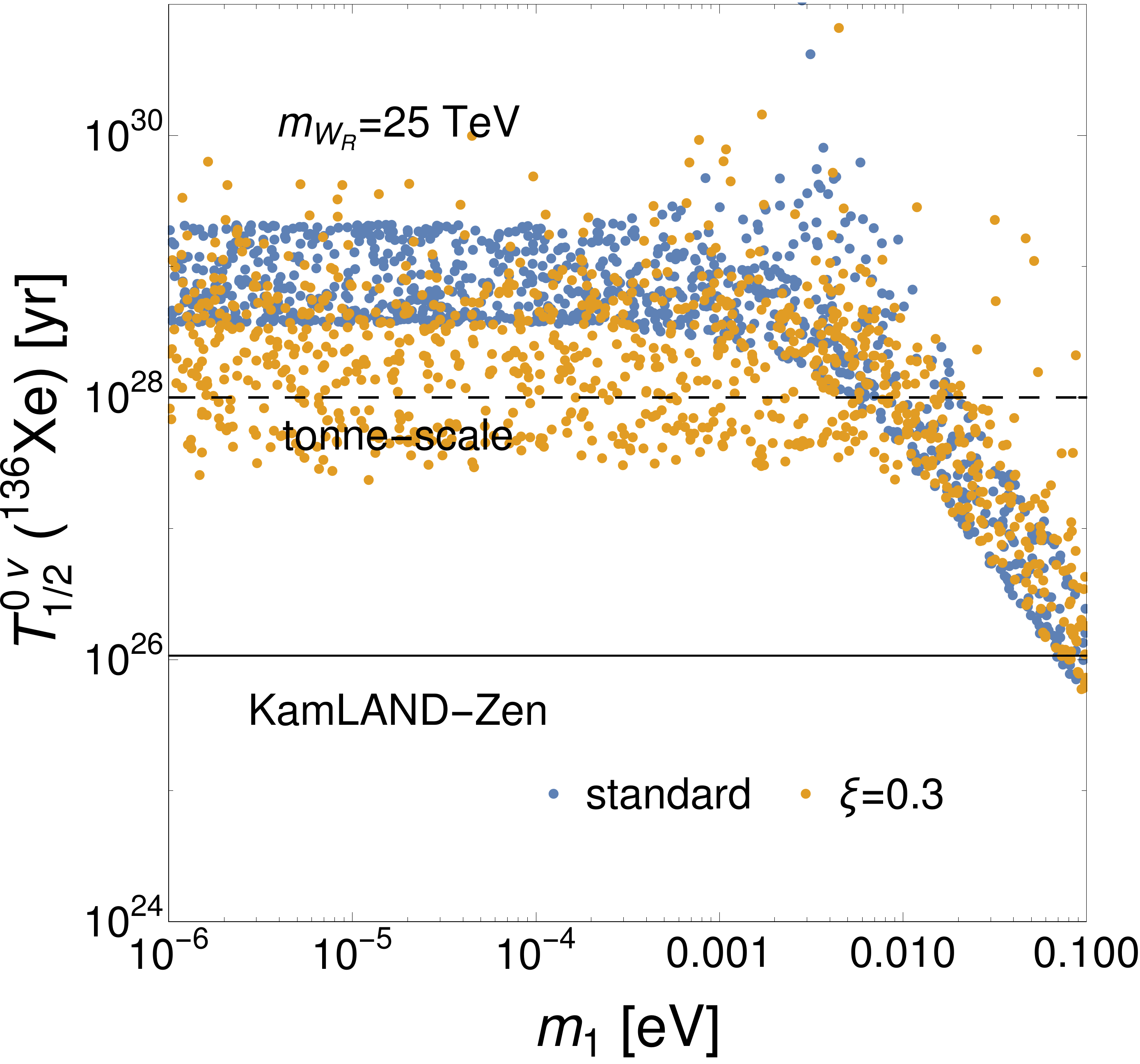}
\caption{
${}^{136}$Xe half-life as a function of the lightest neutrino mass $m_1$ in the type-I seesaw dominance scenario (orange). The active neutrinos are assumed to be in the NH, and the sterile neutrino masses are randomly chosen between $10\mev$ to $1\tev$.
The right-handed gauge boson mass is set to be $m_{W_R}=7\tev$  and $25\tev$ in the upper and lower panels for $\xi=0$ (left) and $0.3$ (right). For comparison, the predicted half-life in the standard mechanism is also shown (blue). The solid and dashed black lines denote the current and expected limits, respectively.
}
\label{fig:half-life_I}
\end{figure}

To compare with the results in type-II seesaw dominance, we assume that the neutrino masses are in the NH, that is, the active neutrino masses satisfy $m_1 < m_2 < m_3$, while
  the sterile neutrino masses $m_4$, $m_5$ and $m_6$ vary in the range from $10\mev$ to $1\tev$. We stress that there is no connection between the sterile neutrino and active neutrino masses and scan over a wide range of sterile neutrino masses. Two cases of left-right mixing are considered:  $\xi = 0$ and $0.3$, and we take two values for the right-handed gauge boson mass $m_{W_R}= 7\tev$ and $25\tev$. The independent parameters are collected in Tab.~\ref{tab:parameters}. Again, in the type-I seesaw scenario, the $\onbb$ decay rates are insensitive to whether $\mathcal{P}$ or $\mathcal{C}$ is chosen, and the value of $\alpha$, for the cases in Tab.~\ref{tab:parameters}, once the other unknown phases $\lambda_{1,2}$ and $\theta_L$ are varied.

The resulting scatter plots  of half-life as a function of the lightest neutrino mass $m_1$ in the mLRSM 
are depicted in Fig.~\ref{fig:half-life_I},
In the upper panels, we see that most of the parameter space is already excluded by existing limits, while next-generation experiments will probe almost all remaining points. There are still some points escaping from future detection due to the cancellation of the contributions from the two lighter sterile neutrinos. Similar cancellations occurred in the type-II scenario with $\mnmax = 10\gev$. The difference is that in the type-I limit, the sterile neutrino masses are unrelated to the active neutrino masses. The cancellation occurs for a broader region of $m_1$, not at specific points. 

Increasing the mass to  $m_{W_R}=25\tev$, the $\onbb$ decay rates are significantly suppressed and a detection is only possible for non-zero left-right mixing $\xi =0.3$. 
In that case, the predicted half-lives can be well below $10^{28}$ yr in the full range of $m_1 < 0.01\ev$.

It is interesting to further dissect the results by studying the individual contributions. The subamplitude $\mathcal{A}_L(m_i)$ $(i=1,\cdots,6)$ is proportional to $U_{ei}^2 m_i/({\bf q}^2 + m_i^2)$. From Eq.~\eqref{eq:Umatrix},  we obtain for $j=1,2,3$
\begin{align}
U_{ej+3} = \left( i U_{\text{PMNS}} \right)_{ej} \sqrt{\frac{m_j}{m_{j+3}}}\;,
\end{align}
such that for the sterile neutrinos
\begin{align}
U_{ej+3}^2 \dfrac{m_{j+3}}{{\bf q}^2 + m_{j+3}^2} \simeq  
\begin{cases}
-  \left( U_{\text{PMNS}} \right)_{ej}^2
\dfrac{m_j}{{\bf q}^2}\;, & \text{ for $m_{j+3}\ll |{\bf q}|$ } \;,\\
\noalign{\vskip9pt}
-  \left( U_{\text{PMNS}} \right)_{ej}^2 \dfrac{m_{j}}{m_{j+3}^2}\;, & \text{ for $m_{j+3}\gg |{\bf q}|$ } \;,
\end{cases}
\end{align}
where $|{\bf q}| \sim m_\pi$ is a typical nuclear scale. There is a cancellation between contributions from active and sterile neutrinos for small sterile neutrino masses, $m_{j+3} \ll |\bf q|$. In the case all of $m_{4,5,6}\ll |{\bf q}|\sim m_\pi$, the total subamplitude vanishes $\sum_i \mathcal{A}_L(m_i)\to 0$~\cite{Blennow:2010th,Dekens:2020ttz}.

The subamplitudes $\mathcal{A}_{M,E,m_e}(m_i)$ depend on the combination
\begin{align}
\left( C_{{\rm VLL}}^{(6)} \right)_{ei}\left( C_{{\rm VLR}/{\rm VRR}}^{(6)} \right)_{ei} \dfrac{1}{{\bf q}^2 + m_i^2}\,.
\end{align}
This combination is proportional to $(PU)_{ei}(P_s U^*)_{ei}$, which is equal to $\left(U_{\text{PMNS}}\right)_{ei} T^*_{ei}$ 
for $i=1,2,3$ and $S_{e(i-3)} \left(  U_{R}\right)_{e(i-3)}^*$ for $i=4,5,6$ as  shown in Eqs.~\eqref{Udef}~\eqref{eq:projector}. 
The sum of contributions from $\nu_j$ and $\nu_{j+3}$ for $j=1,2,3$ is proportional to 
\begin{align}
\left(U_{\text{PMNS}}\right)_{ej} T^*_{ej} \dfrac{1}{{\bf q}^2 + m_j^2} + S_{ej} \left(U_R\right)_{ej}^* \dfrac{1}{{\bf q}^2 + m_{j+3}^2} \;.
\end{align}
From Eq.~\eqref{eq:Umatrix}, we obtain that for $j=1,2,3$
\begin{align}
\mathcal{P}: \quad \left(U_{\text{PMNS}}\right)_{ej} T^*_{ej} &= -i \left(U_{\text{PMNS}}\right)_{ej} \left(U_{\text{PMNS}} R_D\right)_{ej} = - S_{ej} \left(U_R\right)_{ej}^* \;,\nn\\
\mathcal{C}: \quad \left(U_{\text{PMNS}}\right)_{ej} T^*_{ej} &= -i \left(U_{\text{PMNS}}\right)_{ej} \left(U_{\text{PMNS}}^* R_D\right)_{ej} = - S_{ej} \left(U_R\right)_{ej}^* \;.
\end{align}
Therefore, there exists cancellation between the contributions from the exchange of active and sterile neutrinos in Fig.~\ref{fig:typeI}(b)(c). 
The cancellation is substantial for sterile neutrino mass $m_{j+3}\ll |{\bf q}|$ and negligible for $m_{j+3}\gg |{\bf q}|$.

\begin{figure}[t]
\centering
\includegraphics[width=0.35\columnwidth]{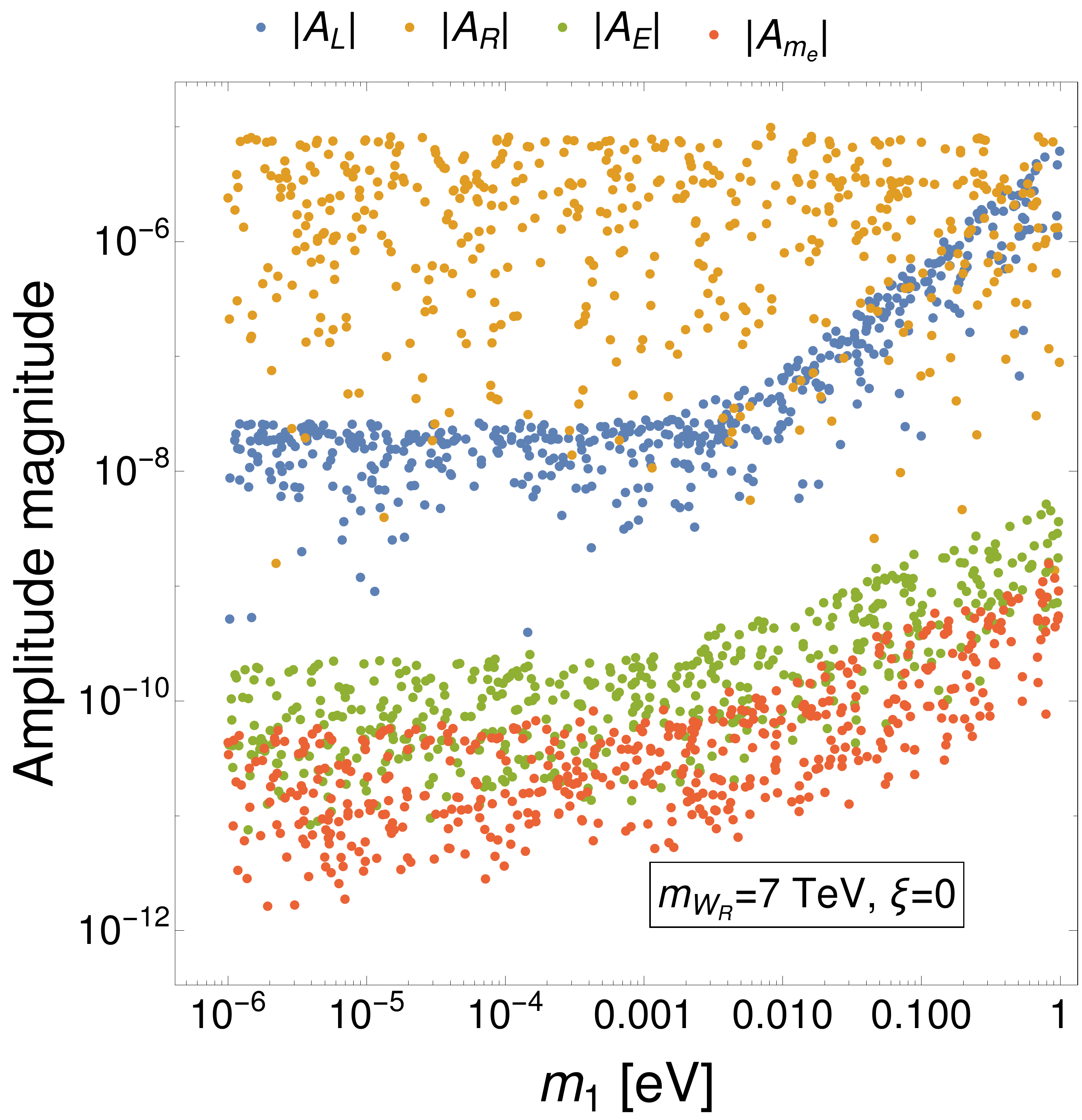}
\includegraphics[width=0.35\columnwidth]{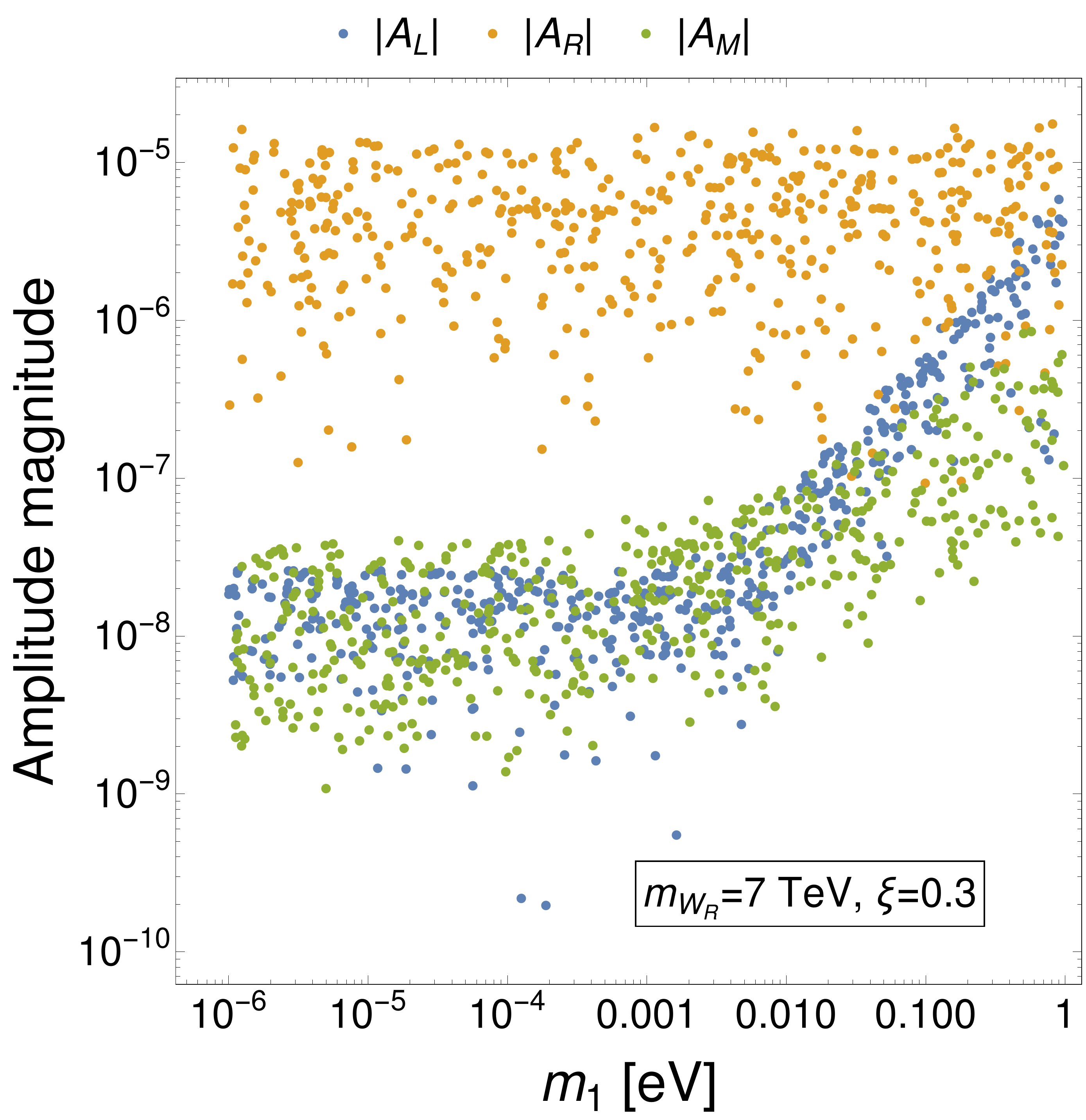}\\
\includegraphics[width=0.35\columnwidth]{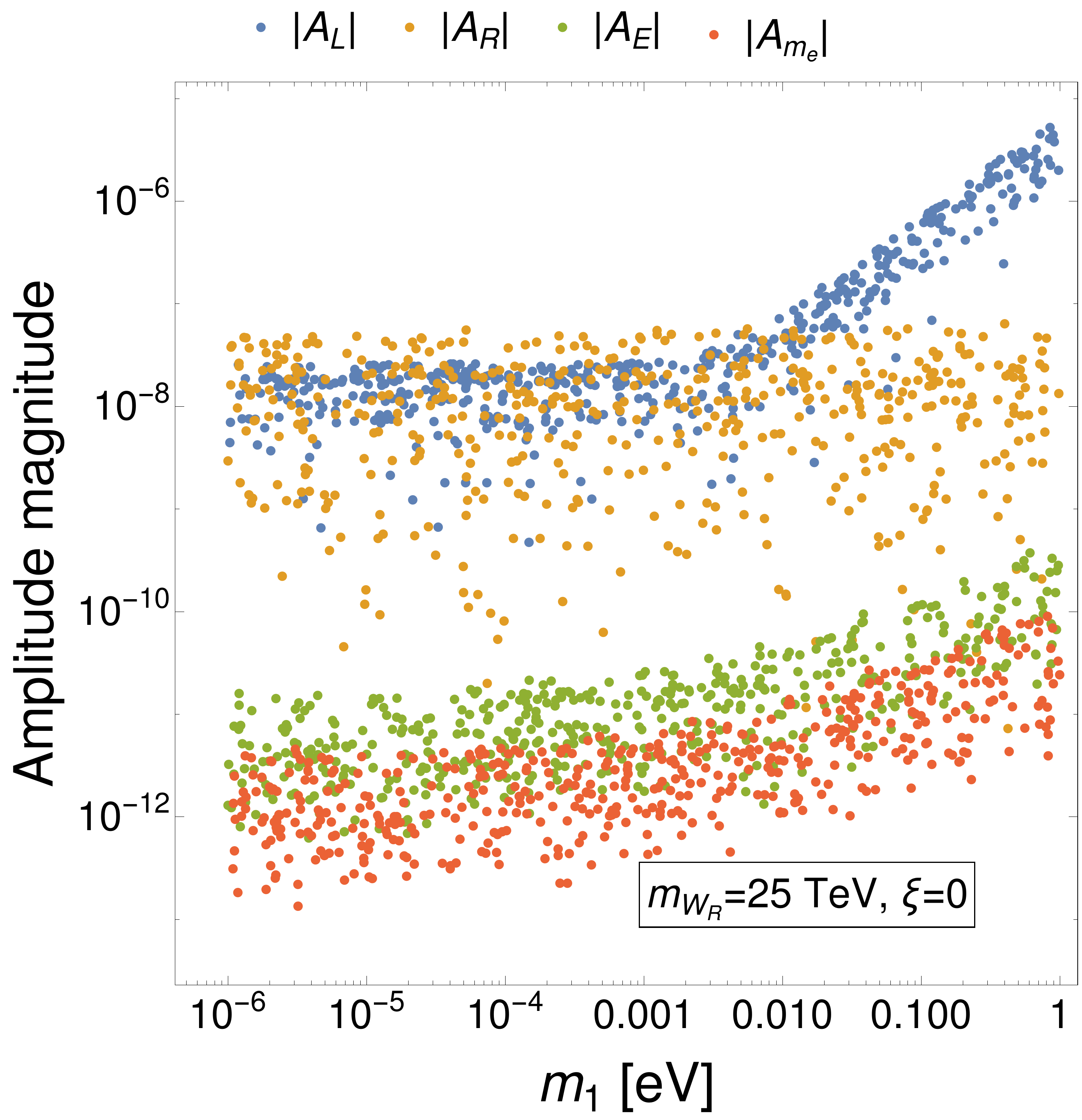}
\includegraphics[width=0.35\columnwidth]{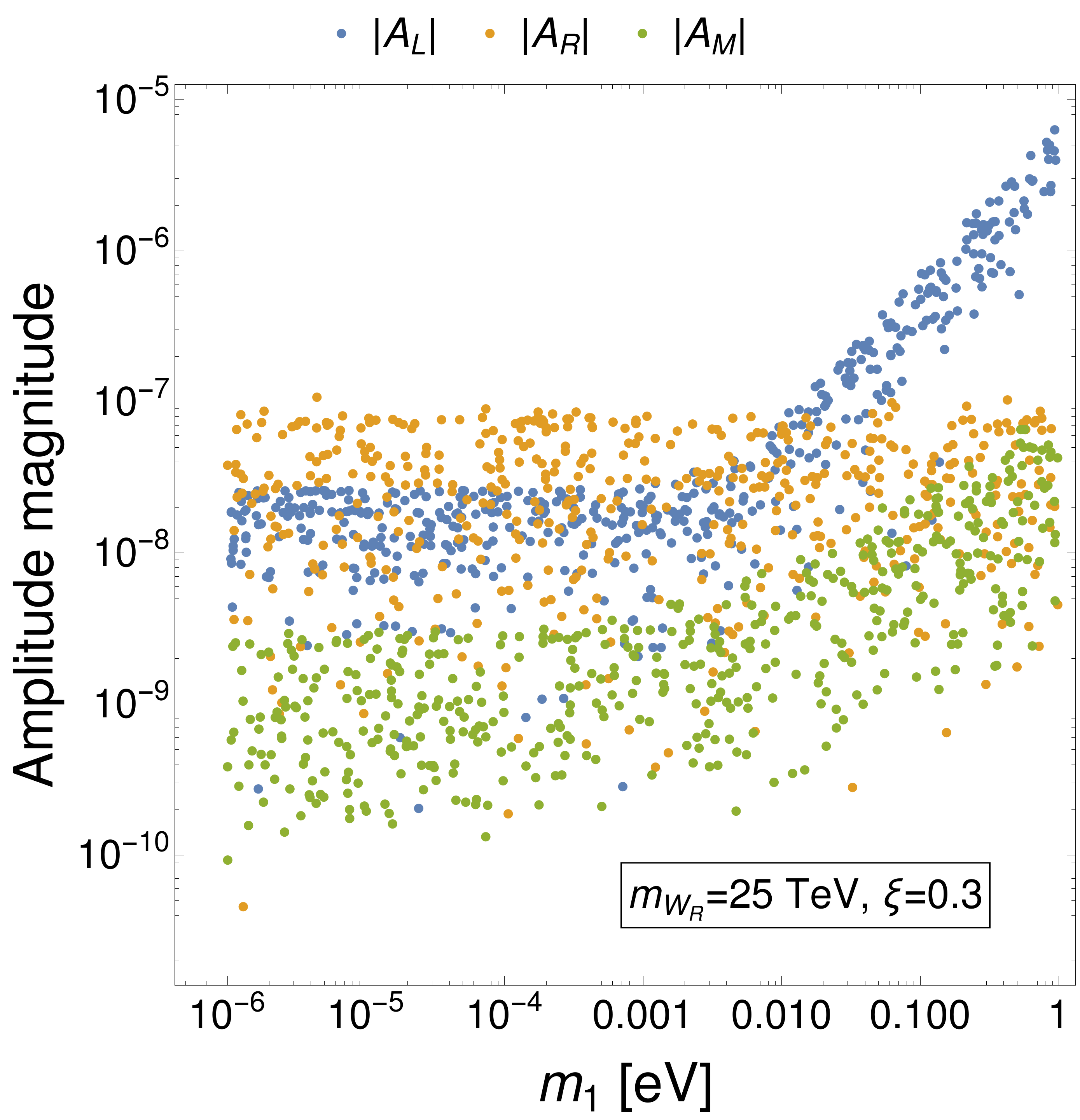}
\caption{
Amplitude magnitudes as a function of the lightest neutrino mass $m_1$ in the type-I seesaw scenario. Active neutrinos are assumed to be in the NH, and the sterile neutrino masses are varied from $10\mev$ to $1\tev$. 
The right-handed gauge bosons mass is set to be $m_{W_R}=7\tev$  ($25\tev$) in the upper (lower) panels with $\xi=0$ (left) and $0.3$ (right). In the left panels, $|\mathcal{A}_L|$, $|\mathcal{A}_R|$, $|\mathcal{A}_E|$ and $|\mathcal{A}_{m_e}|$ are shown in blue, orange, green and red, respectively; in the right panels, $|\mathcal{A}_L|$, $|\mathcal{A}_R|$ and $|\mathcal{A}_M|$ are shown in blue, orange and green, respectively. 
}
\label{fig:amp_I}
\end{figure}

In Fig.~\ref{fig:amp_I} we show the magnitudes of the subamplitudes, including the contributions from all neutrinos using the same parameters as in Fig.~\ref{fig:half-life_I}. In the left panels we set $\xi =0$. $\mathcal{A}_E$ are $\mathcal{A}_{m_e}$ are negligible compared to $\mathcal{A}_L$ and $\mathcal{A}_R$ ( $|\mathcal A_M|$ vanishes for $\xi=0$). $|\mathcal{A}_R|$ is almost always significantly larger than $|\mathcal{A}_L|$ because of the contribution of light sterile neutrinos. For $m_{W_R}=25\tev$, $|\mathcal{A}_R|$ is suppressed but there is still possibility that  $|\mathcal{A}_R|$ exceeds $|\mathcal{A}_L|$  for $m_1 < 0.01\ev$.

 We turn on left-right mixing in the right panels and set $\xi=0.3$. In this case $|A_R|$ grows, and $|\mathcal{A}_M|\gg |\mathcal{A}_{E,m_e}|$ becomes relevant. For $m_{W_R}=7\tev$, $|\mathcal{A}_M|$ is comparable to $|\mathcal{A}_L|$, but both are subleading with respect to $|\mathcal{A}_R|$. For $m_{W_R}=25\tev$, the subamplitudes $|\mathcal{A}_R|$ and $|\mathcal{A}_M|$, decrease accordingly, but $|\mathcal{A}_R|$ still exceeds $|\mathcal{A}_L|$ for a sizable fraction of parameter space.

Therefore, in the mLRSM with type-I seesaw dominance, if we assume the right-handed neutrino mixing matrix $U_R=U_{\rm PMNS}^{(*)}$ as in Tab.~\ref{tab:parameters}, the dominant contributions to $\onbb$ decay always come from $\mathcal{A}_L$ and $\mathcal{A}_R$. 
Contributions from $\mathcal{A}_{M,E,m_e}$ are subleading. In Refs.~\cite{Barry:2013xxa,Dev:2014xea} it has been pointed out that by specifying other choices for the Dirac mass matrix $M_D$, the contributions from the diagrams in Fig.~\ref{fig:typeI}(b)(c)
can be enhanced. Within the EFT framework, this should then be reflected by important contributions from $\mathcal{A}_{M,E,m_e}$\footnote{Rather than imposing the left-right symmetry in the Yukawa sector, in Refs.~\cite{Barry:2013xxa,Dev:2014xea} an {\it ansatz} was made, such that
neither $\mathcal{P}$-symmetric nor $\mathcal{C}$-symmetric mLRSM was studied. 
}.

\subsection{Theta term: explicit examples } \label{CPconnection}
In this subsection we discuss the one-loop contribution to $\bar{\theta}$ in the mLRSM.
In the type-II seesaw dominance scenario, $\bar{\theta}_{\text{loop}}$ vanishes, which is obvious since $M_D = 0$.  For type-I seesaw dominance, we consider $\mathcal{P}$ as the left-right symmetry following Ref.~\cite{Senjanovic:2020int}. If the right-handed leptonic mixing matrix $U_R=U_{\rm PMNS}^*$,
the Dirac mass matrix $M_D$ is explicitly expressed in Eq.~\eqref{MD_P}.
By substituting  it into  Eq.~\eqref{thetaQCDLR}, we obtain
\begin{align}\label{thetaQCDLR_diagonal_lep}
\bar{\theta}_{\text{loop}}  = &   \frac{1}{4 \pi^{2}} \frac{m_{t}}{m_{b}} \frac{1}{v_R^2(\kappa^2-\kappa'^2) } \ln \frac{\sqrt{2} M_{P l}}{v_{R}}\nn\\
& \times \Im{\operatorname{Tr}\bigg( U_{\text{PMNS}} \Big[\widehat{M}_N \widehat{M}_N, \sqrt{-\widehat{M}_N\widehat{M}_\nu }~\Big]   U_{\text{PMNS}}^\dagger \widehat M_l
\bigg)} \;,
\end{align}
where we have chosen the basis of diagonal charged lepton mass matrix $M_\ell = \widehat{M}_{\ell}={\rm diag}(m_e,m_\mu,m_\tau)$ with $m_{e,\mu,\tau}$ the charged lepton masses. Since $\widehat{M}_N$ and $\widehat{M}_\nu$ are diagonal , $\bar\theta_{\text{loop}}=0$. It indicates that the one-loop contribution from the light sterile neutrinos vanishes.

Thus, for this case, there is no direct connection between light neutrinos and the strong CP problem,  in contrast with the result shown in Ref.~\cite{Senjanovic:2020int}. It can also be shown that  choosing $U_R = {\bf 1}$ in Eq.~\eqref{thetaQCDLR} $\bar{\theta}_{\text{loop}}$ vanishes as well. 

The cancellations might be just accidental for the specific textures of $U_R$ chosen here. If so, the estimates taken in Ref.~\cite{Senjanovic:2020int} could be indicative of the order of magnitude of the $\bar{\theta}_{\text{loop}}$ when $U_R\neq (U_{\text{PMNS}}^*,\mathbf{1})$: for $m_{W_R}=10\tev$, the sterile neutrino masses $m_i \lesssim 100\gev$, $i=4,5,6$ in order to satisfy the EDM limits.

Unfortunately, despite earnest efforts, we have not been able to explicitly solve $M_D$ in terms of the heavy and light neutrino mass matrices in the general case. This outstanding task would be beneficial in scrutinizing the, in principle, the exciting connection between $0\nu\beta\beta$ decay and EDMs.

\section{Conclusion}
\label{sec:conclusion}

Light sterile neutrinos are interesting as they elegantly avoid existing experimental constraints, have been motivated to address the strong CP problem,  and can be detected by next-generation $\onbb$ decay experiments. We have investigated the impact of light sterile neutrinos in left-right symmetric models on $\onbb$ decay phenomenology. We have applied an EFT formalism to describe the contributions of a generic sterile neutrino to $\onbb$ decay \cite{Dekens:2020ttz}. 
 Using chiral EFT ensures that the hadronic and nuclear matrix elements are consistent with the symmetries of QCD and that the dependence of these matrix elements on neutrino masses correctly interpolates between the limits of light and heavy neutrinos.

To show the importance of sterile neutrinos for $\onbb$ decay in a realistic BSM model, we investigate $\onbb$ decay rate in the minimal left-right symmetric model with parity $\mathcal{P}$ or charge conjugation $\mathcal{C}$  as the left-right symmetry -- assuming relatively light sterile neutrinos ($\geq 10\mev$).
We consider two specific scenarios that account for active neutrino masses: the type-II and type-I seesaw dominance. In the former scenario, the sterile neutrino masses are related to the active neutrino masses. The right-handed neutrino mixing matrix $U_R$ equals the PMNS matrix $U_{\rm PMNS}$ up to additional phases. In the latter scenario, they are uncorrelated. Therefore, as a representative case study, we assume $U_R$ to take the same form as in the type-II seesaw dominance to reduce the number of free parameters. More general choices can be studied straightforwardly within our setup. The $0\nu\beta\beta$ decay rates depend on the parameters: $W_R$ mass, neutrino masses, and phases as collected in Tab.~\ref{tab:parameters}.

We now summarize the main results of our analysis
\begin{itemize}
\item  We find ample parameter space not excluded by other experiments, where $0\nu\beta\beta$ decay rates are in reach of next-generation experiments even in the normal hierarchy. 
\item For type-II seesaw dominance, there is a one-to-one correspondence between the sterile and active neutrino masses. For relatively small mass $m_{W_R}=7\tev$, the model has essentially been ruled out when all sterile neutrino masses are below 10~GeV apart from narrow regions where cancellations occur. If, however, the mass of the heaviest sterile neutrino is relatively large. There are good prospects for observing a positive signal in next-generation $\onbb$ decay experiments for the lightest sterile neutrino mass above 10~GeV~\cite{Li:2020flq}. For the medium mass $m_{W_R}=15\tev$, a large portion of parameter space can be probed in next-generation $0\nu\beta\beta$ decay experiments. For a heavier mass, $m_{W_R}=25\tev$, the contributions from the exchange of $W_R$ are suppressed, and positive signals can still be expected if the lightest sterile neutrino mass lies below 1~GeV.
\item For type-I seesaw dominance, the analysis is more complicated. New parameters arise since the sterile and active neutrino masses are no longer connected. In the $\mathcal{P}$-symmetric or $\mathcal{C}$-symmetric mLRSM, the Dirac mass term $M_D$ is then determined by the lightest neutrino mass, the PMNS matrix, and three sterile neutrino masses.  
\begin{itemize}
    \item  Compared to type-II dominance, we find a better sensitivity to $\onbb$ decay in tonne-scale experiments since the sterile neutrino masses are not related to the active neutrino masses and can all be  small simultaneously (c.f. Fig.~\ref{fig:half-life_I}).
    \item  If all sterile neutrinos are light (below typical nuclear scales), there are possible cancellations between contributions from active and sterile neutrinos.
    \item Similar to type-II dominance, next-generation $0\nu\beta\beta$ decay experiments can make detection even in the normal hierarchy for 
    significant portions of parameter space for right-handed scales as high as 25 TeV (c.f. lower panels of Fig.~\ref{fig:half-life_I}).
    \item Additional contributions appear due to the non-zero $M_D$, which are subleading for the form of $M_D$ considered in this work (c.f. Fig.~\ref{fig:amp_I}).
\end{itemize}
\end{itemize}

An additional motivation of our study was to scrutinize the possible connection between light sterile neutrinos and the strong CP problem identified in Refs.~\cite{Kuchimanchi:2014ota,Senjanovic:2020int}. To our surprise, we found for certain representative cases, namely $U_R=U_{\text{PMNS}}^*$ and $U_R={\bf 1}$ in the type-I dominance of $\mathcal{P}$-symmetric mLRSM, that the loop contributions to $\bar{\theta}$ vanish, in conflict with the statements of Refs.~\cite{Kuchimanchi:2014ota,Senjanovic:2020int}. It remains to be seen whether this conclusion applies in general.   
Recently, Ref.~\cite{Alves:2022yav} also studied  other leptonic observables such as $\mu\to eee$, $\mu\to e\gamma$, and $\mu\to e$ conversion in the mLRSMs with light sterile neutrinos. It would be interesting to combine them with hadronic observables ($K-\bar K$ mass difference etc.)~\cite{Dekens:2021bro}. 

\section*{Acknowledgments}
GL would like to thank  Dong-Liang Fang, Xiao-Dong Ma, Jiang-Ming Yao, and Jiang-Hao Yu for helpful discussions.   J.d.V acknowledges support from the Dutch Research Council (NWO) in the form of a VIDI grant.  GL, MJRM, and JCV were partially funded under the US Department of Energy contract DE-SC0011095. GL was also supported in part by the Fundamental Research Funds for the Central Universities, China, Sun Yat-sen University. JCV was also supported in part under the US Department of Energy contract DE-SC0015376. MJRM was also supported in part under National Science Foundation of China grant No. 19Z103010239.

\newpage
\bibliographystyle{JHEP}
\bibliography{bibliography}

\providecommand{\href}[2]{#2}\begingroup\raggedright\begin{thebibliography}{100}

\bibitem{Minkowski:1977sc}
P.~Minkowski, \emph{{$\mu \to e\gamma$ at a Rate of One Out of $10^{9}$ Muon
  Decays?}}, \href{http://dx.doi.org/10.1016/0370-2693(77)90435-X}{\emph{Phys.
  Lett. B} {\bf 67} (1977) 421--428}.

\bibitem{Gell-Mann:1979vob}
M.~Gell-Mann, P.~Ramond and R.~Slansky, \emph{{Complex Spinors and Unified
  Theories}}, {\emph{Conf. Proc. C} {\bf 790927} (1979) 315--321},
  [\href{http://arxiv.org/abs/1306.4669}{{\tt 1306.4669}}].

\bibitem{Yanagida:1979as}
T.~Yanagida, \emph{{Horizontal gauge symmetry and masses of neutrinos}},
  {\emph{Conf. Proc. C} {\bf 7902131} (1979) 95--99}.

\bibitem{Glashow:1979nm}
S.~L. Glashow, \emph{{The Future of Elementary Particle Physics}},
  \href{http://dx.doi.org/10.1007/978-1-4684-7197-7_15}{\emph{NATO Sci. Ser. B}
  {\bf 61} (1980) 687}.

\bibitem{Mohapatra:1979ia}
R.~N. Mohapatra and G.~Senjanovic, \emph{{Neutrino Mass and Spontaneous Parity
  Nonconservation}},
  \href{http://dx.doi.org/10.1103/PhysRevLett.44.912}{\emph{Phys. Rev. Lett.}
  {\bf 44} (1980) 912}.

\bibitem{Schechter:1980gr}
J.~Schechter and J.~W.~F. Valle, \emph{{Neutrino Masses in SU(2) x U(1)
  Theories}}, \href{http://dx.doi.org/10.1103/PhysRevD.22.2227}{\emph{Phys.
  Rev. D} {\bf 22} (1980) 2227}.

\bibitem{Drewes:2013gca}
M.~Drewes, \emph{{The Phenomenology of Right Handed Neutrinos}},
  \href{http://dx.doi.org/10.1142/S0218301313300191}{\emph{Int. J. Mod. Phys.
  E} {\bf 22} (2013) 1330019}, [\href{http://arxiv.org/abs/1303.6912}{{\tt
  1303.6912}}].

\bibitem{Dasgupta:2021ies}
B.~Dasgupta and J.~Kopp, \emph{{Sterile Neutrinos}},
  \href{http://dx.doi.org/10.1016/j.physrep.2021.06.002}{\emph{Phys. Rept.}
  {\bf 928} (2021) 1--63}, [\href{http://arxiv.org/abs/2106.05913}{{\tt
  2106.05913}}].

\bibitem{Abdullahi:2022jlv}
A.~M. Abdullahi et~al., \emph{{The Present and Future Status of Heavy Neutral
  Leptons}},  in \emph{{2022 Snowmass Summer Study}}, 3, 2022.
\newblock \href{http://arxiv.org/abs/2203.08039}{{\tt 2203.08039}}.

\bibitem{Boyarsky:2018tvu}
A.~Boyarsky, M.~Drewes, T.~Lasserre, S.~Mertens and O.~Ruchayskiy,
  \emph{{Sterile neutrino Dark Matter}},
  \href{http://dx.doi.org/10.1016/j.ppnp.2018.07.004}{\emph{Prog. Part. Nucl.
  Phys.} {\bf 104} (2019) 1--45}, [\href{http://arxiv.org/abs/1807.07938}{{\tt
  1807.07938}}].

\bibitem{Asaka:2005pn}
T.~Asaka and M.~Shaposhnikov, \emph{{The $\nu$MSM, dark matter and baryon
  asymmetry of the universe}},
  \href{http://dx.doi.org/10.1016/j.physletb.2005.06.020}{\emph{Phys. Lett. B}
  {\bf 620} (2005) 17--26}, [\href{http://arxiv.org/abs/hep-ph/0505013}{{\tt
  hep-ph/0505013}}].

\bibitem{Pati:1974yy}
J.~C. Pati and A.~Salam, \emph{{Lepton Number as the Fourth Color}},
  \href{http://dx.doi.org/10.1103/PhysRevD.10.275,
  10.1103/PhysRevD.11.703.2}{\emph{Phys. Rev.} {\bf D10} (1974) 275--289}.

\bibitem{Mohapatra:1974gc}
R.~Mohapatra and J.~C. Pati, \emph{{A Natural Left-Right Symmetry}},
  \href{http://dx.doi.org/10.1103/PhysRevD.11.2558}{\emph{Phys. Rev. D} {\bf
  11} (1975) 2558}.

\bibitem{Senjanovic:1975rk}
G.~Senjanovic and R.~N. Mohapatra, \emph{{Exact Left-Right Symmetry and
  Spontaneous Violation of Parity}},
  \href{http://dx.doi.org/10.1103/PhysRevD.12.1502}{\emph{Phys. Rev. D} {\bf
  12} (1975) 1502}.

\bibitem{Mohapatra:1980yp}
R.~N. Mohapatra and G.~Senjanovic, \emph{{Neutrino Masses and Mixings in Gauge
  Models with Spontaneous Parity Violation}},
  \href{http://dx.doi.org/10.1103/PhysRevD.23.165}{\emph{Phys. Rev.} {\bf D23}
  (1981) 165}.

\bibitem{ATLAS:2018dcj}
{\scshape ATLAS} collaboration, M.~Aaboud et~al., \emph{{Search for heavy
  Majorana or Dirac neutrinos and right-handed $W$ gauge bosons in final states
  with two charged leptons and two jets at $ \sqrt{s}=13 $ TeV with the ATLAS
  detector}}, \href{http://dx.doi.org/10.1007/JHEP01(2019)016}{\emph{JHEP} {\bf
  01} (2019) 016}, [\href{http://arxiv.org/abs/1809.11105}{{\tt 1809.11105}}].

\bibitem{ATLAS:2019isd}
{\scshape ATLAS} collaboration, M.~Aaboud et~al., \emph{{Search for a
  right-handed gauge boson decaying into a high-momentum heavy neutrino and a
  charged lepton in $pp$ collisions with the ATLAS detector at $\sqrt{s}=13$
  TeV}}, \href{http://dx.doi.org/10.1016/j.physletb.2019.134942}{\emph{Phys.
  Lett. B} {\bf 798} (2019) 134942},
  [\href{http://arxiv.org/abs/1904.12679}{{\tt 1904.12679}}].

\bibitem{CMS:2018agk}
{\scshape CMS} collaboration, A.~M. Sirunyan et~al., \emph{{Search for a heavy
  right-handed W boson and a heavy neutrino in events with two same-flavor
  leptons and two jets at $\sqrt{s}=$ 13 TeV}},
  \href{http://dx.doi.org/10.1007/JHEP05(2018)148}{\emph{JHEP} {\bf 05} (2018)
  148}, [\href{http://arxiv.org/abs/1803.11116}{{\tt 1803.11116}}].

\bibitem{Li:2020wxi}
T.~Li, X.-D. Ma and M.~A. Schmidt, \emph{{Constraints on the charged currents
  in general neutrino interactions with sterile neutrinos}},
  \href{http://dx.doi.org/10.1007/JHEP10(2020)115}{\emph{JHEP} {\bf 10} (2020)
  115}, [\href{http://arxiv.org/abs/2007.15408}{{\tt 2007.15408}}].

\bibitem{Bamert:1994qh}
P.~Bamert, C.~P. Burgess and R.~N. Mohapatra, \emph{{Heavy sterile neutrinos
  and neutrinoless double beta decay}},
  \href{http://dx.doi.org/10.1016/0550-3213(94)00575-Y}{\emph{Nucl. Phys. B}
  {\bf 438} (1995) 3--16}, [\href{http://arxiv.org/abs/hep-ph/9408367}{{\tt
  hep-ph/9408367}}].

\bibitem{Blennow:2010th}
M.~Blennow, E.~Fernandez-Martinez, J.~Lopez-Pavon and J.~Menendez,
  \emph{{Neutrinoless double beta decay in seesaw models}},
  \href{http://dx.doi.org/10.1007/JHEP07(2010)096}{\emph{JHEP} {\bf 07} (2010)
  096}, [\href{http://arxiv.org/abs/1005.3240}{{\tt 1005.3240}}].

\bibitem{Mitra:2011qr}
M.~Mitra, G.~Senjanovic and F.~Vissani, \emph{{Neutrinoless Double Beta Decay
  and Heavy Sterile Neutrinos}},
  \href{http://dx.doi.org/10.1016/j.nuclphysb.2011.10.035}{\emph{Nucl. Phys. B}
  {\bf 856} (2012) 26--73}, [\href{http://arxiv.org/abs/1108.0004}{{\tt
  1108.0004}}].

\bibitem{deGouvea:2011zz}
A.~de~Gouvea and W.-C. Huang, \emph{{Constraining the (Low-Energy) Type-I
  Seesaw}}, \href{http://dx.doi.org/10.1103/PhysRevD.85.053006}{\emph{Phys.
  Rev. D} {\bf 85} (2012) 053006}, [\href{http://arxiv.org/abs/1110.6122}{{\tt
  1110.6122}}].

\bibitem{Barry:2011wb}
J.~Barry, W.~Rodejohann and H.~Zhang, \emph{{Light Sterile Neutrinos: Models
  and Phenomenology}},
  \href{http://dx.doi.org/10.1007/JHEP07(2011)091}{\emph{JHEP} {\bf 07} (2011)
  091}, [\href{http://arxiv.org/abs/1105.3911}{{\tt 1105.3911}}].

\bibitem{Li:2011ss}
Y.~F. Li and S.-s. Liu, \emph{{Vanishing effective mass of the neutrinoless
  double beta decay including light sterile neutrinos}},
  \href{http://dx.doi.org/10.1016/j.physletb.2011.11.054}{\emph{Phys. Lett. B}
  {\bf 706} (2012) 406--411}, [\href{http://arxiv.org/abs/1110.5795}{{\tt
  1110.5795}}].

\bibitem{Ghosh:2013nya}
M.~Ghosh, S.~Goswami, S.~Gupta and C.~S. Kim, \emph{{Implication of a vanishing
  element in the 3+1 scenario}},
  \href{http://dx.doi.org/10.1103/PhysRevD.88.033009}{\emph{Phys. Rev. D} {\bf
  88} (2013) 033009}, [\href{http://arxiv.org/abs/1305.0180}{{\tt 1305.0180}}].

\bibitem{Girardi:2013zra}
I.~Girardi, A.~Meroni and S.~T. Petcov, \emph{{Neutrinoless Double Beta Decay
  in the Presence of Light Sterile Neutrinos}},
  \href{http://dx.doi.org/10.1007/JHEP11(2013)146}{\emph{JHEP} {\bf 11} (2013)
  146}, [\href{http://arxiv.org/abs/1308.5802}{{\tt 1308.5802}}].

\bibitem{Barea:2015zfa}
J.~Barea, J.~Kotila and F.~Iachello, \emph{{Limits on sterile neutrino
  contributions to neutrinoless double beta decay}},
  \href{http://dx.doi.org/10.1103/PhysRevD.92.093001}{\emph{Phys. Rev. D} {\bf
  92} (2015) 093001}, [\href{http://arxiv.org/abs/1509.01925}{{\tt
  1509.01925}}].

\bibitem{Bolton:2019pcu}
P.~D. Bolton, F.~F. Deppisch and P.~S. Bhupal~Dev, \emph{{Neutrinoless double
  beta decay versus other probes of heavy sterile neutrinos}},
  \href{http://dx.doi.org/10.1007/JHEP03(2020)170}{\emph{JHEP} {\bf 03} (2020)
  170}, [\href{http://arxiv.org/abs/1912.03058}{{\tt 1912.03058}}].

\bibitem{Jha:2021oxl}
T.~Jha, S.~Khan, M.~Mitra and A.~Patra, \emph{{Zooming in on eV-MeV Scale
  Sterile Neutrinos in light of Neutrinoless Double Beta Decay}},
  \href{http://arxiv.org/abs/2107.03807}{{\tt 2107.03807}}.

\bibitem{Asaka:2020wfo}
T.~Asaka, H.~Ishida and K.~Tanaka, \emph{{Hiding neutrinoless double beta decay
  in the minimal seesaw mechanism}},
  \href{http://dx.doi.org/10.1103/PhysRevD.103.015014}{\emph{Phys. Rev. D} {\bf
  103} (2021) 015014}, [\href{http://arxiv.org/abs/2012.12564}{{\tt
  2012.12564}}].

\bibitem{Dekens:2020ttz}
W.~Dekens, J.~de~Vries, K.~Fuyuto, E.~Mereghetti and G.~Zhou, \emph{{Sterile
  neutrinos and neutrinoless double beta decay in effective field theory}},
  \href{http://dx.doi.org/10.1007/JHEP06(2020)097}{\emph{JHEP} {\bf 06} (2020)
  097}, [\href{http://arxiv.org/abs/2002.07182}{{\tt 2002.07182}}].

\bibitem{Prezeau:2003xn}
G.~Prezeau, M.~Ramsey-Musolf and P.~Vogel, \emph{{Neutrinoless double beta
  decay and effective field theory}},
  \href{http://dx.doi.org/10.1103/PhysRevD.68.034016}{\emph{Phys. Rev. D} {\bf
  68} (2003) 034016}, [\href{http://arxiv.org/abs/hep-ph/0303205}{{\tt
  hep-ph/0303205}}].

\bibitem{Li:2020flq}
G.~Li, M.~Ramsey-Musolf and J.~C. Vasquez, \emph{{Left-Right Symmetry and
  Leading Contributions to Neutrinoless Double Beta Decay}},
  \href{http://dx.doi.org/10.1103/PhysRevLett.126.151801}{\emph{Phys. Rev.
  Lett.} {\bf 126} (2021) 151801}, [\href{http://arxiv.org/abs/2009.01257}{{\tt
  2009.01257}}].

\bibitem{Cirigliano:2018yza}
V.~Cirigliano, W.~Dekens, J.~de~Vries, M.~L. Graesser and E.~Mereghetti,
  \emph{{A neutrinoless double beta decay master formula from effective field
  theory}}, \href{http://dx.doi.org/10.1007/JHEP12(2018)097}{\emph{JHEP} {\bf
  12} (2018) 097}, [\href{http://arxiv.org/abs/1806.02780}{{\tt 1806.02780}}].

\bibitem{Cirigliano:2017djv}
V.~Cirigliano, W.~Dekens, J.~de~Vries, M.~L. Graesser and E.~Mereghetti,
  \emph{{Neutrinoless double beta decay in chiral effective field theory:
  lepton number violation at dimension seven}},
  \href{http://dx.doi.org/10.1007/JHEP12(2017)082}{\emph{JHEP} {\bf 12} (2017)
  082}, [\href{http://arxiv.org/abs/1708.09390}{{\tt 1708.09390}}].

\bibitem{Kuchimanchi:2014ota}
R.~Kuchimanchi, \emph{{Leptonic CP problem in left-right symmetric model}},
  \href{http://dx.doi.org/10.1103/PhysRevD.91.071901}{\emph{Phys. Rev. D} {\bf
  91} (2015) 071901}, [\href{http://arxiv.org/abs/1408.6382}{{\tt 1408.6382}}].

\bibitem{Senjanovic:2020int}
G.~Senjanovic and V.~Tello, \emph{{Strong CP violation: problem or blessing?}},
   \href{http://arxiv.org/abs/2004.04036}{{\tt 2004.04036}}.

\bibitem{Senjanovic:1978ev}
G.~Senjanovic, \emph{{Spontaneous Breakdown of Parity in a Class of Gauge
  Theories}}, \href{http://dx.doi.org/10.1016/0550-3213(79)90604-7}{\emph{Nucl.
  Phys. B} {\bf 153} (1979) 334--364}.

\bibitem{Maiezza:2010ic}
A.~Maiezza, M.~Nemevsek, F.~Nesti and G.~Senjanovic, \emph{{Left-Right Symmetry
  at LHC}}, \href{http://dx.doi.org/10.1103/PhysRevD.82.055022}{\emph{Phys.
  Rev. D} {\bf 82} (2010) 055022}, [\href{http://arxiv.org/abs/1005.5160}{{\tt
  1005.5160}}].

\bibitem{Maiezza:2014ala}
A.~Maiezza and M.~Nemev\v{s}ek, \emph{{Strong P invariance, neutron electric
  dipole moment, and minimal left-right parity at LHC}},
  \href{http://dx.doi.org/10.1103/PhysRevD.90.095002}{\emph{Phys. Rev. D} {\bf
  90} (2014) 095002}, [\href{http://arxiv.org/abs/1407.3678}{{\tt 1407.3678}}].

\bibitem{Ramsey-Musolf:2020ndm}
M.~J. Ramsey-Musolf and J.~C. Vasquez, \emph{{Left-right symmetry and electric
  dipole moments. A global analysis}},
  \href{http://dx.doi.org/10.1016/j.physletb.2021.136136}{\emph{Phys. Lett. B}
  {\bf 815} (2021) 136136}, [\href{http://arxiv.org/abs/2012.02799}{{\tt
  2012.02799}}].

\bibitem{Deshpande:1990ip}
N.~G. Deshpande, J.~F. Gunion, B.~Kayser and F.~I. Olness, \emph{{Left-right
  symmetric electroweak models with triplet Higgs}},
  \href{http://dx.doi.org/10.1103/PhysRevD.44.837}{\emph{Phys. Rev. D} {\bf 44}
  (1991) 837--858}.

\bibitem{Schechter:1981cv}
J.~Schechter and J.~W.~F. Valle, \emph{{Neutrino Decay and Spontaneous
  Violation of Lepton Number}},
  \href{http://dx.doi.org/10.1103/PhysRevD.25.774}{\emph{Phys. Rev. D} {\bf 25}
  (1982) 774}.

\bibitem{Xing:2011zza}
Z.-z. Xing and S.~Zhou, \emph{{Neutrinos in particle physics, astronomy and
  cosmology}}.
\newblock 2011.

\bibitem{Barry:2013xxa}
J.~Barry and W.~Rodejohann, \emph{{Lepton number and flavour violation in
  TeV-scale left-right symmetric theories with large left-right mixing}},
  \href{http://dx.doi.org/10.1007/JHEP09(2013)153}{\emph{JHEP} {\bf 09} (2013)
  153}, [\href{http://arxiv.org/abs/1303.6324}{{\tt 1303.6324}}].

\bibitem{Senjanovic:2011zz}
G.~Senjanovic, \emph{{Neutrino mass: From LHC to grand unification}},
  \href{http://dx.doi.org/10.1393/ncr/i2011-10061-8}{\emph{Riv. Nuovo Cim.}
  {\bf 34} (2011) 1--68}.

\bibitem{Nemevsek:2012iq}
M.~Nemevsek, G.~Senjanovic and V.~Tello, \emph{{Connecting Dirac and Majorana
  Neutrino Mass Matrices in the Minimal Left-Right Symmetric Model}},
  \href{http://dx.doi.org/10.1103/PhysRevLett.110.151802}{\emph{Phys. Rev.
  Lett.} {\bf 110} (2013) 151802}, [\href{http://arxiv.org/abs/1211.2837}{{\tt
  1211.2837}}].

\bibitem{Senjanovic:2016vxw}
G.~Senjanovi\'c and V.~Tello, \emph{{Probing Seesaw with Parity Restoration}},
  \href{http://dx.doi.org/10.1103/PhysRevLett.119.201803}{\emph{Phys. Rev.
  Lett.} {\bf 119} (2017) 201803}, [\href{http://arxiv.org/abs/1612.05503}{{\tt
  1612.05503}}].

\bibitem{Korner:1992zk}
J.~G. Korner, A.~Pilaftsis and K.~Schilcher, \emph{{Leptonic CP asymmetries in
  flavor changing H0 decays}},
  \href{http://dx.doi.org/10.1103/PhysRevD.47.1080}{\emph{Phys. Rev. D} {\bf
  47} (1993) 1080--1086}, [\href{http://arxiv.org/abs/hep-ph/9301289}{{\tt
  hep-ph/9301289}}].

\bibitem{Grimus:2000vj}
W.~Grimus and L.~Lavoura, \emph{{The Seesaw mechanism at arbitrary order:
  Disentangling the small scale from the large scale}},
  \href{http://dx.doi.org/10.1088/1126-6708/2000/11/042}{\emph{JHEP} {\bf 11}
  (2000) 042}, [\href{http://arxiv.org/abs/hep-ph/0008179}{{\tt
  hep-ph/0008179}}].

\bibitem{Hettmansperger:2011bt}
H.~Hettmansperger, M.~Lindner and W.~Rodejohann, \emph{{Phenomenological
  Consequences of sub-leading Terms in See-Saw Formulas}},
  \href{http://dx.doi.org/10.1007/JHEP04(2011)123}{\emph{JHEP} {\bf 04} (2011)
  123}, [\href{http://arxiv.org/abs/1102.3432}{{\tt 1102.3432}}].

\bibitem{Fernandez-Martinez:2016lgt}
E.~Fernandez-Martinez, J.~Hernandez-Garcia and J.~Lopez-Pavon, \emph{{Global
  constraints on heavy neutrino mixing}},
  \href{http://dx.doi.org/10.1007/JHEP08(2016)033}{\emph{JHEP} {\bf 08} (2016)
  033}, [\href{http://arxiv.org/abs/1605.08774}{{\tt 1605.08774}}].

\bibitem{Lehman:2014jma}
L.~Lehman, \emph{{Extending the Standard Model Effective Field Theory with the
  Complete Set of Dimension-7 Operators}},
  \href{http://dx.doi.org/10.1103/PhysRevD.90.125023}{\emph{Phys. Rev. D} {\bf
  90} (2014) 125023}, [\href{http://arxiv.org/abs/1410.4193}{{\tt 1410.4193}}].

\bibitem{Liao:2016qyd}
Y.~Liao and X.-D. Ma, \emph{{Operators up to Dimension Seven in Standard Model
  Effective Field Theory Extended with Sterile Neutrinos}},
  \href{http://dx.doi.org/10.1103/PhysRevD.96.015012}{\emph{Phys. Rev. D} {\bf
  96} (2017) 015012}, [\href{http://arxiv.org/abs/1612.04527}{{\tt
  1612.04527}}].

\bibitem{Beall:1981ze}
G.~Beall, M.~Bander and A.~Soni, \emph{{Constraint on the Mass Scale of a
  Left-Right Symmetric Electroweak Theory from the K(L) K(S) Mass Difference}},
  \href{http://dx.doi.org/10.1103/PhysRevLett.48.848}{\emph{Phys. Rev. Lett.}
  {\bf 48} (1982) 848}.

\bibitem{Bertolini:2014sua}
S.~Bertolini, A.~Maiezza and F.~Nesti, \emph{{Present and Future K and B Meson
  Mixing Constraints on TeV Scale Left-Right Symmetry}},
  \href{http://dx.doi.org/10.1103/PhysRevD.89.095028}{\emph{Phys. Rev. D} {\bf
  89} (2014) 095028}, [\href{http://arxiv.org/abs/1403.7112}{{\tt 1403.7112}}].

\bibitem{Dekens:2021bro}
W.~Dekens, L.~Andreoli, J.~de~Vries, E.~Mereghetti and F.~Oosterhof, \emph{{A
  low-energy perspective on the minimal left-right symmetric model}},
  \href{http://dx.doi.org/10.1007/JHEP11(2021)127}{\emph{JHEP} {\bf 11} (2021)
  127}, [\href{http://arxiv.org/abs/2107.10852}{{\tt 2107.10852}}].

\bibitem{Senjanovic:2014pva}
G.~Senjanovi\'c and V.~Tello, \emph{{Right Handed Quark Mixing in Left-Right
  Symmetric Theory}},
  \href{http://dx.doi.org/10.1103/PhysRevLett.114.071801}{\emph{Phys. Rev.
  Lett.} {\bf 114} (2015) 071801}, [\href{http://arxiv.org/abs/1408.3835}{{\tt
  1408.3835}}].

\bibitem{Senjanovic:2015yea}
G.~Senjanovi\'c and V.~Tello, \emph{{Restoration of Parity and the Right-Handed
  Analog of the CKM Matrix}},
  \href{http://dx.doi.org/10.1103/PhysRevD.94.095023}{\emph{Phys. Rev. D} {\bf
  94} (2016) 095023}, [\href{http://arxiv.org/abs/1502.05704}{{\tt
  1502.05704}}].

\bibitem{Horoi:2017gmj}
M.~Horoi and A.~Neacsu, \emph{{Towards an effective field theory approach to
  the neutrinoless double-beta decay}},
  \href{http://arxiv.org/abs/1706.05391}{{\tt 1706.05391}}.

\bibitem{Stoica:2013lka}
S.~Stoica and M.~Mirea, \emph{{New calculations for phase space factors
  involved in double-$\beta$ decay}},
  \href{http://dx.doi.org/10.1103/PhysRevC.88.037303}{\emph{Phys. Rev. C} {\bf
  88} (2013) 037303}, [\href{http://arxiv.org/abs/1307.0290}{{\tt 1307.0290}}].

\bibitem{Cirigliano:2018hja}
V.~Cirigliano, W.~Dekens, J.~De~Vries, M.~L. Graesser, E.~Mereghetti,
  S.~Pastore et~al., \emph{{New Leading Contribution to Neutrinoless
  Double-\ensuremath{\beta} Decay}},
  \href{http://dx.doi.org/10.1103/PhysRevLett.120.202001}{\emph{Phys. Rev.
  Lett.} {\bf 120} (2018) 202001}, [\href{http://arxiv.org/abs/1802.10097}{{\tt
  1802.10097}}].

\bibitem{Hyvarinen:2015bda}
J.~Hyv\"arinen and J.~Suhonen, \emph{{Nuclear matrix elements for
  $0\nu\beta\beta$ decays with light or heavy Majorana-neutrino exchange}},
  \href{http://dx.doi.org/10.1103/PhysRevC.91.024613}{\emph{Phys. Rev. C} {\bf
  91} (2015) 024613}.

\bibitem{Menendez:2017fdf}
J.~Men\'endez, \emph{{Neutrinoless $\beta\beta$ decay mediated by the exchange
  of light and heavy neutrinos: The role of nuclear structure correlations}},
  \href{http://dx.doi.org/10.1088/1361-6471/aa9bd4}{\emph{J. Phys. G} {\bf 45}
  (2018) 014003}, [\href{http://arxiv.org/abs/1804.02105}{{\tt 1804.02105}}].

\bibitem{Barea:2015kwa}
J.~Barea, J.~Kotila and F.~Iachello, \emph{{$0\nu\beta\beta$ and
  $2\nu\beta\beta$ nuclear matrix elements in the interacting boson model with
  isospin restoration}},
  \href{http://dx.doi.org/10.1103/PhysRevC.91.034304}{\emph{Phys. Rev. C} {\bf
  91} (2015) 034304}, [\href{http://arxiv.org/abs/1506.08530}{{\tt
  1506.08530}}].

\bibitem{Cirigliano:2019vdj}
V.~Cirigliano, W.~Dekens, J.~De~Vries, M.~L. Graesser, E.~Mereghetti,
  S.~Pastore et~al., \emph{{Renormalized approach to neutrinoless double-
  $\beta$ decay}},
  \href{http://dx.doi.org/10.1103/PhysRevC.100.055504}{\emph{Phys. Rev. C} {\bf
  100} (2019) 055504}, [\href{http://arxiv.org/abs/1907.11254}{{\tt
  1907.11254}}].

\bibitem{Richardson:2021xiu}
T.~R. Richardson, M.~R. Schindler, S.~Pastore and R.~P. Springer,
  \emph{{Large-$N_c$ analysis of two-nucleon neutrinoless double-$\beta$ decay
  and charge-independence-breaking contact terms}},
  \href{http://dx.doi.org/10.1103/PhysRevC.103.055501}{\emph{Phys. Rev. C} {\bf
  103} (2021) 055501}, [\href{http://arxiv.org/abs/2102.02184}{{\tt
  2102.02184}}].

\bibitem{Cirigliano:2020dmx}
V.~Cirigliano, W.~Dekens, J.~de~Vries, M.~Hoferichter and E.~Mereghetti,
  \emph{{Toward Complete Leading-Order Predictions for Neutrinoless Double
  $\beta$ Decay}},
  \href{http://dx.doi.org/10.1103/PhysRevLett.126.172002}{\emph{Phys. Rev.
  Lett.} {\bf 126} (2021) 172002}, [\href{http://arxiv.org/abs/2012.11602}{{\tt
  2012.11602}}].

\bibitem{Cirigliano:2021qko}
V.~Cirigliano, W.~Dekens, J.~de~Vries, M.~Hoferichter and E.~Mereghetti,
  \emph{{Determining the leading-order contact term in neutrinoless double
  $\beta$ decay}}, \href{http://dx.doi.org/10.1007/JHEP05(2021)289}{\emph{JHEP}
  {\bf 05} (2021) 289}, [\href{http://arxiv.org/abs/2102.03371}{{\tt
  2102.03371}}].

\bibitem{Wirth:2021pij}
R.~Wirth, J.~M. Yao and H.~Hergert, \emph{{Ab~Initio Calculation of the Contact
  Operator Contribution in the Standard Mechanism for Neutrinoless Double Beta
  Decay}}, \href{http://dx.doi.org/10.1103/PhysRevLett.127.242502}{\emph{Phys.
  Rev. Lett.} {\bf 127} (2021) 242502},
  [\href{http://arxiv.org/abs/2105.05415}{{\tt 2105.05415}}].

\bibitem{Jokiniemi:2021qqv}
L.~Jokiniemi, P.~Soriano and J.~Men\'endez, \emph{{Impact of the leading-order
  short-range nuclear matrix element on the neutrinoless double-beta decay of
  medium-mass and heavy nuclei}},
  \href{http://dx.doi.org/10.1016/j.physletb.2021.136720}{\emph{Phys. Lett. B}
  {\bf 823} (2021) 136720}, [\href{http://arxiv.org/abs/2107.13354}{{\tt
  2107.13354}}].

\bibitem{Nicholson:2018mwc}
A.~Nicholson et~al., \emph{{Heavy physics contributions to neutrinoless double
  beta decay from QCD}},
  \href{http://dx.doi.org/10.1103/PhysRevLett.121.172501}{\emph{Phys. Rev.
  Lett.} {\bf 121} (2018) 172501}, [\href{http://arxiv.org/abs/1805.02634}{{\tt
  1805.02634}}].

\bibitem{Tuo:2022hft}
X.-Y. Tuo, X.~Feng and L.-C. Jin, \emph{{Lattice QCD calculation of light
  sterile neutrino contribution in $0\nu2\beta$ decay}},
  \href{http://arxiv.org/abs/2206.00879}{{\tt 2206.00879}}.

\bibitem{Graf:2022lhj}
L.~Gr\'af, M.~Lindner and O.~Scholer, \emph{{Unraveling the
  0\ensuremath{\nu}\ensuremath{\beta}\ensuremath{\beta} decay mechanisms}},
  \href{http://dx.doi.org/10.1103/PhysRevD.106.035022}{\emph{Phys. Rev. D} {\bf
  106} (2022) 035022}, [\href{http://arxiv.org/abs/2204.10845}{{\tt
  2204.10845}}].

\bibitem{Bertolini:2019out}
S.~Bertolini, A.~Maiezza and F.~Nesti, \emph{{Kaon CP violation and neutron EDM
  in the minimal left-right symmetric model}},
  \href{http://dx.doi.org/10.1103/PhysRevD.101.035036}{\emph{Phys. Rev. D} {\bf
  101} (2020) 035036}, [\href{http://arxiv.org/abs/1911.09472}{{\tt
  1911.09472}}].

\bibitem{Nemevsek:2018bbt}
M.~Nemev\v{s}ek, F.~Nesti and G.~Popara, \emph{{Keung-Senjanovi\'c process at
  the LHC: From lepton number violation to displaced vertices to invisible
  decays}}, \href{http://dx.doi.org/10.1103/PhysRevD.97.115018}{\emph{Phys.
  Rev. D} {\bf 97} (2018) 115018}, [\href{http://arxiv.org/abs/1801.05813}{{\tt
  1801.05813}}].

\bibitem{ATLAS:2017jbq}
{\scshape ATLAS} collaboration, M.~Aaboud et~al., \emph{{Search for a new heavy
  gauge boson resonance decaying into a lepton and missing transverse momentum
  in 36 fb$^{-1}$ of $pp$ collisions at $\sqrt{s} =$ 13 TeV with the ATLAS
  experiment}},
  \href{http://dx.doi.org/10.1140/epjc/s10052-018-5877-y}{\emph{Eur. Phys. J.
  C} {\bf 78} (2018) 401}, [\href{http://arxiv.org/abs/1706.04786}{{\tt
  1706.04786}}].

\bibitem{CMS:2022yjm}
{\scshape CMS} collaboration, A.~Tumasyan et~al., \emph{{Search for new physics
  in the lepton plus missing transverse momentum final state in proton-proton
  collisions at $\sqrt{s} =$ 13 TeV}},
  \href{http://dx.doi.org/10.1007/JHEP07(2022)067}{\emph{JHEP} {\bf 07} (2022)
  067}, [\href{http://arxiv.org/abs/2202.06075}{{\tt 2202.06075}}].

\bibitem{Seng:2018yzq}
C.-Y. Seng, M.~Gorchtein, H.~H. Patel and M.~J. Ramsey-Musolf, \emph{{Reduced
  Hadronic Uncertainty in the Determination of $V_{ud}$}},
  \href{http://dx.doi.org/10.1103/PhysRevLett.121.241804}{\emph{Phys. Rev.
  Lett.} {\bf 121} (2018) 241804}, [\href{http://arxiv.org/abs/1807.10197}{{\tt
  1807.10197}}].

\bibitem{Li:2022cuq}
G.~Li, M.~J. Ramsey-Musolf and J.~C. Vasquez, \emph{{Unraveling the left-right
  mixing using 0\ensuremath{\nu}\ensuremath{\beta}\ensuremath{\beta} decay and
  collider probes}},
  \href{http://dx.doi.org/10.1103/PhysRevD.105.115021}{\emph{Phys. Rev. D} {\bf
  105} (2022) 115021}, [\href{http://arxiv.org/abs/2202.01789}{{\tt
  2202.01789}}].

\bibitem{Czakon:1999ga}
M.~Czakon, J.~Gluza and M.~Zralek, \emph{{Low-energy physics and left-right
  symmetry: Bounds on the model parameters}},
  \href{http://dx.doi.org/10.1016/S0370-2693(99)00567-5}{\emph{Phys. Lett. B}
  {\bf 458} (1999) 355--360}, [\href{http://arxiv.org/abs/hep-ph/9904216}{{\tt
  hep-ph/9904216}}].

\bibitem{Nemevsek:2011aa}
M.~Nemevsek, F.~Nesti, G.~Senjanovic and V.~Tello, \emph{{Neutrinoless Double
  Beta Decay: Low Left-Right Symmetry Scale?}},
  \href{http://arxiv.org/abs/1112.3061}{{\tt 1112.3061}}.

\bibitem{Sabti:2020yrt}
N.~Sabti, A.~Magalich and A.~Filimonova, \emph{{An Extended Analysis of Heavy
  Neutral Leptons during Big Bang Nucleosynthesis}},
  \href{http://dx.doi.org/10.1088/1475-7516/2020/11/056}{\emph{JCAP} {\bf 11}
  (2020) 056}, [\href{http://arxiv.org/abs/2006.07387}{{\tt 2006.07387}}].

\bibitem{Boyarsky:2020dzc}
A.~Boyarsky, M.~Ovchynnikov, O.~Ruchayskiy and V.~Syvolap, \emph{{Improved big
  bang nucleosynthesis constraints on heavy neutral leptons}},
  \href{http://dx.doi.org/10.1103/PhysRevD.104.023517}{\emph{Phys. Rev. D} {\bf
  104} (2021) 023517}, [\href{http://arxiv.org/abs/2008.00749}{{\tt
  2008.00749}}].

\bibitem{Carpentier:2010ue}
M.~Carpentier and S.~Davidson, \emph{{Constraints on two-lepton, two quark
  operators}},
  \href{http://dx.doi.org/10.1140/epjc/s10052-010-1482-4}{\emph{Eur. Phys. J.
  C} {\bf 70} (2010) 1071--1090}, [\href{http://arxiv.org/abs/1008.0280}{{\tt
  1008.0280}}].

\bibitem{Barbieri:1988av}
R.~Barbieri and R.~N. Mohapatra, \emph{{Limits on Right-handed Interactions
  From {SN1987A} Observations}},
  \href{http://dx.doi.org/10.1103/PhysRevD.39.1229}{\emph{Phys. Rev. D} {\bf
  39} (1989) 1229}.

\bibitem{CMS:2017fhs}
{\scshape CMS} collaboration, A.~M. Sirunyan et~al., \emph{{Observation of
  electroweak production of same-sign W boson pairs in the two jet and two
  same-sign lepton final state in proton-proton collisions at $\sqrt{s} = $ 13
  TeV}}, \href{http://dx.doi.org/10.1103/PhysRevLett.120.081801}{\emph{Phys.
  Rev. Lett.} {\bf 120} (2018) 081801},
  [\href{http://arxiv.org/abs/1709.05822}{{\tt 1709.05822}}].

\bibitem{ATLAS:2017xqs}
{\scshape ATLAS} collaboration, M.~Aaboud et~al., \emph{{Search for doubly
  charged Higgs boson production in multi-lepton final states with the ATLAS
  detector using proton\textendash{}proton collisions at $\sqrt{s}=13\,\text
  {TeV}$}}, \href{http://dx.doi.org/10.1140/epjc/s10052-018-5661-z}{\emph{Eur.
  Phys. J. C} {\bf 78} (2018) 199},
  [\href{http://arxiv.org/abs/1710.09748}{{\tt 1710.09748}}].

\bibitem{ATLAS:2021jol}
{\scshape ATLAS} collaboration, G.~Aad et~al., \emph{{Search for doubly and
  singly charged Higgs bosons decaying into vector bosons in multi-lepton final
  states with the ATLAS detector using proton-proton collisions at $
  \sqrt{\mathrm{s}} $ = 13 TeV}},
  \href{http://dx.doi.org/10.1007/JHEP06(2021)146}{\emph{JHEP} {\bf 06} (2021)
  146}, [\href{http://arxiv.org/abs/2101.11961}{{\tt 2101.11961}}].

\bibitem{Dev:2018sel}
P.~S.~B. Dev, M.~J. Ramsey-Musolf and Y.~Zhang, \emph{{Doubly-Charged Scalars
  in the Type-II Seesaw Mechanism: Fundamental Symmetry Tests and High-Energy
  Searches}}, \href{http://dx.doi.org/10.1103/PhysRevD.98.055013}{\emph{Phys.
  Rev. D} {\bf 98} (2018) 055013}, [\href{http://arxiv.org/abs/1806.08499}{{\tt
  1806.08499}}].

\bibitem{Maiezza:2016ybz}
A.~Maiezza, G.~Senjanovi\'c and J.~C. Vasquez, \emph{{Higgs sector of the
  minimal left-right symmetric theory}},
  \href{http://dx.doi.org/10.1103/PhysRevD.95.095004}{\emph{Phys. Rev. D} {\bf
  95} (2017) 095004}, [\href{http://arxiv.org/abs/1612.09146}{{\tt
  1612.09146}}].

\bibitem{Esteban:2020cvm}
I.~Esteban, M.~C. Gonzalez-Garcia, M.~Maltoni, T.~Schwetz and A.~Zhou,
  \emph{{The fate of hints: updated global analysis of three-flavor neutrino
  oscillations}}, \href{http://dx.doi.org/10.1007/JHEP09(2020)178}{\emph{JHEP}
  {\bf 09} (2020) 178}, [\href{http://arxiv.org/abs/2007.14792}{{\tt
  2007.14792}}].

\bibitem{Joshipura:2001ui}
A.~S. Joshipura, E.~A. Paschos and W.~Rodejohann, \emph{{A Simple connection
  between neutrino oscillation and leptogenesis}},
  \href{http://dx.doi.org/10.1088/1126-6708/2001/08/029}{\emph{JHEP} {\bf 08}
  (2001) 029}, [\href{http://arxiv.org/abs/hep-ph/0105175}{{\tt
  hep-ph/0105175}}].

\bibitem{Halprin:1983ez}
A.~Halprin, S.~T. Petcov and S.~P. Rosen, \emph{{Effects of Light and Heavy
  Majorana Neutrinos in Neutrinoless Double Beta Decay}},
  \href{http://dx.doi.org/10.1016/0370-2693(83)91296-0}{\emph{Phys. Lett. B}
  {\bf 125} (1983) 335--338}.

\bibitem{Leung:1984vy}
C.~N. Leung and S.~T. Petcov, \emph{{On the Possibility of Destructive
  Interference Between Light and Heavy Majorana Neutrinos in Neutrinoless
  Double beta Decay}},
  \href{http://dx.doi.org/10.1016/0370-2693(84)90071-6}{\emph{Phys. Lett. B}
  {\bf 145} (1984) 416--420}.

\bibitem{Xing:2009ce}
Z.-z. Xing, \emph{{Low-energy limits on heavy Majorana neutrino masses from the
  neutrinoless double-beta decay and non-unitary neutrino mixing}},
  \href{http://dx.doi.org/10.1016/j.physletb.2009.07.040}{\emph{Phys. Lett. B}
  {\bf 679} (2009) 255--259}, [\href{http://arxiv.org/abs/0907.3014}{{\tt
  0907.3014}}].

\bibitem{Doi:1982dn}
M.~Doi, T.~Kotani, H.~Nishiura and E.~Takasugi, \emph{{DOUBLE BETA DECAY}},
  \href{http://dx.doi.org/10.1143/PTP.69.602}{\emph{Prog. Theor. Phys.} {\bf
  69} (1983) 602}.

\bibitem{Vergados:1985pq}
J.~D. Vergados, \emph{{The Neutrino Mass and Family, Lepton and Baryon
  Nonconservation in Gauge Theories}},
  \href{http://dx.doi.org/10.1016/0370-1573(86)90088-8}{\emph{Phys. Rept.} {\bf
  133} (1986) 1}.

\bibitem{Pantis:1996py}
G.~Pantis, F.~Simkovic, J.~D. Vergados and A.~Faessler, \emph{{Neutrinoless
  double beta decay within QRPA with proton - neutron pairing}},
  \href{http://dx.doi.org/10.1103/PhysRevC.53.695}{\emph{Phys. Rev. C} {\bf 53}
  (1996) 695--707}, [\href{http://arxiv.org/abs/nucl-th/9612036}{{\tt
  nucl-th/9612036}}].

\bibitem{Suhonen:1998ck}
J.~Suhonen and O.~Civitarese, \emph{{Weak-interaction and nuclear-structure
  aspects of nuclear double beta decay}},
  \href{http://dx.doi.org/10.1016/S0370-1573(97)00087-2}{\emph{Phys. Rept.}
  {\bf 300} (1998) 123--214}.

\bibitem{Hirsch:1996qw}
M.~Hirsch, H.~V. Klapdor-Kleingrothaus and O.~Panella, \emph{{Double beta decay
  in left-right symmetric models}},
  \href{http://dx.doi.org/10.1016/0370-2693(96)00185-2}{\emph{Phys. Lett. B}
  {\bf 374} (1996) 7--12}, [\href{http://arxiv.org/abs/hep-ph/9602306}{{\tt
  hep-ph/9602306}}].

\bibitem{Chakrabortty:2012mh}
J.~Chakrabortty, H.~Z. Devi, S.~Goswami and S.~Patra, \emph{{Neutrinoless
  double-$\beta$ decay in TeV scale Left-Right symmetric models}},
  \href{http://dx.doi.org/10.1007/JHEP08(2012)008}{\emph{JHEP} {\bf 08} (2012)
  008}, [\href{http://arxiv.org/abs/1204.2527}{{\tt 1204.2527}}].

\bibitem{Dev:2014xea}
P.~S. Bhupal~Dev, S.~Goswami and M.~Mitra, \emph{{TeV Scale Left-Right Symmetry
  and Large Mixing Effects in Neutrinoless Double Beta Decay}},
  \href{http://dx.doi.org/10.1103/PhysRevD.91.113004}{\emph{Phys. Rev. D} {\bf
  91} (2015) 113004}, [\href{http://arxiv.org/abs/1405.1399}{{\tt 1405.1399}}].

\bibitem{Stefanik:2015twa}
D.~Stefanik, R.~Dvornicky, F.~Simkovic and P.~Vogel, \emph{{Reexamining the
  light neutrino exchange mechanism of the $0\nu\beta\beta$ decay with left-
  and right-handed leptonic and hadronic currents}},
  \href{http://dx.doi.org/10.1103/PhysRevC.92.055502}{\emph{Phys. Rev. C} {\bf
  92} (2015) 055502}, [\href{http://arxiv.org/abs/1506.07145}{{\tt
  1506.07145}}].

\bibitem{Yang:2021ueh}
J.~L. Yang, C.-H. Chang and T.-F. Feng, \emph{{Nuclear
  0\ensuremath{\nu}2\ensuremath{\beta} decays in B-L symmetric SUSY model and
  in TeV scale left\textendash{}right symmetric model}},
  \href{http://dx.doi.org/10.1088/1572-9494/ac7781}{\emph{Commun. Theor. Phys.}
  {\bf 74} (2022) 085202}, [\href{http://arxiv.org/abs/2107.01367}{{\tt
  2107.01367}}].

\bibitem{Alves:2022yav}
G.~F.~S. Alves, C.~S. Fong, L.~P.~S. Leal and R.~Z. Funchal, \emph{{Exploring
  the Neutrino Sector of the Minimal Left-Right Symmetric Model}},
  \href{http://arxiv.org/abs/2208.07378}{{\tt 2208.07378}}.

\end{thebibliography}\endgroup

\end{document}